\documentclass[a4paper,11pt]{article}
\pdfoutput=1 
\usepackage{jheppub} 
\usepackage[T1]{fontenc} 
\usepackage{amsmath,amssymb,amsfonts,cases}
\usepackage{amsthm}
\usepackage{enumitem}
\usepackage{adjustbox}
\usepackage{graphicx,caption,subcaption}
\usepackage[T1]{fontenc} 
\usepackage{enumitem}  
\usepackage{empheq}
\usepackage{dsfont}
\usepackage{hyperref}
\usepackage[mathscr]{euscript}
\usepackage{dcolumn}
\usepackage{bm}
\usepackage{bbm}
\usepackage{slashed}
\usepackage{wrapfig}
\usepackage{mathtools}
\usepackage{tikz}

\newcommand{\vol}[1]{\text{vol}(#1)~}
\newcommand{\beq}{\begin{equation}}
\newcommand{\eeq}{\end{equation}}
\newcommand{\ba}{\begin{align}}
\newcommand{\ea}{\end{align}}
\newcommand{\bea}{\begin{eqnarray}}
\newcommand{\eea}{\end{eqnarray}}
\newcommand{\bi}{\begin{itemize}}
\newcommand{\ei}{\end{itemize}}
\newcommand{\ben}{\begin{enumerate}}
\newcommand{\een}{\end{enumerate}}

\def\Tr{{\rm Tr}}

\renewcommand{\d}{\partial}

\newcommand{\braketbis}[1]{\left< #1 \right>}

\newcommand{\rhop}{\rho^\prime}

\renewcommand{\a}{\alpha}

\renewcommand{\d}{\delta}
\newcommand{\e}{\epsilon}

\renewcommand{\r}{\rho}

\def\a{\alpha}

\def\d{\delta}

\def\r{\rho}                                     

\newcommand{\Delslash}{\nabla\!\!\!\!\slash\,}
\newcommand{\Dslash}{D\!\!\!\!\slash\,}

\newcommand{\eps}{\varepsilon}

\newcommand\eq[1]{eq.~(\ref{eq:#1})}
\newcommand\tbl[1]{table~\ref{tab:#1}}
\newcommand{\sn}[1]{section~\ref{sec:#1}}
\newcommand{\fig}[1]{figure~\ref{fig:#1}}
\newcommand{\App}[1]{appendix~\ref{app:#1}}

\newcommand\mc[1]{{\mathcal{#1}}}

\newcommand\CD{{\mc{D}}}

\newcommand\CM{{\mc{M}}}
\newcommand\CN{{\mc{N}}}
\newcommand\CO{{\mc{O}}}

\newcommand\CR{{\mc{R}}}

\newcommand\vphi{\varphi}

\newcommand\vsig{\varsigma}

\newcommand\II{{\rm I\hspace{-0.02cm} I}}
\newcommand\oII{\mathring{\II}}

\definecolor{cardinal}{rgb}{0.6,0,0}
\definecolor{darkgreen}{rgb}{0,0.5,0}
\definecolor{golden}{rgb}{0.92, 0.7, 0}
\definecolor{midnight}{rgb}{0, 0, 0.5}
\definecolor{darkblue}{rgb}{0.2, 0, 0.8}

\allowdisplaybreaks[4]

\title{\LARGE Monodromy Defects in Free Field Theories}  

\preprint{UUITP- 16/21} 
   
\author[a,b]{Lorenzo Bianchi,}
\author[c]{Adam Chalabi,}
\author[d]{Vladim\'ir Proch\'azka,}
\author[e]{Brandon Robinson,}
\author[c]{and Jacopo Sisti}

\affiliation[a]{Universit\`a di Torino, Dipartimento di Fisica,
Via P. Giuria 1, I-10125 Torino, Italy}
\affiliation[b]{I.N.F.N. - sezione di Torino,
Via P. Giuria 1, I-10125 Torino, Italy}
\affiliation[c]{STAG Research Centre, Physics and Astronomy, University of Southampton, Highfield, Southampton SO17 1BJ, UK}
\affiliation[d]{Department of Physics and Astronomy, Uppsala University, Box 516, SE-75120, Uppsala, Sweden}
\affiliation[e]{Instituut voor Theoretische Fysica, K.U. Leuven, Celestijnenlaan 200D, BE-3001 Leuven, Belgium}
\emailAdd{lorenzo.bianchi@unito.it}
\emailAdd{a.chalabi@soton.ac.uk}
\emailAdd{vladimir.prochazka@physics.uu.se}
\emailAdd{brandon.robinson@kuleuven.be}
\emailAdd{j.sisti@soton.ac.uk}

\abstract{We study co-dimension two monodromy defects in theories of conformally coupled scalars and free Dirac fermions in arbitrary $d$ dimensions.  We characterise this family of conformal defects by computing the one-point functions of the stress-tensor and conserved current for Abelian flavour symmetries as well as two-point functions of the displacement operator.  In the case of $d=4$, the normalisation of these correlation functions are related to defect Weyl anomaly coefficients, and thus provide crucial information about the defect conformal field theory.  We provide explicit checks on the values of the defect central charges by calculating the universal part of the defect contribution to entanglement entropy, and further, we use our results to extract the universal part of the vacuum R\'enyi entropy. Moreover, we leverage the non-supersymmetric free field results to compute a novel defect Weyl anomaly coefficient in a $d=4$ theory of free $\CN=2$ hypermultiplets. Including singular modes in the defect operator product expansion of fundamental fields, we identify notable relevant deformations in the singular defect theories and show that they trigger a renormalisation group flow towards an IR fixed point with the most regular defect OPE.  We also study Gukov-Witten defects in free $d=4$ Maxwell theory and show that their central charges vanish. }

\begin{document}
\maketitle
\flushbottom

\section{Introduction}\label{sec:introduction}

Quantum field theories (QFTs) allow for deformations by extended operators, called defects, which enrich the dynamics and extend the algebra of local operators. Defects play an essential role in understanding the complete spectra of QFTs, and may provide a basis for a more robust classification scheme of QFTs and phases of matter. Despite defects in generic QFTs remaining poorly characterised, much progress can be made by imposing restrictive symmetries. A relatively well-understood class of highly-symmetric theories are conformal field theories (CFTs). The study of conformal defects in CFT (dCFT) has seen tremendous progress in recent years in a wide range of contexts (see e.g.~\cite{Andrei:2018die} for a survey).

CFTs in $d$ dimensions naturally arise as the endpoints of renormalisation group (RG) flows. In the Wilsonian picture, RG flows are associated with a coarse-graining of the ultraviolet (UV) degrees of freedom along the flow to the infrared (IR). One can define and compute certain observables called central charges that quantify the degrees of freedom in a CFT.  For a CFT in even dimensions, central charges can be defined through the coefficients in the Weyl anomaly, i.e. the trace anomaly of the stress tensor.  Even though there is no Weyl anomaly for a CFT in odd dimensions, a central charge is identified with the sphere free energy, $F$.  In both even and odd dimensions, the central charges of unitary CFTs obey positivity constraints. Moreover, the universal part of the sphere free energy is expected to decrease under RG flows triggered by relevant deformations, obeying $c$-theorems in even dimensions~\cite{Zamolodchikov:1986gt, Cardy:1988cwa, Komargodski:2011vj} and $F$-theorems in odd dimensions~\cite{Jafferis:2011zi, Klebanov:2011gs}. Hence, it is understood to provide a count of degrees of freedom of the CFT.  In the special case of $d=2$, positivity follows from ground state normalisability, and a strong version of the $c$-theorem -- where the central charge $c$ follows a monotonic, gradient descent along the flow to the IR -- has been proven~\cite{Zamolodchikov:1986gt}. Moreover, the central charges of a CFT appear in physical observables such as correlation functions of the stress-tensor and other conserved currents, thermal entropy~\cite{Cardy:1986ie, Affleck:1986bv}, and entanglement entropy (EE)~\cite{Holzhey:1994we}.

However, dCFTs are quite different. Introducing a $p$-dimensional defect in a Lorentzian $d$-dimensional CFT breaks the ambient $SO(d,2)$ conformal group to at most $SO(p,2) \times SO(d-p)_N \subset SO(d,2)$.  Due to the broken translational symmetries normal to the embedded submanifold supporting the defect, a dCFT does not contain a unique $p$-dimensional spin-2 conserved current, i.e. there is no conserved stress-tensor intrinsic to the defect.  Even so, one can study the bulk stress tensor of the theory in the presence of a defect, and use this information to define central charges through the defect contribution to the trace anomaly~\cite{Deser:1993yx,Graham:1999pm}.  For example, unlike standard $d=2$ CFTs where there is a single central charge $c$, $p=2$ dCFTs have three trace anomaly coefficients labelled $b$, $d_1$, and $d_2$, which play the role of putative central charges.\footnote{If parity is broken along the defect, there are two more independent trace anomaly coefficients \cite{Jensen:2018rxu}.} The absence of a conserved stress tensor makes proving statements about positivity~\cite{Jensen:2018rxu} and $c$-theorems~\cite{Jensen:2015swa,Kobayashi:2018lil, Nishioka:2021uef, Sato:2021eqo,Wang:2020xkc,Wang:2021mdq} for defect central charges more difficult, though not impossible.  This is particularly salient for $p=2$ where one does not generally expect an enhancement from a global $SO(2,2)$ conformal symmetry to a full Virasoro symmetry unless the defect completely decouples or the ambient theory is topological.

In addition to many of the numerical~\cite{Liendo:2012hy,Billo:2013jda,Gaiotto:2013nva} and perturbative~\cite{Soderberg:2017oaa} methods available in ordinary dCFTs, supersymmetry (SUSY) often provides additional powerful analytical tools to compute physical observables that characterise defects, and prove exact results about them~\cite{Drukker:2010jp,Gukov:2014gja,Cordova:2017mhb,Bianchi:2018zpb, Bianchi:2019umv,Bianchi:2019dlw, Chalabi:2020iie, Wang:2020xkc}. These include extremisation principles along defect RG flows~\cite{Wang:2020xkc, Wang:2021mdq} analogous to~\cite{Intriligator:2003jj,Benini:2012cz}.

In the present work, we focus on one of the simplest types of defects that can be introduced into a QFT: a monodromy defect. Monodromy defects can be thought of as surface operators on which co-dimension 1 topological domain walls that implement flavour symmetry rotations can end.  Note that a topological defect on its own does not host any interesting physics in that it does not effect any ambient correlation functions. However, a co-dimension 2 conformal surface defect that is charged under the flavour symmetry is a non-trivial deformation of the theory.\footnote{Here we mean ``non-trivial'' in the colloquial sense, i.e. there are physical quantities that depend on the parameters associated to the defect. This should be distinguished from the technical way that the authors of~\cite{Lauria:2020emq} use the term ``non-trivial'' to mean that a defect in a free field theory does not simply arise as a boundary or singularity condition on ambient fields and so supports ambient-defect couplings consistent with conformal symmetry.} This type of construction of defects in QFTs fits into a larger class of topological defects effecting generalised global symmetry transformations~\cite{Gaiotto:2014kfa}. 

We will study monodromy defects in free field theories. The simple nature of these defects allows us to obtain exact analytic results for certain one- and two-point correlation functions in $d$ ambient dimensions without requiring SUSY. We will, however, briefly comment on and compare our results to the analogous defect in simple superconformal field theories (SCFTs). In the special case where $d=4$, we will be able to directly relate the normalisation of these correlation functions to defect Weyl anomaly coefficients~\cite{Bianchi:2015liz}.  The results of these computations are novel, and they are neatly summarised in the first two rows of \tbl{central-charges}.

\begin{table}[h b]
\small
\begin{adjustbox}{center}
\begin{tabular}{| c || c | c | }
\hline
Free CFT& $b$ & $d_1=d_2$\\\hline\rule{0pt}{1.05em}
Scalar & $\frac{1}{2}(1-\alpha)^2\alpha^2+2\xi\alpha^3+2\tilde\xi (1-\a)^3$ &  $\frac{3}{2}(1-\alpha)^2\alpha^2+6\xi\alpha^3+6\tilde\xi(1-\a)^3$\\\hline\rule{0pt}{1.05em}
Fermion &$\alpha^2(2-\alpha^2 -2\xi(3-2\alpha)) +\xi$ &$3\alpha(1-\alpha)(\alpha(1+\alpha)+2\xi(1-2\alpha))$\\\hline\rule{0pt}{1.05em}
Hyper& $3 \alpha^2 $ & $6\alpha^2$\\\hline
\end{tabular}
\end{adjustbox}
\caption{Central charges of monodromy defects in $d=4$ with parameter $\alpha$ of free scalars, free fermions, and free $\CN=2$ hypermultiplets.  From eqs.~\eqref{eq:d2-h-4d},~\eqref{eq:d1-CD}, and~\eqref{eq:CD-h-q=2}, $d_1=d_2$ is expected for monodromy defects and is verified explicitly in the calculations below. The parameters $\xi$ and $\tilde\xi$ are discussed in detail in \sn{scalar} and take values in the range $\xi,\tilde\xi\in[0,1]$. }
\label{tab:central-charges}
\end{table}

In the last line on the right, we quote known results for the 1/2-BPS monodromy defect in the free $\CN=2$ free hypermultiplet theory.  This theory consists of two complex scalars and one Dirac fermion. The 1/2-BPS monodromy defect is created with respect to a $U(1)$ flavour group under which the two scalars are oppositely charged (see e.g.~\cite{Cordova:2017mhb}). Consistency with SUSY also requires that one of the two scalars has a singular defect operator product expansion (OPE)~\cite{Bianchi:2019sxz}. This eventuality is accounted for in our analysis and therefore, by combining the contribution of the free scalars and free fermions  in \tbl{central-charges}, we can extract the value of the central charges for the free hyper. Specifically, taking the values in \tbl{central-charges} for one complex scalar with $\xi=0, \tilde \xi=0$, another complex scalar for $\xi=1, \, \tilde \xi =0$ and a Dirac fermion with $\xi=0$ gives exactly $6\alpha^2$, which matches the value for $d_1$ for the free hyper reported in~\cite{Bianchi:2019sxz}.\footnote{Note our conventions for both the defect Weyl anomaly and periodicity of $\alpha$ differ slightly from those used in~\cite{Bianchi:2019sxz}.  Their defect Weyl anomaly is normalised with a factor of $1/2\pi$, whereas our is $1/24\pi$.  Their monodromy parameter has range $\alpha\in[0,2)$, whereas we use $\alpha\in[0,1)$.  This results in $d_1^{\rm Here} = 48 d_1^{\rm There}$.} While we do not yet have a direct method for computing $b$ for the monodromy defect in the theory of $d=4$ $\CN=2$ free hypermultiplets, following the logic for $d_1$, we surmise a value of $b=3\alpha^2$ from the non-SUSY free field results. Following from the methods for computing $b$ in non-SUSY cases, we will be led in \sn{discussion} to propose a way to extract $b$ using spectral flow in the chiral algebra description of these $1/2$-BPS monodromy defects \cite{Beem:2013sza,Cordova:2017mhb, Bianchi:2019sxz}.

Due to the simple construction of monodromy defects in free fields, we will be able to utilise the data in \tbl{central-charges} to compute other physical observables.  In particular considering the monodromy defect as a conical singularity in the branched $n$-sheeted cover of the ambient space, we will use our results to calculate and extract the universal part of the R\'enyi entropy for free fields in $d=4$. We will recover previously known expressions for the case of $d=4$ free scalars and fermions \cite{Casini:2010kt,Dowker:2010bu,Klebanov:2011uf,Fursaev:2012mp,Lee:2014xwa,Bianchi:2015liz}. Furthermore, an immediate consequence of our  results is that for free theories the stress tensor one-point function in the presence of a twist operator is related to the displacement two-point function as was conjectured in \cite{Bianchi:2015liz}.\footnote{This relation is known to be violated holographically \cite{Dong:2016wcf, Bianchi:2016xvf, Chu:2016tps} and it is expected to hold only for the case of supersymmetric R\'enyi entropies \cite{Bianchi:2019sxz}.} We will also see that our defect central charges satisfy certain relations amongst themselves \cite{Lewkowycz:2014jia} and the ambient $d=4$ central charges $a_{4d}$ and $c_{4d}$ \cite{Fursaev:2013fta}. 

Even with the ability to efficiently compute all of the central charges for the theories in \tbl{central-charges}, there are still some open questions involving monodromy defects in CFTs.  It has been pointed out in several places recently that monodromy defects in free $\CN=2$ hypermultiplets in $d=4$~\cite{Bianchi:2019sxz} and theories of free scalars in arbitrary $d$~\cite{Giombi:2021uae} have parameters that we label $\xi$ and $\tilde{\xi}$ that are associated with particular terms in the defect OPE.  From the perspective of the ambient theory, $\xi$ and $\tilde \xi$ are parameters that can be seen as controlling mildly singular terms of the mode expansion of the ambient field near the defect. Due to the appearance of $\xi$ (or $\tilde{\xi}$) in the central charge $b$, it cannot be associated with a defect marginal coupling but may be related to a combination of defect relevant or bulk marginal parameters~\cite{Herzog:2019rke,Bianchi:2019umv}.  In this work we will shed some light on the monodromy defect OPE, the role of $\xi$ and $\tilde \xi$ in characterising monodromy defects, and study defect RG flows in the presence of these singular modes.

In particular, we will examine RG flows sourced by relevant defect operators associated to these mildly singular modes in the theories of free scalars and free fermions in arbitrary dimensions.  In the case of free scalars, we will find that regardless of the UV value of $\xi$ or $\tilde \xi$, the IR fixed point of the flow will be a defect with $\xi =\tilde{\xi}=0$.  Using conformal perturbation theory we will be able to compute the $\beta$-function for the defect coupling to all orders in $\alpha$.  For an ambient theory of free fermions, the analysis is more subtle as defects with $\alpha\in(0,\frac{1}{2})$ flow to an IR fixed point with $\xi=0$, while defects with $\alpha \in (\frac{1}{2},1)$ flow to an IR fixed point with a defect having $\xi=1$. Of particular interest in both cases is $d=4$, where we will be able to use the UV and IR values of $\xi$ to demonstrate an explicit, non-trivial test of the $b$-theorem~\cite{Jensen:2015swa}.

We also investigate a closely related defect in pure $d=4$ Maxwell theory, i.e. pure abelian gauge theory: the non-SUSY analogue of the Gukov-Witten defect that was initially studied in $\CN=4$ SUSY Yang-Mills (SYM) theory~\cite{Gukov:2006jk}. As for the monodromy defect, we will also compute the Weyl anomaly coefficients of the Gukov-Witten defect. As can be expected from the fact that Gukov-Witten defects in pure $d=4$ Maxwell theory are known to be topological, we will explicitly show that $b=d_1=d_2=0$.

The structure of the paper is as follows.  In \sn{review}, we review the salient features of dCFTs.  In sections~\ref{sec:scalar} and \ref{sec:fermion}, we will compute one-point functions of the stress tensor and flavour current, and the two-point function of the displacement operator for monodromy defects in free field CFTs in $d$-dimensions. We will then use their data to compute the defect central charges of monodromy defects in $d=4$. In \sn{maxwell}, we will consider Gukov-Witten defects in $d=4$ free Maxwell theory and show that the central charges vanish identically. Lastly, in \sn{rg-flows}, we will study the behaviour of monodromy defects in theories of free scalars and free fermions under defect RG flows.  Some of the details of the computations in these sections are expanded upon in the Appendix.\\

While in the process of finishing the writing of this paper, \cite{Giombi:2021uae} appeared, which has overlap with some of the computations done in \sn{scalar}.  However, the connection of the correlation functions computed in $d=4$ to defect Weyl anomaly coefficients, the discussion of defects with arbitrary values of $\xi,\tilde\xi\in [0,1]$, details of the defect RG flow including the $\beta$-function for the defect coupling, and the resulting comparison to the expectation from the $b$-theorem~\cite{Jensen:2015swa} are novel. 

\section{Background}\label{sec:review}

In this section we review some facts about conformal defects that we will need for our discussion of the monodromy defect in free theories.

\subsection*{Central charges of 2d defects}

Consider an ambient CFT on an arbitrary $d$-dimensional background $\CM_d$ with global conformal symmetry group $SO(2,d)$ (or $SO(1,d+1)$ in Euclidean signature). A conformal defect supported on an immersed co-dimension $q=d-p$ submanifold $\Sigma\hookrightarrow\CM_d$ preserves at most a $SO(2,p)\times SO(q)_N$ subgroup of the ambient conformal symmetry, where $SO(2,p)$ is the global conformal symmetry on a $p$-dimensional manifold and $SO(q)_N$ is the symmetry group of rotations in the directions normal to the defect.  Throughout this work we consider conformal defects of dimension $p=d-2$, i.e. $q=2$. However, we will keep the dependence on $p$ and $q$ explicit in this section.

Consider a dCFT on $\Sigma\hookrightarrow\CM_d$, and let $\{\sigma^a\}$ with $a=1,\ldots,p$ denote the coordinates on $\Sigma$ and $X^\mu (\sigma)$ with $\mu=1,\ldots,d$ denote the embedding functions in the ambient space. The generating functional of connected correlation functions $W[g_{\mu\nu}, X^\mu(\sigma)]\equiv -i \log Z$, where $Z$ is the dCFT partition function, is a function of the ambient metric $g_{\mu\nu}$ and the embedding functions $X^\mu(\sigma)$. Varying $W$ we define two important quantities: the stress tensor $T_{\mu\nu}$ and the displacement operator $\mc{D}_{\mu}$ as
\begin{align}\label{eq:W_variation}
\delta W = \frac{1}{2}\int d^d x\sqrt{|g|} \langle T^{\mu\nu}\rangle \delta g_{\mu\nu} + \int_\Sigma d^p\sigma \sqrt{|\gamma|}\langle \mc{D}_\mu\rangle \delta X^\mu(\sigma)\,,
\end{align}
where $\gamma_{ab}\equiv \partial_a X^\mu(\sigma) \partial_b X^\nu(\sigma) g_{\mu\nu}$ is the induced metric on $\Sigma$. Reparametrisation invariance of $W$ has two implications.  First, invariance under $\sigma^a$ reparametrisations implies that the displacement operator $\mc{D}_\mu$ has no non-trivial components parallel to $\Sigma$ \cite{Billo:2016cpy}.  Second, invariance under $x^\mu$ reparametrisations gives rise to the broken Ward identities for translations normal to the defect 
\begin{equation}
\label{eq:displacement}
\nabla_\mu  T^{\mu i}=  \delta(\Sigma)\mc{D}^i\,,
\end{equation}
where $\nabla$ is the covariant derivative on $\mc{M}_d$, $\delta(\Sigma) \equiv \delta^{(q)}(x_\perp)$ localises to the defect submanifold, and $i=p+1,\ldots,d$ is an index labelling the transverse directions, $x_\perp^i$.

For $p=2$ and $d\geq3$, the defect trace anomaly takes the form
\beq
\label{eq:defecttrace}
T^{\mu}_{~\mu}\Big|_{\Sigma} = - \frac{1}{24\pi} \left(b \, \mathcal{E}_2 + d_1\,\oII^\mu_{ab}\oII_\mu^{ab} - d_2 \, W_{ab}{}^{ab} \right),
\eeq
where $\mc{E}_2 = \mc{R}_\Sigma$ is the 2d Euler density built out of $\gamma$, ${\oII^\mu_{ab} =\II^\mu_{ab} - \frac{1}{2}\gamma_{ab} \gamma^{cd}\II^\mu_{cd} }$ is the trace-free second fundamental form, and $W_{ab}{}^{ab}$ is the trace of the pullback of the ambient Weyl tensor to $\Sigma$.  Note that for the special case of $d=3$, the Weyl tensor vanishes identically, and so the $d_2$ term only exists for $d\geq 4$. 

The coefficients $b$, $d_1$, and $d_2$ in \eq{defecttrace} are often referred to as defect central charges. Arguably, this is a slight abuse of terminology. Central charges are expected to obey certain properties such as positivity and having lower values at IR fixed points as compared to UV values. However, none of the coefficients in \eq{defecttrace} seem to satisfy both of those conditions.  Along defect RG flows, $b$ has been shown to obey a weak version of a $c$-theorem called the $b$-theorem \cite{Jensen:2015swa} and for superconformal defects obeys a $b$-extremisation \cite{Wang:2020xkc}. Yet, a simple free field boundary CFT (bCFT) computation in a $d=3$ theory of a free scalar with Dirichlet boundary conditions shows that $b$ is not necessarily positive (see e.g. \cite{Solodukhin:2015eca}) and the question of whether a lower bound exists remains open \cite{Herzog:2017kkj, Herzog:2020lel}.  Further, since $d_1$ and $d_2$ are associated with $B$-type anomalies in the classification of \cite{Deser:1993yx}, they can depend on bulk and defect marginal parameters and thus are unlikely to satisfy any version of a defect $c$-theorem.  However, it is clear that in a unitary dCFT $d_1 \geq 0$ as it is related to the coefficient of a two-point function of the displacement operator. It was shown in \cite{Jensen:2018rxu} that $d_2\geq0$ if the averaged null energy condition (ANEC) holds in the presence of a $p=2$ conformal defect.  

In the subsequent sections, we will compute the defect central charges for monodromy defects in $d=4$ free field theories.  To that end, we will need to briefly review the physical observables that we will be calculating and their relations to $b$, $d_1$, and $d_2$. 

First, let us quote the form of the one point function of the stress tensor for a flat $p\geq2$ dimensional defect $\gamma_{ab} = \eta_{ab}$ embedded in $\mathbb{R}^d$ 
\begin{align}\label{eq:defect-T-one-point-function}
\langle T^{ab} \rangle = - \frac{h}{2\pi} \frac{(d-p-1)\delta^{ab}}{|x_\perp|^d}\, ,\qquad \langle T^{ij} \rangle =\frac{h}{2\pi}\frac{(p+1)\delta^{ij} - d \frac{x^i_\perp x_\perp^j}{|x_\perp|^2}}{|x_\perp|^d}\,.
\end{align}
In the particular case where $p=2$, a $d$-dimensional generalisation of the results of \cite{Lewkowycz:2014jia,Bianchi:2015liz} establishes a relation between $h$ and $d_2$
\begin{align}\label{eq:d2-h-general-d}
h \equiv \frac{1}{3(d-1)\vol{\mathbb{S}^{d-3}}}d_2\,,
\end{align}
where $\vol{\mathbb{S}^{d-3}}$ is the volume of a unit $\mathbb{S}^{d-3}$ \cite{Jensen:2018rxu}.  Since $p=d-2$ for monodromy defects, we will need $d=4$ in order to exploit the relation between $d_2$ and $h$, in which case
\begin{align}\label{eq:d2-h-4d}
d_2 = 18\pi h\,.
\end{align}

Of the remaining central charges in \eq{defecttrace}, $d_1$ is also generically computable through the normalisation of a correlation function.  Namely, we begin with the two-point function of the displacement operator, which takes the form
\begin{align}\label{eq:defect-D-two-point-function}
\langle \mc{D}^i \mc{D}^j \rangle = \frac{C_{\mc{D}}}{2|\sigma^a|^{2(d-1)}}\eta^{ij}\,.
\end{align}
For the monodromy defects we are studying, where $p=2$ and $d=4$ \cite{Bianchi:2015liz},
\begin{align}\label{eq:d1-CD}
d_1 = \frac{3\pi^2}{4}C_{\mc{D}}\,.
\end{align}
In \cite{Bianchi:2015liz}, it was conjectured that for co-dimension $q=2$ defects, $C_\CD$ and $h$ (and thus $d_1$ and $d_2$) are proportional to one another,
\begin{align}\label{eq:CD-h-q=2}
C_\CD =2^{d-1} d \frac{\Gamma(\frac{d+1}{2})}{\pi^{\frac{d-1}{2}}}h\,,
\end{align}
where $\Gamma(z)$ is the Euler Gamma function. It was also conjectured in \cite{Bianchi:2019sxz} that for $p$-dimensional superconformal defects with co-dimension $q$
\begin{align}\label{eq:CD-h-general-p-q}
C_\CD = \frac{2^{p}(q+p-1)(p+2)}{(q-1)\pi^{\frac{p-q+3}{2}}}\frac{\Gamma(\frac{p+1}{2})}{\Gamma(\frac{q}{2})}h\,.
\end{align}
We will see explicitly that non-supersymmetric monodromy defects in free field theories satisfy \eq{CD-h-q=2}.

Lastly, the defect central charge $b$ shows up in a number of places.  The most familiar one is the free energy of a spherical defect. Let the defect submanifold $\Sigma = \mathbb{S}^2\hookrightarrow\CM_d$ be smoothly embedded in the ambient geometry and denote the curvature scale on the defect by $L$.  Let $Z_0$ be the partition function of the ambient CFT on $\CM_d$ without the defect insertion.  For a CFT on an $\mathbb{S}^{2k}$, the free energy has a log divergence proportional to the integrated Weyl anomaly, and in the case of a CFT on $\mathbb{S}^{2k+1}$ the free energy is finite and unambiguous.  Setting $\CM_d = \mathbb{S}^{d}$, it is clear from \eq{defecttrace} that for a defect wrapping a round, equatorial $\mathbb{S}^2\subset \mathbb{S}^{d}$, only the integrated Euler density contributes. Denoting the dCFT partition function by $Z$, one finds
\begin{align}\label{eq:defect-sphere-F-b}
F = -\log\frac{Z}{Z_{0}} = -\frac{b}{3}\log \Lambda L\,,
\end{align}
where $\Lambda$ is a UV scale. 

There is another physical observable that can be used to derive $b$ if $d_2$ is known (and vice versa): the defect contribution to EE of a region $A$. Starting from a flat $p=2$ dimensional defect embedded in $\mathbb{R}^d$, it was shown in \cite{Jensen:2018rxu} that by using the Casini-Huerta-Myers prescription \cite{Casini:2011kv} together with eqs.~\eqref{eq:defect-T-one-point-function} and \eqref{eq:defect-sphere-F-b} the change in EE for a region with $\partial A = \mathbb{S}^{d-2}$ due to the defect insertion is given by
\begin{align}\label{eq:defect-EE-general}
\Delta S_A  = \frac{1}{3}\left(b - \frac{d-3}{d-1} d_2\right)\log \frac{L}{\eps} + \CO(\eps^0) \,,
\end{align}
where $\varepsilon$ is a short-distance cut-off. In particular, the relative sign in $\Delta S_A$ shows that the universal coefficient of the defect EE can change sign along RG flows (see e.g. \cite{Rodgers:2018mvq}), which unlike ordinary CFTs limits the use of defect EE as a central charge.

Although we will not make use of holography in this work, we should mention the fact that the defect contribution to EE can be computed efficiently for CFTs which admit an holographic dual through the Ryu-Takayanagi formula \cite{Ryu:2006bv,Ryu:2006ef}. This is the case for holographic theories dual to BCFTs \cite{Takayanagi:2011zk} where boundary central charges can be extracted by employing \eq{defect-sphere-F-b}  \cite{FarajiAstaneh:2017hqv,Seminara:2017hhh,Seminara:2018pmr} and in theories with defects as done in \cite{Estes:2014hka,Jensen:2018rxu,Rodgers:2018mvq}.

\subsubsection*{R\'enyi Entropies}

The R\'enyi entropies are more refined quantities than EE. Consider a unitary QFT defined on $\CM_d$ in a generic quantum state described by a density matrix $\rho$. If the QFT allows for a bipartite factorisation of the Hilbert space \(\mc{H}\) into a subspace $A$ and its complement $\bar{A}$, one can compute the reduced density matrix on \(A\), defined as  $\rho_A \equiv \Tr_{\bar{A}} \rho$, by tracing over the subsystem $\bar A$. The \(n\)-th R\'enyi entropy is then defined as
\begin{equation}\label{eq:n-Renyi}
S_A^{(n)} \equiv \frac{1}{1-n} \log \Tr_A \rho_A^n\,.
\end{equation}

In QFT, one often considers a geometric bipartition defined on a Cauchy hypersurface, so that $A$ and its complement are regions in space. The trace $\Tr_A \rho_A^n$ is the partition function \(Z\) of $n$ copies of the original theory glued together along \(A\). This is equivalent to the partition function of the original theory on a manifold $\mathcal{M}_n$ with a conical deficit of $2\pi /n$ at the entangling surface separating \(A\) from \(\bar{A}\), i.e. $\Tr_A \rho_A^n = Z[\mathcal{M}_n]/Z[\mathcal{M}_1]^n$. 

Taking the limit $n\to1$ of \eq{n-Renyi}, we return to the original background $\CM_d$ without a conical deficit, and the R\'enyi entropy reduces to the EE associated with the subspace $A$,
\begin{equation}
\label{eq:eelim}
S_A\equiv \lim_{n\rightarrow 1} S_A^{(n)} = - \lim_{n\rightarrow 1}\partial_n \Tr_A \rho_A^n = - \Tr_A \rho_A \log \rho_A\,.
\end{equation} 
This procedure is known as the replica trick~\cite{Callan:1994py}, and has been particularly successful when applied to $d=2$ CFTs~\cite{Holzhey:1994we,Calabrese:2004eu, Calabrese:2009qy, Calabrese:2009ez,Calabrese:2010he}. If the initial state \(\r\) is a pure state, which we will assume in the following sections, then $S_A = S_{\bar{A}}$, and the EE is a good measure of the amount of quantum entanglement between $A$ and $\bar{A}$.

It will be helpful for later calculations to define the generating functional on the $n$-sheeted cover of $\CM_d$ as $W[\mathcal{M}_n]\equiv-\log Z[\mathcal{M}_n]$, which allows us to re-express eq.~\eqref{eq:eelim} in a more immediately useful form
\begin{equation}
\label{eq:EE_repl_trick_action}
S_A = \lim_{n\rightarrow 1} (\partial_n-1)W[\mathcal{M}_n]\,.
\end{equation}
This final expression for EE will be particularly useful as it allows us to start from the path integral of the free field CFTs that we study and employ heat kernel methods to efficiently calculate defect EE, and hence, defect central charges.

\subsection*{Monodromy defects}

Here we review our conventions for monodromy defects that will be used in all subsequent computations.  In addition, we will collect some results for monodromy defects that will allow us to compute their central charges in $d=4$. 

We begin with an ambient free field CFT with a global flavour symmetry group containing a $U(1)_f$ subgroup. Let $I_{CFT}$ be its action on $\CM_d = \mathbb{R}^{1,d-1}$ with coordinates $x^\mu  = \{t,\,\vec{\sigma},\,\rho,\,\theta\}$ and metric $g_{\mu\nu}$,
\beq\label{eq:background-metric-R-d}
ds^2 =g_{\mu\nu}dx^\mu dx^\nu = -dt^2 + d\vec{\sigma}^2+d\rho^2+\rho^2 d\theta\,.
\eeq
In most computations, we will need to Wick rotate to Euclidean signature, i.e. $\mathbb{R}^d$, by taking $t\to -i\tau$.  We insert a monodromy defect along $\sigma^a = \{t,\vec{\sigma}\}$ located at $\rho=0$ in the transverse $\{\rho,\theta\}$-plane\footnote{Note that the cylindrical coordinates \eq{background-metric-R-d} make manifest the preserved $SO(2)_N\simeq U(1)_N$ rotational symmetry around the defect submanifold $\Sigma \hookrightarrow \CM_d$.} by turning on a constant background gauge field for the $U(1)_f$ flavour symmetry along
\beq
\label{eq:A-monodromy}
A = \alpha  d\theta\,.
\eeq
The gauge background \eq{A-monodromy} is a closed but not exact form since is singular at $\rho=0$. In particular, we may perform a singular gauge transformation to gauge away $A$. This transformation affects any field $\Psi(x)$ minimally coupled to $A$ with unit charge in the following way
\begin{equation}\label{eq:gauge_transf}
\Psi (x) \rightarrow e^{-i \alpha \theta}\Psi (x).
\end{equation}
Thus, the introduction of the potential \eqref{eq:A-monodromy} is equivalent to prescribing a non-trivial monodromy to $\Psi(x)$. 
The conserved current sourced by $A$, $J_\mu \equiv\delta I_{CFT}/\delta A^\mu$, has a non-trivial one-point function
\begin{equation}
\label{eq:CJ-d}
\left< J^{\theta} (x) \right> = \frac{C_J}{\rho^d}\,,
\end{equation}
where the coefficient $C_J$ is a function of the monodromy parameter $C_J \equiv C_J(\alpha)$.  

From the relationship between the generating functional and the Weyl anomaly, which led to \eq{defect-sphere-F-b}, it is natural to expect that $C_J$ in $d=4$ be related to defect central charges for monodromy defects.  However, naively computing 
\begin{align}
\label{eq:log-Z-J}
-\frac{d}{d \alpha} \log Z[\alpha] = \int d^4 x \, \left< J^\theta (x)\right>
\end{align}
we find no log divergences, only power law divergences. This is due to the fact that the defect we constructed above is flat and the integrated trace anomaly vanishes identically. 

In order to obtain a non-trivial result in \eq{log-Z-J}, we can modify \eq{A-monodromy} to include a non-trivial shape function
\begin{equation} \label{eq:bc-gauge-field}
A_\mu = \alpha \, f_\mu (x)\,,
\end{equation}
where $f_\mu (x)$ are the components of a certain closed but not exact form, i.e. it is singular on a co-dimension $q=2$ submanifold.  Repeating the computation above, we have
\begin{equation}
\label{eq:log-generic-defect}
-\frac{d}{d \alpha} \log Z[\alpha] = \int d^4 x \, \left< J^\mu (x)\right>_f f_\mu(x)\,,
\end{equation}
where $\langle\cdot\rangle_f$ denotes the expectation value in the presence of a defect of generic shape.  As discussed above \eq{defect-sphere-F-b}, a spherical defect in the Euclidean theory obeys ${\int \sqrt{g} \left< T^\mu_{~\mu} \right>=-b/3}$. Therefore, after inserting in \eq{log-generic-defect} the function $f_{\mu}(x)$ associated to a spherical defect profile, a straightforward computation outlined in \App{spherical} shows that
\begin{equation}
\label{eq:b-CJ-relation}
\frac{d}{d \alpha} b(\alpha) = 12 \pi^2 C_J \left( \alpha \right).
\end{equation}
Thus, we can easily derive $b$ whenever the normalisation of the current one-point function is known.
More generally, from the results in \App{spherical}, we find a relation connecting the one-point function of the current $C_J(\alpha)$ with the defect free energy defined in \eq{defect-sphere-F-b}, which reads
\begin{equation}\label{eq:F_deriv}
\frac{d}{d\alpha} F = -  C_J(\alpha) \frac{2 \pi ^{\frac{d}{2}+1}}{\Gamma \left(\frac{d}{2}\right) \sin \left( \frac{\pi}{2} d \right)} \,.
\end{equation}
The above equation was originally found in \cite{Herzog:2019rke,Bianchi:2019umv} where the authors studied how the defect free energy depends on bulk marginal couplings. The present case is slightly different from theirs since here the marginal operator $J^\theta(x)$ has explicit space-time dependence. 
We notice that \eq{F_deriv} has a simple pole when $d$ is even. This reflects the fact that the sphere free energy contains a logarithmic divergence which corresponds to the A-type defect anomaly. In particular, defining the integrated anomaly as ${\int \sqrt{g} \left< T^\mu_{~\mu} \right>\equiv -\mathcal A}$, we have
\begin{equation}
\label{eq:A-anom-der}
\frac{d}{d\alpha} \mathcal A = (-1)^{d/2} C_J(\alpha) \frac{4  \pi^{d/2}}{\Gamma \left(\frac{d}{2}\right)}\,,
\end{equation} 
which reduces to \eq{b-CJ-relation} when $d=4$ ($\mathcal A=b/3$).

Due to the relatively simple construction of monodromy defects through \eq{A-monodromy}, higher point correlation functions of $J^\mu$ will be related to other important physical observables.  Consider the stress tensor of a field theory coupled to $A$. With the insertion of a monodromy defect, $T_{\mu\nu}$ is no longer conserved at the location of the defect, and
\begin{equation}\label{eq:stress-conv-flux}
\nabla^\mu T_{\mu\nu}=J^\mu F_{\mu\nu}\,.
\end{equation} 
From \eq{A-monodromy}, $F_{\mu\nu}$ is proportional to a Dirac delta function at $\rho =0$. This is most clearly seen in Cartesian coordinates in the transverse space to the defect $x = \rho \cos \theta,\, y = \rho \sin\theta$, where now $F_{xy}= 2\pi \alpha \delta^2(x,y)$. Comparing \eq{stress-conv-flux} to \eq{displacement}, we identify the displacement operator as
\begin{equation}
\mc{D}_x=\left.  -2\pi \alpha J_y \right|_{x,y=0}, \qquad \mc{D}_y=\left.  2\pi \alpha J_x \right|_{x,y=0}.
\end{equation}
In the following sections it will be convenient to use complex coordinates ${z=x+i y}$, $ {\bar z=x-iy}$ in the transverse space. In these coordinates we have
\begin{equation}
\mc{D}_z=\left.  -2\pi  i \alpha J_z \right|_{z,\bar z=0}, \qquad \mc{D}_{\bar z}=\left.  2\pi i \alpha J_{\bar z} \right|_{z, \bar z=0}.
\end{equation}
This brief computation has demonstrated a second important use for correlators of $J_i$.\footnote{Here we use orthogonal indices because, as explained below \eq{W_variation}, only those components of the displacement operator are non-trivial.} Since $\CD_i \propto J_i$, we see that the displacement operator two-point function is computable through $\langle J_i J_j\rangle$. This in turn means that the defect limit of the current two-point function is controlled by $C_\CD$, which is proportional to $d_1$ in $d=4$. 

One important caveat to this statement arises when non-trivial sources for relevant defect operators in the defect OPE are included. These modes will be central to the analysis in the subsequent sections. When they are turned on, the relationship between $J_i$ and $\CD_i$ for monodromy defects no longer holds, and we will need to resort to other techniques to compute $C_\CD$.

  \section{Free Scalar}\label{sec:scalar}

In this section, we study the monodromy defect in the theory of a free, conformally coupled complex scalar field, $\varphi(x)$, in $d$ dimensions. As explained in \sn{review}, we engineer this defect by turning on a constant background gauge field for the $U(1)_f$ global symmetry. We take the Euclidean action of the theory to be
\begin{equation}
\label{eq:scalar_action}
I_{\mathrm{scalar}}=\int d^d x \, \sqrt{g} \left[    D^\mu\varphi (D_\mu \varphi)^\dagger + \frac{d-2}{4(d-1)} \mathcal{R} \left| \varphi\right|^2 \right],
\end{equation}
where the coupling to the scalar curvature $\mathcal{R}$ is needed to have a conformal and Weyl invariant action. We also define the gauge covariant derivative $D_\mu \equiv \nabla_\mu -i eA_\mu$.  

Varying $I_{\rm scalar}$ with respect to the gauge field $A_\mu$ gives the conserved $U(1)_f$-current
\begin{equation}
\label{eq:scalar_current}
J_\mu = \frac{1}{\sqrt{g}}\frac{\delta I_{\mathrm{scalar}}}{\delta A^\mu} = - i \left( \varphi \nabla_\mu \varphi^\dagger -\nabla_\mu \varphi \varphi^\dagger  + 2 i e A_\mu \left|\varphi \right|^2\right),
\end{equation} 
while the variation with respect to the metric $g_{\mu\nu}$ produces the stress energy-momentum tensor
\begin{equation}
\begin{split}
T_{\mu \nu}   
&= \frac{2}{\sqrt{g}} \frac{\delta I_{\mathrm{scalar}}}{\delta{g^{\mu\nu}}}  \\ 
&= D_\mu \varphi  (D_\nu\varphi)^\dagger + (D_\mu\varphi)^\dagger D_\nu  \varphi  -\frac{d-2}{2(d-1)} \left[ \nabla_\mu \nabla_\nu  +\frac{ g_{\mu\nu}}{d-2} \nabla^2 - \CR_{\mu\nu}  \right]\left|\varphi \right|^2,
\end{split}
\end{equation}
where $\CR_{\mu\nu}$ is the Ricci tensor for the background geometry.
\subsection{Mode expansion and propagator}
In order to introduce a monodromy defect, we set $A_\mu$ to be as in \eq{A-monodromy}, and for simplicity we set the charge $e=1$. 
In appendix \ref{app:scalar} we provide the detailed derivation of the mode expansion and the propagator. Here we report and discuss the results.
In what follows we will restrict $\alpha\in (0,1)$, and we will treat the limits $\a\to0,1$ carefully.  We will also adopt complex coordinates transverse to the defect, i.e. $z = \rho e^{i \theta}$ and $\bar z = \rho e^{-i \theta}$.
The mode expansion can be written as 
\begin{equation}
\label{eq:mod_exp_salar_defect}
\varphi =  \varphi_{-\a} \,  z^{-\alpha} +\varphi_{\a-1} \, \bar z^{\alpha-1} +  \sum_{m=1}^{\infty}  \varphi_{m-\alpha}  z^{m-\alpha} +\sum_{m=0}^{\infty} \varphi_{m+\alpha} \bar z^{m+\alpha} \,.
\end{equation}
The modes $\varphi_{m+\a}$ are defined for $m\geq 0$ and $\varphi_{m-\a}$  for $m\geq 1$ as follows:
\begin{equation}
\label{eq:oper_O}
  \varphi_{m\pm\a} \equiv \int d k_\rho \int d^{d-3} \vec k \left[ f ( k)  a_{\mp m}( k)  + f^* ( k)  b_{\mp m}^\dagger( k)      \right]  \frac{J_{m\pm\alpha}(k_\rho \rho)}{\rho^{m\pm\alpha}}\,,
\end{equation}
where $J_\nu (\zeta)$ is the Bessel function of the first kind, and 
\begin{equation}
f(k) = \frac{\sqrt{k_\rho}}{(\sqrt{2\pi})^{d-2}\sqrt{2\omega}}e^{-i \omega t+i {\vec k\cdot \vec \sigma}}\,,
\end{equation}
with $\omega^2=k_{\r}^2+\vec k^2$. Throughout we use the shorthand $k=(k_\rho,\vec k)$.
The modes $\varphi_{\a}$ and $\varphi_{1-\a}$ are special because they are naturally paired with two singular modes $\varphi_{-\a}$ and $\varphi_{\a-1}$, respectively. These modes can be included if one allows for divergences milder than $\CO\left(\rho^{-1}\right)$ as $\rho\to 0$. In principle, we can introduce these four modes independently but the canonical commutation relation for $\varphi$ fixes their coefficients in terms of two free parameters $\xi,\tilde \xi \in [0,1]$ so that we have
\begin{subequations}\label{eq:zero_mode_scalar}
\begin{align}
\varphi_{\a}= \sqrt{1-\xi} &\int d k_\rho \int d^{d-3} \vec k  \left[ f (k)  a_0^{(+)}( k )  + f^* ( k)  b_0^{(+)\dagger}( k )      \right]  \frac{J_{\alpha}(k_\rho \rho)}{\rho^{\alpha}}\,, \\
\varphi_{-\a}=\sqrt{\xi} &\int d k_\rho \int d^{d-3} \vec k  \left[ f (k)  a_0^{(-)}(k )  + f^* (k)  b_0^{(-)\dagger}( k )      \right]  \frac{J_{-\alpha}(k_\rho \rho)}{\rho^{-\alpha}}\,, \\
\varphi_{1-\a}= \sqrt{1-\tilde \xi} &\int d k_\rho \int d^{d-3} \vec k \left[ f (k)  a_1^{(+)}( k )  + f^* (k)  b_1^{(+)\dagger}( k )      \right]  \frac{J_{1-\alpha}(k_\rho \rho)}{\rho^{1-\alpha}}\,, \\
\varphi_{\a-1}= \sqrt{\tilde \xi}& \int d k_\rho \int d^{d-3} \vec k  \left[ f ( k)  a_1^{(-)}( k )  + f^* (k)  b_1^{(-)\dagger}( k )      \right]  \frac{J_{\alpha-1}(k_\rho \rho)}{\rho^{\alpha-1}}\, .
\end{align}
\end{subequations}
Similar modes have been already discussed from an abstract defect CFT perspective in \cite{Bianchi:2019sxz, Lauria:2020emq}. In order to make contact with these works it is useful to match our mode expansion with the defect OPE of the bulk field $\varphi$. The latter allows to expand any bulk operator in terms of defect primaries $\hat O_m$ and their descendants. The coefficients of this expansion are the bulk to defect couplings. For a monodromy defect, the allowed defect operators in the defect OPE of a bulk scalar in our conventions must have orthogonal spin $s\in \mathbb{Z}-\alpha$. Furthermore, in a free theory, the equations of motion for $\varphi$ allow for two sets of defect operators in the defect OPE of $\varphi$. The dimensions of these two sets of operators were denoted in \cite{Lauria:2020emq} as $\hat{\Delta}^{+}_s=\frac{d}{2}-1+|s|$ and $\hat{\Delta}^{-}_s=\frac{d}{2}-1-|s|$. While the former are always allowed, the latter violate the unitarity bound for a co-dimension 2 defect in $d > 4 $ unless $|s|<1$. For $d\leq 4$, unitarity requires $|s|<\frac{d-2}{2}$. The defect OPE then reads
\begin{equation}\label{eq:defect_OPE}
\begin{split}
\varphi =\;& \sum_{s\in \mathbb{Z}-\a} c_{\varphi \hat O_s^+} \rho^{|s|} e^{is\theta} \mathcal{C}^+_{s}\left(\rho^2 \partial^2_{\sigma}\right) \hat O^+_{s}(\sigma)\\
&+c_{\varphi \hat O_{-\a}^-}\frac{e^{-i\a\theta}}{\rho^{\a}} \mathcal{C}^-_{s}\left(\rho^2 \partial^2_{\sigma}\right) \hat O^-_{-\a}(\sigma)+c_{\varphi \hat O_{1-\a}^-}\frac{e^{i(1-\a)\theta}}{\rho^{1-\a}} \mathcal{C}^-_{s}\left(\rho^2 \partial^2_{\sigma}\right) \hat O^-_{1-\a}(\sigma)\, .
\end{split}
\end{equation}
The differential operators $\mathcal{C}^{\pm}_{s}(\rho^2 \partial_{\sigma}^2)$ resum the contribution of all the conformal descendants, and are fixed by conformal invariance to be
\begin{equation}\label{eq:diff_op}
\mathcal{C}^{\pm}_{s}\left(\rho^2 \partial^2_{\sigma}\right) \equiv \sum_{k=0}^{+\infty} \frac{(-4)^{-k}(\rho^2 \partial^2_{\sigma})^k}{k!(1\pm|s|)_k}\,,
\end{equation}
where $(a)_k \equiv a(a+1)\dots(a+k-1)$ if $k \ne 0$ and $(a)_0\equiv 1$ is the Pochhammer symbol.
Comparing this expression with the small $\rho$ expansion of eq.~\eqref{eq:mod_exp_salar_defect} after Wick rotation, one finds a one-to-one correspondence between the mode expansion and the defect OPE, thus establishing that each mode in eq.~\eqref{eq:mod_exp_salar_defect} creates a conformal family of defect operators. Including modes that are less singular than $\CO(\rho^{-1})$ is equivalent to allowing for defect operators with dimension $\hat \Delta^{-}_s$ above the unitarity bound. We will determine the bulk to defect couplings $c_{\varphi \hat O}$ by comparing the propagator with the defect block expansion.

The propagator is computed in appendix \ref{app:scalar} where we find that the result consists precisely of a sum over defect blocks.  A bulk two-point function can be expressed in terms of two cross ratios: the relative angle $\theta$ and the combination
\begin{equation}\label{eq:crossratio-def}
\eta\equiv \frac{2 \rho \rho'}{\rho^2+\rho'^2 +  |\sigma^a|^2} \,,
\end{equation}
where we have set $\sigma'{}^a=0$ by translational invariance along the defect. A defect block is a function of these cross ratios:
\begin{equation}\label{defectblock}
 F_{{\hat \Delta}, s}(\eta,\theta)=\left(\frac{\eta}{2} \right)^{\hat{\Delta}_s} \phantom{}_2 F_1\left( \frac{\hat{\Delta}_s}{2},\frac{\hat \Delta+1}{2}; \hat{\Delta}_s+2-\frac{d}{2}; \eta^2  \right) e^{i s\theta}\,,
\end{equation}
where ${}_2 F_1 (a,b;c;z)$ is the ordinary hypergeometric function. The propagator then takes the form
\begin{align}\label{eq:two-point function_gaug_transf}
	\left< \varphi(x)\varphi^\dagger(0,\rho') \right> = \left(\frac{1}{\rho\rhop}\right)^{\frac{d}{2}-1} \Bigg(\sum_{s\in \mathbb{Z}-\a} c_s^+ \,    F_{\hat{\Delta}^+, s}(\eta,\theta)		+  c_{-\a}^- \,    F_{\hat{\Delta}^-, -\a}(\eta,\theta)+c_{1-\a}^- F_{\hat{\Delta}^-, 1-\a}(\eta,\theta)\Bigg)\,,
\end{align}
with
\begin{equation}\label{bulkdefcoeff}
c_s^+= \frac{\Gamma\left( \frac{d}{2}-1+|s|  \right)}{4\pi^{d/2}\Gamma\left(1+ |s| \right)} \qquad \text{for   } s\neq-\a,1-\a\,,
\end{equation}
and special cases
\begin{subequations}
\begin{align}
 c_{-\a}^+&= (1-\xi)\frac{\Gamma\left( \frac{d}{2}-1+\a  \right)}{4\pi^{d/2}\Gamma\left(1+ \a \right)}\,, & c_{-\a}^-&= \xi\frac{\Gamma\left( \frac{d}{2}-1-\a  \right)}{4\pi^{d/2}\Gamma\left(1- \a \right)}\,, \label{bulkdefcoeff1}\\
 c_{1-\a}^+&= (1-\tilde \xi)\frac{\Gamma\left( \frac{d}{2}-\a  \right)}{4\pi^{d/2}\Gamma\left(2- \a \right)}\,, & c_{1-\a}^-&=\tilde  \xi\frac{\Gamma\left( \frac{d}{2}-2+\a  \right)}{4\pi^{d/2}\Gamma\left( \a \right)}\,. \label{bulkdefcoeff2} 
\end{align}
\end{subequations}
In \sn{fermion}, we will find it advantageous to adopt an alternative notation for the scalar propagator in \eq{two-point function_gaug_transf}, $G_{S,\alpha,\xi,\tilde \xi}(x,x') \equiv\left< \varphi(x)\varphi^\dagger(0,\rho') \right>$. By matching the defect block expansion with the defect OPE \eq{defect_OPE} and using the normalisation 
\begin{align}\label{defdefprop}
 \left<\hat O_{s}^{\pm}(\sigma)\hat O_{s'}^{\dagger \pm}(0)\right> =\frac{\d_{s, - s'}}{|\sigma^a|^{d-2\pm2|s|}}
\end{align}
for the defect operators, we immediately identify $c_s^{\pm}=c_{\varphi \hat O_{s}^{\pm}}c_{\varphi^{\dagger} \hat O_{s}^{\dagger \pm}}$. Reflection positivity imposes that $c_s^{\pm}>0$, which determines the range of the parameters $\xi$ and $\tilde \xi$ to be
\begin{equation}\label{eq:range_xi}
 0\leq\xi,\tilde\xi\leq1\,.
\end{equation}

In $d<4$, the unitarity bounds above \eq{defect_OPE} impose further conditions on the range of $\a$ at non-zero $\xi$ or $\tilde\xi$. As an example consider $d=3$, for which the unitarity bound reads $|s|<\frac{1}{2}$. This requires that either $\a\in \left[0,\frac{1}{2}\right)$ if $\xi\neq 0$ and $\tilde\xi =0$, or $\a\in \left(\frac{1}{2},1\right]$ if $\xi= 0$ and $\tilde\xi \neq 0$. Notice that since the two ranges of $\a$ don't overlap, one cannot turn on both deformations $\xi,\tilde\xi\neq 0$ without breaking unitarity.

At this point, we are ready to discuss what happens for the limiting values $\a\to0$ and $\a\to 1$. In the absence of the divergent modes, i.e. for $\xi=\tilde\xi=0$, the defect OPE of $\varphi$ simply reduces to the Taylor expansion of the free field $\varphi$ around the co-dimension two surface $\rho=0$. In other words, the defect reduces to the trivial defect as one would expect in the absence of a monodromy. The singular modes, instead, lead to a singular behaviour of \eq{defect_OPE} either at $\a=0$ or at $\a=1$. For $\tilde \xi=0$ and $\xi\neq0$ the limit $\a\to 0$ is perfectly well-defined and leads again to the free field Taylor expansion, while the limit $\a\to 1$ is singular. This seems to be in contrast with our definition of the monodromy defect, which should reduce to the trivial defect for integer $\a$. Nevertheless, a glance at \eqref{bulkdefcoeff1} shows that, at least for $d>4$, the singular mode decouples from the bulk at $\a=1$ since $c^-_{-1}=0$. This leads to a two-dimensional theory, which is decoupled from the bulk free scalar. This is still not enough to affirm that the limit $\a\to 1$ is well-defined. Even though the singular mode decouples, the remaining bulk propagator still depends on $\xi$ as the contribution of the $\varphi_\a$ mode gives a factor of $1-\xi$ to one of the terms in the propagator. Thus, the propagator does not reduce to that of a free complex scalar but has an extra term proportional to $-\xi$ consisting of a single defect block. This bulk two-point function is not crossing invariant and therefore it does not lead to a consistent defect CFT. Therefore, we have to conclude that the limit $\a\to 1$ cannot be smooth for a constant value of $\xi$. Either $\xi$ is a function of $\a$ or some discontinuous behaviour must be introduced at $\a=1$ so that the periodicity in $\a$ is reinstated. In the following, we do not make any assumption on $\xi$ and we will mention explicitly the places where we will assume that it is not a function of $\a$. More generally, we keep an abstract point of view on this issue, assuming there could be a dynamical mechanism which causes the decoupling of this mode for $\a=1$. For the case of $\xi=0$ but non-vanishing $\tilde \xi$ as $\a\to0$ an identical discussion applies.

\subsection{Correlation functions and central charges}
\label{sec:scalar_corr_func}
We start by computing some relevant one-point functions by taking a suitable coincident limit of the propagator. 
A generic one-point function of a composite operator can be found by Wick contracting the fundamental fields and then taking the coincident limit, carefully regularising the short distance divergences. In the following, we consider only one-point functions of operators quadratic in the fundamental field $\varphi$. In this case, we only need to take the appropriate combinations of derivatives of the propagator, and then take the coincident limit.

To this end, it is convenient to start from the non-singular propagator ($\xi=\tilde \xi=0$) for which we can use the form in \eq{heat_kernel_prop_no_A}, which after a change of variables $\zeta= 2 /(s \rho \rho')$ becomes
\begin{equation}
\label{eq:heat_kern_scal_gen_d}
\left<\varphi(x)\varphi^\dagger(x')\right>_{\xi=\tilde \xi=0} = \frac{1}{2 (2\pi)^{d/2}} \frac{1}{(\rho \rho')^{d/2-1}} \int^{+\infty}_{\frac{2 \varepsilon^2}{\rho \rho'}} d\zeta \, e^{-\frac{1}{\eta \zeta}} \zeta^{-d/2} \sum_m  e^{i (m-\alpha)\theta} I_{|m-\alpha|} \left(\frac{1}{\zeta}\right),
\end{equation}
where $I_\nu(\zeta)$ is the modified Bessel function of the first kind. The integral is divergent in the coincident ($\eta\to1$) limit. For this reason we introduce the UV cut-off $\varepsilon$. The one-point function will be a power expansion in terms of $\varepsilon$. If the divergent term is independent of $\alpha$, the divergences can be consistently removed by subtracting the one-point function with $\alpha=0$. After computing the contribution from the regular modes of $\varphi$, we will add the singular parts, which are proportional to $\xi$ and $\tilde \xi$.

\subsubsection{One-point function of $\boldsymbol{|\varphi|^2}$}
Let us start with the simplest case: the one-point function of $|\varphi|^2$. Taking the coincident limit of \eq{heat_kern_scal_gen_d}  we obtain 
\begin{equation}
\label{eq:heat_kern_one_point_func}
\left<\varphi(x)\varphi^\dagger(x') \right>_{\xi=\tilde \xi=0} = \frac{1}{2 (2\pi)^{d/2}} \frac{1}{\rho^{d-2}} \int^{+\infty}_{\frac{2 \varepsilon^2}{\rho^2}} d\zeta \, e^{-\frac{1}{\zeta}} \zeta^{-d/2}\mathcal{I}^{(1)}_\alpha \left(\frac{1}{\zeta}\right),
\end{equation}
where we defined\footnote{In the second equality we used \eq{Bessel-sums-1}.}
\begin{equation}
\label{eq:I1_sum_m}
\begin{split}
\mathcal{I}^{(1)}_\alpha (\zeta) \equiv\sum_{m} I_{|m -\alpha | }(\zeta) &=  \frac{1}{2 \alpha}  \left[ e^{\zeta}  \int_0^\zeta e^{-x} I_{-\alpha }(x)\, dx - \zeta I_{-\alpha }(\zeta) - \zeta I_{1-\alpha}(\zeta)  \right]  \\
& +  \frac{1}{2(1-\alpha)}  \left[ e^{\zeta}  \int_0^\zeta e^{-x} I_{\alpha-1}(x)\, dx - \zeta I_{\alpha-1}(\zeta) - \zeta I_{\alpha}(\zeta)  \right] .
\end{split}
\end{equation}
The integral over $\zeta$ in \eq{heat_kern_one_point_func} can be performed, and indeed one can see that it diverges for $\varepsilon \rightarrow 0$. However, the divergences are independent of $\alpha$ and can be unambiguously subtracted. The final result is 
\begin{equation}
\label{eq:phisq_d_dim}
\left< |\varphi(x)|^2 \right>_{\xi=\tilde \xi=0} = - \frac{ 
	\Gamma \left(\frac{d}{2}-\alpha \right) \Gamma
	\left(\frac{d}{2}+\alpha -1 \right)\sin (\pi  \alpha )}{2^{d-1} \pi^{\frac{d+1}{2}}(d-2) \Gamma
	\left(\frac{d-1}{2}\right)} \frac{1}{\rho^{d-2}} \,.
\end{equation}
We observe that the one-point function vanishes both when $\alpha=0$ and $\alpha=1$. The former is obvious since it corresponds to the absence of a defect, while the latter reflects the fact that the flux is defined modulus integers.
We now compute the contribution of the singular modes. The part of the propagator proportional to $\xi$ can be deduced from eq.~\eqref{eq:two-point function_gaug_transf} and it reads
\begin{equation}\label{propxi}
 \left< \varphi(x)\varphi^\dagger(x') \right>_\xi
		=\xi \left(\tfrac{1}{\rho\rhop}\right)^{\frac{d}{2}-1} \left(-\tfrac{\Gamma\left( \frac{d}{2}-1+\a  \right)}{4\pi^{d/2}\Gamma\left(1+ \a \right)}  \,    F_{\hat{\Delta}^+, -\a}(\eta,\theta)+  \tfrac{\Gamma\left( \frac{d}{2}-1-\a  \right)}{4\pi^{d/2}\Gamma\left(1- \a \right)} \,    F_{\hat{\Delta}^-, -\a}(\eta,\theta)\right).
\end{equation}
We are interested in the $\theta\to0$ and $\eta\to 1$ limit. The former simply eliminates the exponential in eq.~\eqref{defectblock}, while the latter gives a singular limit for the hypergeometric function in eq.~\eqref{defectblock}. Actually, each defect block is logarithmically divergent in the limit $\eta\to 1$. However, these logarithms cancel in the combination eq.~\eqref{propxi}, leaving us only with power law divergences. After subtracting them we get
\begin{equation}
\begin{split}\label{eq:phisq_d_dim-sing}
 \left< |\varphi(x)|^2 \right>_{\xi}&  
= \xi \frac{  \Gamma \left(\frac{d}{2}-\alpha -1 \right) \Gamma \left(\frac{d}{2}+\alpha -1
	\right)\sin (\pi  \alpha )}{2^{d-1} \pi^{\frac{d+1}{2}}\Gamma \left(\frac{d-1}{2}\right)} \frac{1}{\rho^{d-2}}\,.
\end{split}
\end{equation}
The result proportional to $\tilde \xi$ can be simply obtained by replacing $\a\to1-\a$:
\begin{equation}
 \left< |\varphi(x)|^2 \right>_{\tilde \xi}= \tilde \xi \frac{  \Gamma \left(\frac{d}{2}+\alpha -2 \right) \Gamma \left(\frac{d}{2}-\alpha
	\right)\sin (\pi  \alpha )}{2^{d-1} \pi^{\frac{d+1}{2}}\Gamma \left(\frac{d-1}{2}\right)} \frac{1}{\rho^{d-2}}\,.
\end{equation}
Putting everything together we get the final result
\begin{equation}\label{oneptphi2full}
 \left< |\varphi(x)|^2 \right>= \frac{ \Gamma (\frac{d}{2}-\alpha ) \Gamma
	(\frac{d}{2}+\alpha -1 ) \sin (\pi  \alpha )}{2^{d-1} \pi^{\frac{d+1}{2}}\Gamma \left(\frac{d-1}{2}\right)} \frac{1}{\rho^{d-2}} \left(-\frac{1}{d-2}+\frac{\xi}{\frac{d}{2}-\a-1}+\frac{\tilde \xi}{\frac{d}{2}+\a-2}\right).
\end{equation}

As a special case we note that when $d=4$, the one-point function becomes
\begin{equation}\label{eq:scalar-one-point-function-4d}
\left< |\varphi(x)|^2 \right> =  -\frac{(1-\alpha ) \alpha -2 \xi \alpha -2 \tilde \xi (1-\a) }{8 \pi ^2 \rho^2} \,.
\end{equation}

\subsubsection{One-point function of $\boldsymbol{T_{\mu\nu}}$}
The one-point function of the stress-energy tensor in the presence of a $p=d-2$ dimensional conformal defect is fixed by conformal symmetry to be of the form in \eq{defect-T-one-point-function}. Thus, we only need to compute the coefficient $h$ to determine the full stress tensor one-point function. To do this, we can choose a particular component, and we pick 
\begin{equation}
\label{eq:stress_scalar_rhorho}
\left< T_{\rho \rho} \right>= 2\left<  \partial_\rho \varphi \partial_{\rho}\varphi^\dagger\right>- \frac{1}{2 (d-1)} \left[ (d-1) \partial_\rho^2 + \frac{1}{\rho} \partial_\rho \right]\left<|\varphi|^2 \right>.
\end{equation}
Having already found $\left< |\varphi|^2 \right>$ above, we only need to compute the first term in eq.~\eqref{eq:stress_scalar_rhorho}. We need to take the coincident limit of the propagator after taking derivatives with respect to $\rho$. Once again, we start from the regular part of the propagator ($\xi=\tilde \xi=0$), for which we use the representation \eq{heat_kern_scal_gen_d},
\begin{equation}
\label{eq:heat_kern_one_point_func_der}
\left< \partial_\rho \varphi(x) \partial_{\rho}\varphi^\dagger(x) \right>_{\xi=\tilde\xi=0} = \frac{1}{8(2\pi)^{d/2}}\frac{1}{\rho^d}\int^{+\infty}_{\frac{2 \varepsilon^2}{\rho^2}} d\zeta \, e^{-1/\zeta } \frac{(d-2)^2 \xi +4}{\zeta ^{\frac{d}{2}+1}}\, \mathcal{I}^{(1)}_\alpha \left(\frac{1}{\zeta}\right) .
\end{equation}
Just as above, we find that the divergences in $\varepsilon$ are independent of $\alpha$ and can thus be subtracted unambiguously. The finite part reads 
\begin{equation}
\label{eq:dphidphi_gen_d}
\left< \partial_\rho \varphi \partial_{\rho}\varphi^\dagger(x)\right>_{\xi=\tilde\xi=0} =-\frac{((d-2) d^2+4 (1-\alpha) \alpha ) \sin (\pi  \alpha ) \Gamma
	\left(\frac{d}{2}-\alpha \right) \Gamma \left(\frac{d}{2}+\alpha
	-1\right)}{ 2^{d+1} \pi^{ \frac{d+1}{2}} d\left(2d - 1\right)  \Gamma
	\left(\frac{d-1}{2}\right)}\frac{1}{\rho^d}\,.
\end{equation}
Performing the same analysis for the singular contributions we find
\begin{align}
\label{eq:dphidphi_gen_d_sing}
 \left< \partial_\rho \varphi \partial_{\rho}\varphi^\dagger(x)\right> _{\xi} &=\xi \frac{((d-2)^2 d-8 \alpha ^2)\Gamma\left(\frac{d}{2}-\alpha-1 \right) \Gamma \left(\frac{d}{2}+\alpha-1
	\right)\sin(\pi \alpha)}{2^{d+2}\pi^{\frac{d+1}{2}}((d-2)^2-4 \alpha ^2)\Gamma\left(\frac{d+1}{2} \right)}\frac{1}{\rho^d}\,, \\ \label{eq:dphidphi_gen_d_sing2}
\left< \partial_\rho \varphi \partial_{\rho}\varphi^\dagger(x)\right> _{\tilde \xi} &=\tilde \xi \frac{((d-2)^2 d-8 (1-\alpha) ^2)\Gamma\left(\frac{d}{2}+\alpha-2 \right) \Gamma \left(\frac{d}{2}-\alpha
	\right)\sin(\pi \alpha)}{2^{d+2}\pi^{\frac{d+1}{2}}((d-2)^2-4 (1-\alpha) ^2)\Gamma\left(\frac{d+1}{2} \right)}\frac{1}{\rho^d}\,.
\end{align}

Thus, by plugging eqs.~\eqref{oneptphi2full}, \eqref{eq:dphidphi_gen_d}, \eqref{eq:dphidphi_gen_d_sing} and \eqref{eq:dphidphi_gen_d_sing} into \eq{stress_scalar_rhorho}, we find the full one-point function of the stress tensor 
\begin{align}
\langle T_{\rho\rho}(x)\rangle &=-\frac{ \Gamma \left(\frac{d}{2}-\alpha \right) \Gamma \left(\frac{d}{2}+\alpha -1\right)\sin (\pi  \alpha )  \left(\frac{\alpha  (1-\alpha )}{d}+\frac{\alpha ^2 \xi }{ \frac{d}{2}-\alpha-1}+\frac{(1-\alpha )^2 \tilde{\xi
   }}{\frac{d}{2}+\alpha -2}\right)}{2^{d-1}\pi^{\frac{d+1}{2}}\Gamma \left(\frac{d+1}{2}\right)}\frac{1}{\rho^d}\,,
\end{align}
which is vanishing for $\alpha=0$ and $\alpha=1$ when $d>4$, as expected. The contribution with $\xi=\tilde{\xi}=0$ was previously computed in~\cite{Dowker:1987mn}. Comparing to \eq{defect-T-one-point-function}, we find that $h$ is expressed as
\begin{equation}
\label{eq:h_scalar}
h =   \frac{ \Gamma \left(\frac{d}{2}-\alpha \right) \Gamma \left(\frac{d}{2}+\alpha -1\right)\sin (\pi  \alpha )  \left(\frac{\alpha  (1-\alpha )}{d}+\frac{\alpha ^2 \xi }{\frac{d}{2}-\alpha-1}+\frac{(1-\alpha )^2 \tilde{\xi
   }}{\frac{d}{2}+\alpha -2}\right)}{2^{d-2}\pi^{\frac{d-1}{2}}\Gamma \left(\frac{d+1}{2}\right)}\,.
\end{equation}
Note that $h\geq 0$ for the ranges $\a \in (0,1)$, and $\xi,\tilde{\xi} \in [0,1]$ in $d\geq 4$. For $d<4$, unitarity requires that the range of $\a$ be restricted in the presence of singular modes, as discussed below \eq{range_xi}. In that case $h$ is manifestly non-negative.

Specialising again to $d=4$ in order to connect to defect central charges and using \eq{d2-h-4d}, we find  
\begin{equation}
\label{eq:d2-free-scalar}
d_2 = \frac{3}{2} \left( (1-\alpha)^2 \alpha^2 + 4 \xi \alpha^3+ 4 \tilde \xi (1-\alpha)^3 \right).
\end{equation}
Since $h\geq 0$, so is $d_2$. This is in agreement with the expectation that $d_2\geq 0$ if the ANEC holds in the presence of a $p=2$ dimensional defect~\cite{Jensen:2018rxu}. 

\subsubsection{One-point function of $\boldsymbol{J_\theta}$}
In this subsection, we consider the one-point function of the current $\langle J_\mu\rangle$.  By computing the coefficient $C_J$, we can leverage \eq{b-CJ-relation} in $d=4$ to compute the defect central charge~$b$.

We start again from the regular part of the propagator in the form of \eq{heat_kern_scal_gen_d}, and consider the following expectation value 
\begin{equation}
\begin{split}\label{eq:scalar-J-theta-one-point-function}
\left< J_\theta \right>_{\xi=\tilde\xi=0}  =- 2 i \langle\varphi\, \partial_\theta \varphi^\dagger\rangle
= -\frac{1}{ (2\pi)^{d/2}} \frac{1}{\rho^{d-2}} \int^{+\infty}_{\frac{2 \varepsilon^2}{\rho^2}} d\zeta \, e^{-\frac{1}{ \zeta}} \zeta^{-d/2} \mathcal{I}^{(2)}_\alpha \left(\frac{1}{\zeta}\right) , 
\end{split}
\end{equation}
where we have taken the coincident limit and defined the sum\footnote{This follows immediately from \eq{Bessel-sums-2}.}
\begin{equation}
\begin{split}\label{eq:I2-sum}
\mathcal{I}^{(2)}_\alpha \left(\zeta\right)&\equiv\sum_m  \left( m - \alpha \right) I_{|m-\alpha|} \left(\zeta \right) \\
& = \frac{\zeta}{2} \left[ I_{1-\alpha} (\zeta) + I_{-\alpha} (\zeta) - I_{1+\alpha}(\zeta) - I_{\alpha} (\zeta)    \right]  - \alpha I_{\alpha}(\zeta)\,.
\end{split}
\end{equation}
Inserting the result of the sum in \eq{I2-sum} into \eq{scalar-J-theta-one-point-function}, we find that the $\zeta$-integral is convergent in the limit $\varepsilon \rightarrow 0$.  For the regular modes, computing the $\zeta$-integral and removing the UV cutoff gives
\begin{equation}
\label{eq:current_scalar_d_dim}
\langle J_\theta \rangle_{\xi=\tilde\xi=0}= \frac{ (1-2 \alpha )  \Gamma \left(\frac{d}{2}-\alpha \right) \Gamma
	\left(\frac{d}{2}+\alpha -1\right)\sin
	(\pi  \alpha )}{2^{d} \pi ^{\frac{d+1}{2}}\Gamma
	\left(\frac{d+1}{2}\right)} \frac{1}{\rho^{d-2}}\,.
\end{equation}
We note that $\langle J_\theta\rangle=0$ both for $\alpha=0,1$, and for $\alpha=1/2$. The former cases are expected as the monodromy becomes trivial, while the latter follows from symmetry considerations. More specifically, the Lagrangian in \eq{scalar_action} is manifestly invariant under the transformation $\theta \rightarrow -\theta$ provided that $\alpha \rightarrow -\alpha$. If $\varphi$ is regular in the limit $\rho \rightarrow 0$, then $-\alpha$ and $1-\alpha$ are identified by gauge invariance. This implies that $\langle J_\theta \rangle$ is odd under $\alpha \rightarrow 1-\alpha$ in the range $\a\in[0,1] $, and thus it vanishes when $\alpha=1/2$. 
The contribution of the singular modes is
\begin{subequations}
\begin{align}
\label{eq:current_scalar_d_dim_sing_xi}
 \langle J_\theta \rangle _{\xi}=& 2\xi  \frac{ \alpha \,  \Gamma \left(\frac{d}{2}-\alpha -1 \right) \Gamma \left(\frac{d}{2}+\alpha -1
	\right)\sin (\pi  \alpha )}{2^{d-1} \pi^{\frac{d+1}{2}}\Gamma \left(\frac{d-1}{2}\right)} \frac{1}{\rho^{d-2}}\,, \\
	\label{eq:current_scalar_d_dim_sing_xitilde}
	\langle J_\theta \rangle _{\tilde \xi}=& 2 \tilde \xi  \frac{ (1-\alpha) \,  \Gamma \left(\frac{d}{2}+\alpha -2 \right) \Gamma \left(\frac{d}{2}-\alpha 
	\right)\sin (\pi  \alpha )}{2^{d-1} \pi^{\frac{d+1}{2}}\Gamma \left(\frac{d-1}{2}\right)} \frac{1}{\rho^{d-2}}\,.
\end{align}
\end{subequations}

Summing all the contributions we have
\begin{equation}
\langle J_\theta \rangle= \frac{ \Gamma \left(\frac{d}{2}-\alpha \right) \Gamma \left(\frac{d}{2}+\alpha -1\right) \sin (\pi  \alpha )\left(1-2 \alpha+\frac{2 \alpha  (d-1) \xi }{ \frac{d}{2}-\alpha-1}+\frac{2 (1-\alpha ) (d-1)
   \tilde{\xi }}{\frac{d}{2}+\a-2}\right)}{2^d\pi^{\frac{d+1}{2}}\Gamma \left(\frac{d+1}{2}\right) } \frac{1}{\rho^{d-2}}\,.
\end{equation}
We observe that when $\xi, \tilde \xi \ne 0$ the current is not vanishing at $\alpha = 1/2$. This is due to the singular modes, which break the invariance of the theory under the shift $\alpha \rightarrow \alpha + \mathbb{Z}$. We also notice that the integral of this function with respect to $\a$ leads to the universal part of the sphere free energy.\footnote{For instance, for $d=3$ we can compare with the results obtained in Appendix A of~\cite{Belin:2013uta}, finding perfect agreement upon the identification of the flux parameters ($\mu=2\pi \alpha$).}

For $d=4$, the current one-point function, including the contribution from the singular modes, reads
\begin{equation}\label{currentonept}
\langle J_\theta (x) \rangle=\left(\frac{\alpha  (1-2 \alpha ) (1-\alpha )}{12 \pi ^2}+\frac{\alpha ^2 \xi }{2 \pi ^2}+\frac{(1-\alpha )^2 \tilde{\xi }}{2 \pi ^2}\right)\frac{1}{\rho^{d-2}}\,.
\end{equation}
By applying \eq{b-CJ-relation} and integrating over the flux $\alpha$, we obtain the defect central charge
\begin{equation}
\label{eq:b-free-scalar}
b =  \frac{ (1-\alpha)^2 \alpha ^2 + 4 \xi \alpha^3+4 \tilde \xi (1-\alpha)^3}{2} \,.
\end{equation}

A few comments about this result are in order. First of all, when integrating~\eqref{currentonept}, we assumed that $\xi$ and $\tilde \xi$ are not functions of $\alpha$. Were $\xi$ or $\tilde \xi$ function of $\alpha$, the dependence of $b$ on $\alpha$ would be affected by the dynamical source of $\xi$ and $\tilde \xi$ and this unknown source would contribute to the derivative in \eq{b-CJ-relation}. Another important comment is that \eq{b-CJ-relation} involves an integration constant which can be a function of $\xi$ and $\tilde \xi$. This constant can be fixed by requiring that $b$ is vanishing at $\alpha=0$ if $\tilde \xi =0$, and at $\alpha =1$ if $\xi =0$, together with the requirement that its dependence on $\xi$ and $\tilde \xi$ be linear. These requirements fix the integration constant to be $-2\tilde \xi$, giving precisely \eq{b-free-scalar}. As a check of \eq{b-free-scalar}, we will compute in section \ref{sec:scalar-EE} the EE in the presence of the monodromy defect, and we will show that it vanishes for any value of $\a$, $\xi$ and $\tilde \xi$. This implies, in particular, that $b=\frac{d_2}{3}$, which is consistent with eqs.~\eqref{eq:b-free-scalar} and \eqref{eq:d2-free-scalar}. 

\subsubsection{Two-point function of $\boldsymbol{\mathcal{D}_i}$}
In this subsection we study the displacement operator of the monodromy defect of the complex scalar. Our final goal will be to identify the coefficient of its two-point function, which in $4d$ is related to the defect central charge $d_1$. 

Given the defect OPE of the fundamental field $\varphi$ \eq{defect_OPE} and of its conjugate $\varphi^{\dagger}$, we look for the displacement operator in the fusion of defect fields $\hat O^{\pm}_{m-\a}$ and $\hat O^{\dagger \pm}_{m+\a}$. We are looking for an operator of dimension $\Delta_{\mathcal{D}}=d-1$ and spin $s=1$. To be precise, there are two operators $\CD_z$ and $\CD_{\bar z}$ with the same dimension and opposite spin associated to the two broken translations in complex coordinates. The combination of defect operators fulfilling these requirements is
\begin{align}\label{ansatzdisp}
 \mathcal{D}_z =A\, \hat{O}^+_{1-\a}\hat{O}^{\dagger +}_{\a}+ B\, \hat{O}^-_{-\a}\hat{O}^{\dagger +}_{1+\a} +C\, \hat{O}^+_{2-\a}\hat{O}^{\dagger -}_{-1+\a}+ D\, [\hat{O}^{\dagger -}_{\a}\hat{O}^{ -}_{1-\a}]_2\,,
\end{align}
and their conjugates for the operator with spin $s=-1$. The last operator in eq.~\eqref{ansatzdisp} is the conformal primary built out of $\hat{O}^{\dagger -}_{\a}$, $\hat{O}^{ -}_{1-\a}$ and two derivatives, 
\begin{align}
 [\hat{O}^{\dagger -}_{\a}\hat{O}^{ -}_{1-\a}]_2\equiv&\left(\tfrac{1}{2(d-2)}-\tfrac{1}{4\alpha}\right)\hat{O}^{\dagger -}_{\a}\partial_{\sigma}^2 \hat{O}^{ -}_{1-\a} +\tfrac{1}{d-2} \partial_{\sigma} \hat{O}^{\dagger -}_{\a} \partial_{\sigma} \hat{O}^{ -}_{1-\a} \nonumber\\
 &+ \left(\tfrac{1}{2(d-2)}-\tfrac{1}{4(1-\alpha)}\right)\partial_{\sigma}^2\hat{O}^{\dagger -}_{\a} \hat{O}^{ -}_{1-\a}\,.
\end{align}
In eq.~\eqref{ansatzdisp} the coefficients $A$, $B$, $C$ and $D$ are implicitly functions of $\a$, $\xi$ and $\tilde \xi$.
Notice that only the first operator is present for $\xi=\tilde \xi=0$. The second operator includes a mode $\hat{O}^-_{-\a}$ so it must vanish for $\xi= 0$, the third one includes $\hat{O}^{\dagger -}_{-1+\a}$ and it must not appear for $\tilde \xi=0$. The last term appears only when both $\xi$ and $\tilde \xi$ are non-vanishing. Using this piece of information and consistency with the Ward identity
\begin{equation}
\label{eq:ward_ident_scalar}
\int d^{d-2} {\sigma}  \left< |\varphi (z,\bar z, 0)|^2 \mathcal{D}_z (\sigma) \right> = \partial_z \left< |\varphi(z,\bar z ,0)|^2 \right>,
\end{equation}
we can fix the form of the displacement operator. Indeed, the two-point function in eq.~\eqref{eq:ward_ident_scalar} is fixed by conformal invariance to be of the form
\begin{equation}\label{phi2disp}
\left< |\varphi (z,\bar z, 0)|^2 \mathcal{D}_z (\sigma) \right>  = c_{\varphi^2\CD}(\alpha,\xi,\tilde \xi) \frac{\bar z}{  \left(|\sigma^a|^2+  z \bar z\right)^{d-1}}\,,
\end{equation}
and \eq{ward_ident_scalar} fixes
\begin{equation}
 c_{\varphi^2\CD}(\alpha,\xi,\tilde \xi)=\frac{ \Gamma (\frac{d}{2}-\alpha ) \Gamma
	(\frac{d}{2}+\alpha -1 ) \sin (\pi  \alpha )}{4 \pi^{d}} \left(1-\frac{(d-2)\xi}{\frac{d}{2}-\a-1}-\frac{(d-2)\tilde \xi}{\frac{d}{2}+\a-2}\right) .
\end{equation}
Inserting the ansatz eq.~\eqref{ansatzdisp} into the two-point function eq.~\eqref{phi2disp} we find an equation for the coefficients in eq.~\eqref{ansatzdisp}, which has the natural solution
\begin{align}
 A &=-4\pi \a (1-\a) c_{\varphi^{\dagger} \hat{O}^{\dagger +}_{\a}}c_{\varphi\hat{O}^{+}_{1-\a}}, & B&= -4\pi \a (1+\a) c_{\varphi^{\dagger} \hat{O}^{\dagger +}_{1+\a}}c_{\varphi\hat{O}^{-}_{-\a}}, \\ C&= -4\pi (1-\a)(2-\a)  c_{\varphi^{\dagger} \hat{O}^{\dagger -}_{-1+\a}}c_{\varphi\hat{O}^{+}_{2-\a}}, & D&= -4\pi \a (1-\a) c_{\varphi^{\dagger} \hat{O}^{\dagger -}_{\a}}c_{\varphi\hat{O}^{-}_{1-\a}},
\end{align}
where the $c$ coefficients are the bulk to defect couplings introduced in eq.~\eqref{eq:diff_op}.  Using this data it is easy to see that the displacement two-point function
\begin{equation}
\label{eq:DD_scalar}
\begin{split}
\langle{\mathcal{D}_z(\sigma) \mathcal{D}_{\bar z}(0)}\rangle & =   \frac{C_\CD}{2|\sigma^a|^{2d-2}}
\end{split}
\end{equation}
is determined by a linear combination of the squared OPE coefficients in eqs.~\eqref{bulkdefcoeff}, \eqref{bulkdefcoeff1} and \eqref{bulkdefcoeff2}
\begin{align}
 C_\CD&=A^2  + B^2 +C^2 -D^2 \frac{(\frac{d}{2}-1-\a)(\frac{d}{2}-2+\a)}{\a(1-\a)} \nonumber \\
 &= 2\pi^{1-d} \Gamma \left(\tfrac{d}{2}-\alpha \right) \Gamma \left(\tfrac{d}{2}+\alpha -1\right)\sin (\pi  \alpha )  \left(\alpha  (1-\alpha )+\tfrac{d \alpha ^2 \xi }{\frac{d}{2}-\alpha-1}+\tfrac{d(1-\alpha )^2 \tilde{\xi}}{\frac{d}{2}+\alpha -2}\right),\label{eq:C_D_scalar}
\end{align}
and satisfies the relation \eqref{eq:CD-h-q=2}, when compared with the explicit value of $h$ in eq.~\eqref{eq:h_scalar}, and so $C_\CD \geq 0$ whenever $h\geq 0$. Specialising to $d=4$ and using the normalisation \eq{d1-CD}, we find for a monodromy defect created by $\alpha$ units of flux in a 4d theory of free scalars
\begin{align}
d_1 =   \frac{3}{2} \left( (1-\alpha)^2 \alpha^2 + 4 \xi \alpha^3+ 4 \tilde \xi (1-\alpha)^3 \right).
\end{align}
Note that $ d_1 = d_2$, which agrees with \eq{CD-h-q=2}.

As discussed in section \ref{sec:review}, when 
$\xi=\tilde \xi =0$ the displacement operator $\CD_i$ can be identified as the regular term in the defect OPE of the orthogonal components of the current $J_i$. However, we emphasise that the relation \eq{stress-conv-flux} is not gauge invariant under the shift $\alpha \rightarrow \alpha + \mathbb{Z}$. Thus, expressing $\CD_i$ in terms of $J_i$ makes sense only after we fix a specific gauge, which reflects in the choice of the range of $\alpha$. If we choose the range to be $\alpha \in [0,1)$ as above, we need to change the definition of the current to be 
\begin{equation}\begin{split}
 &\alpha J_z \rightarrow \mathcal{J}_z \equiv i \left( (1-\alpha) \varphi \partial_z \varphi^\dagger + \alpha \varphi^\dagger \partial_z \varphi     \right), \\& \alpha J_{\bar z} \rightarrow  \mathcal{J}_{\bar z} \equiv -i \left( (1- \alpha) \varphi^\dagger \partial_{\bar z} \varphi + \alpha \varphi \partial_{\bar z} \varphi^\dagger     \right).
\end{split}\end{equation}
The displacement operator for $\xi=\tilde \xi=0$ is then
\begin{equation}
\label{eq:disp_scalar}
\CD_z = \left.-2 \pi i \mathcal{J}_z \right|_{z,\bar z =0}, \qquad  \CD_{\bar z} = \left. 2 \pi i\mathcal{J}_{\bar z} \right|_{z,\bar z =0}.
\end{equation}  
Inserting in these expressions the defect OPE of $\varphi$ and $\varphi^{\dagger}$, one finds agreement with the first term of eq.~\eqref{ansatzdisp}.

\subsection{Conical singularities}\label{sec:scalar-renyi}
Here we consider the theory of $n$ copies of the free complex scalar glued along a $(d-1)$-dimensional region $A$, which has a conical singularity at $\partial A$. In a quadratic theory the replicated theory can be conveniently described by the sum of $n$ quadratic field theories with different monodromies~\cite{Casini:2005rm}. In order to see this we may rewrite the theory on the $n$-sheeted manifold as a theory of $n$ independent fields with Lagrangian
\begin{equation}
\mathcal{L}=\sum_{j=1}^{n}\mathcal{L}[\tilde \varphi_j(\vec x, \tau)]\,,
\end{equation}
such that $\tilde\varphi_{j}(\vec x, 0^+)=\tilde \varphi_{j+1}(\vec x, 0^-)$ where $\vec x$ are coordinates along $A$, and $\tau$ parametrises the transverse direction. The index $j$ labels the $n$ fields, and $i=n+1$ is identified with $i=1$. For a bosonic theory, such a boundary condition can be diagonalised by defining 
\begin{equation}
\varphi_k = \sum_{j=1}^{n} e^{2\pi i\frac{k}{n}}\tilde\varphi_j\,, \qquad k=0,\dots,n-1 \, .
\end{equation}
Thus, we see that the field $\varphi_k$ does not shift when one crosses the region $A$ but acquires a phase given by $\alpha = k/n$.
If the theory is quadratic, the new Lagrangian simply becomes
\begin{equation}
	\mathcal{L} = \sum_{k=0}^{n-1}\mathcal{L}[ \varphi_k (\vec x, \tau)]\,.
\end{equation}
This means that in order to obtain the coefficients $h$, $C_J$ and $C_\CD$ for a conical singularity we simply need to sum the ones of the monodromy defect with those specific values of $\alpha$. In order to perform the sums it is crucial to assume that $\xi$ and $\tilde \xi$ are independent of $\a$. Furthermore, since $\tilde \xi$ would give a bad $\a\to0$ limit, including that parameter in the sum will lead to correlation functions that are not vanishing at $n=1$. For this reason, in the following we are going to set $\tilde \xi=0$ and show results as a function of $n$ and $\xi$.

The sum over the monodromies behaves qualitatively different for even and odd $d$. For specific even $d$ it is straightforward to resum the expressions of the defect CFT data, although it is harder to find a generic expression as a function of $d$. For odd $d$ instead the sum is quite involved. Since we are mostly interested in the central charges of 2d defects, we focus on $d=4$. 
For $\langle{|\varphi(x)|^2 }\rangle$ we obtain
\begin{equation}\label{phi2renyi}
\langle{|\varphi(x)|^2 }\rangle=-\frac{n^2-1 -6 (n-1) n \xi  }{48 \pi ^2 n \rho ^2}\,,
\end{equation}
which is in agreement with equation (C.16) of ref.~\cite{Bianchi:2015liz} for $\xi=0$.
While the central charges read\footnote{These functions are usually denoted as $f_a(n)$, $f_b(n)$ and $f_c(n)$, see e.g. \cite{Bianchi:2015liz},  and the precise relation is given by $ b= 12 f_a(n)(n-1)$, $d_1=12f_b(n) (n-1)$ and $d_2= 12f_c(n) (n-1)$.}
\begin{equation}\label{eq:d1-d2-conical-scalar-n}
d_1 =d_2 = \sum_{k=0}^{n-1} \frac{3}{2}\left[\left(1-\frac{k}{n}\right)^2\frac{k^2}{n^2} +4 \xi \frac{k^3}{n^3} \right] = \frac{n^4-1 + 30 (n-1)^2  n^2 \, \xi}{20\, n^3}\,,
\end{equation}
and analogously
\begin{equation}\label{eq:b-conical-scalar-n}
b = \frac{n^4-1 + 30 (n-1)^2  n^2 \, \xi}{60 \, n^3} \,.
\end{equation}
In the limit $n \to 1$ we find
\begin{equation}
\label{eq:scala_limt_n1}
d_1 = d_2 = 3 b = \frac{n-1}{5}+\frac{15 \, \xi -3}{10} (n-1)^2 + \mathcal{O}(n-1)^3.
\end{equation}
Notice that in this limit the term proportional to $\xi$ contributes only at order $\CO(n-1)^2$ as $n\to1$. Therefore, the universal part of the vacuum entanglement entropy, i.e. the linear order $\CO(n-1)$ as $n\to 1$ of the central charges, is unaffected by $\xi$. There is however another consistency check we can make on our result. In \cite{Lewkowycz:2013laa} it was proven that the Weyl anomaly coefficients associated to the R\'enyi entropy must satisfy the following relation (translated in our notation)
\begin{equation}
 d_2=n\left(12 a_{4d} -\partial_n b\right)
\end{equation}
where $a_{4d}=\frac{1}{180}$ for a free complex scalar. We can easily check that our results satisfy this relation only when $\xi=0$. This suggests that $\xi=0$ is the correct boundary condition to be chosen for a conical singularity associated to the replica trick, i.e. to the computation of the R\'enyi entropy. 

We conclude this section by showing that our results are in perfect agreement with ref.~\cite{Fursaev:2013fta}. The ambient Weyl anomaly in $d=4$ reads
\begin{equation}
\left< T^{\mu}_{~\mu} \right> = \frac{1}{16 \pi^2} \int_{\mathcal{M}} d^4 x \sqrt{g} \left[ a_{4d} \mathcal{E}_4 - c_{4d} W^2\right]\,,
\end{equation}
where $\mathcal{E}_4$ and $W^2$ are, respectively, the Euler density and the square of the Weyl tensor. In terms of the curvature tensor they read
\begin{subequations}
\begin{eqnarray}
&&\mathcal E_4 = \mathcal{R}_{\mu\nu\alpha\beta}\mathcal{R}^{\mu\nu\alpha\beta}- 4 \mathcal{R}_{\mu\nu} R^{\mu\nu}+ \mathcal{R}^2\,, \\
&& W^2 = \mathcal{R}_{\mu\nu\alpha\beta}\mathcal{R}^{\mu\nu\alpha\beta} - 2 \mathcal{R}_{\mu\nu} \mathcal{R}^{\mu\nu}+ \frac{1}{3} \mathcal{R}^2\,.
\end{eqnarray}
\end{subequations}
In ref.~\cite{Fursaev:2013fta}, it was shown that in the presence of a conical singularity
\begin{subequations}
\begin{eqnarray}
&& \mathcal E_4 \longrightarrow n \mathcal E_4 + 8\pi(1-n) \mathcal{R}_\Sigma \, \delta(\Sigma) + \mathcal{O}(1-n)^2\,, \\
&& W^2 \longrightarrow n W^2 + 8\pi(1-n) \left[ \oII^\mu_{ab}\oII_\mu^{ab} - d_2 \, W_{ab}{}^{ab} \right] \, \delta(\Sigma) + \mathcal{O}(1-n)^2 \,,
\end{eqnarray}
\end{subequations}
which implies
\begin{equation}\label{eq:conical-trace-anomaly-n}
\left<T^{\mu}_{~\mu} \right>\Big|_{\Sigma} = - \frac{1}{2\pi} (n-1) \left( a_{4d} \, \mathcal{E}_2 + c_{4d}\,\oII^\mu_{ab}\oII_\mu^{ab} - c_{4d} \, W_{ab}{}^{ab} \right) + \mathcal{O}(n-1)^2\,.
\end{equation}
Substituting the known values of $a_{4d}= 1/180$ and $c_{4d}= 3/180$, we find perfect agreement with our results in \eq{scala_limt_n1}. 

\subsection{Entanglement entropy}\label{sec:scalar-EE}
\begin{figure}
	\hspace{0.25cm}
	\begin{subfigure}[b]{0.45\textwidth}
		\includegraphics[width=\textwidth]{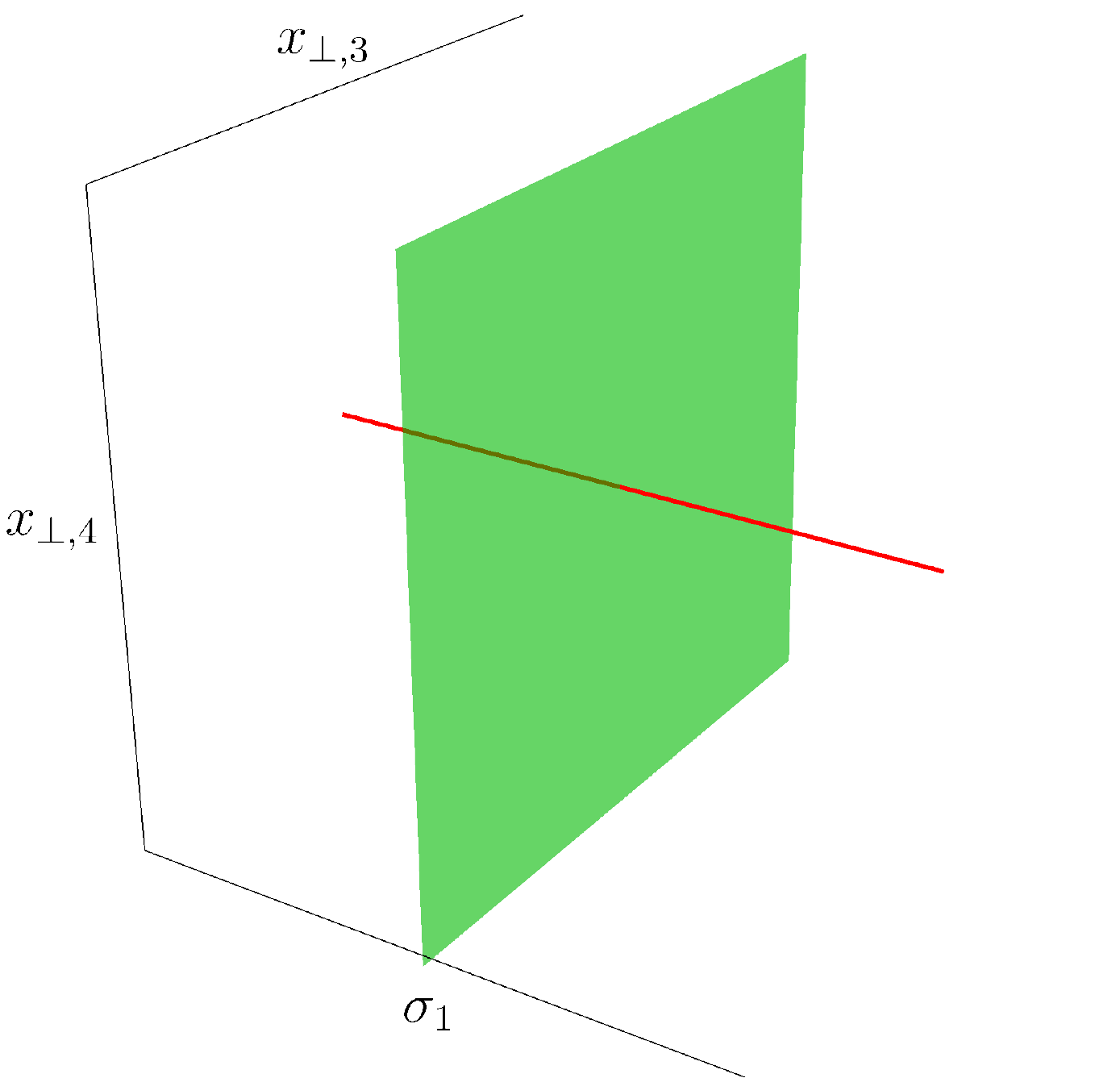}
	\end{subfigure}
\hspace{1.cm}
	\begin{subfigure}[b]{0.45\textwidth}
		\includegraphics[width=\textwidth]{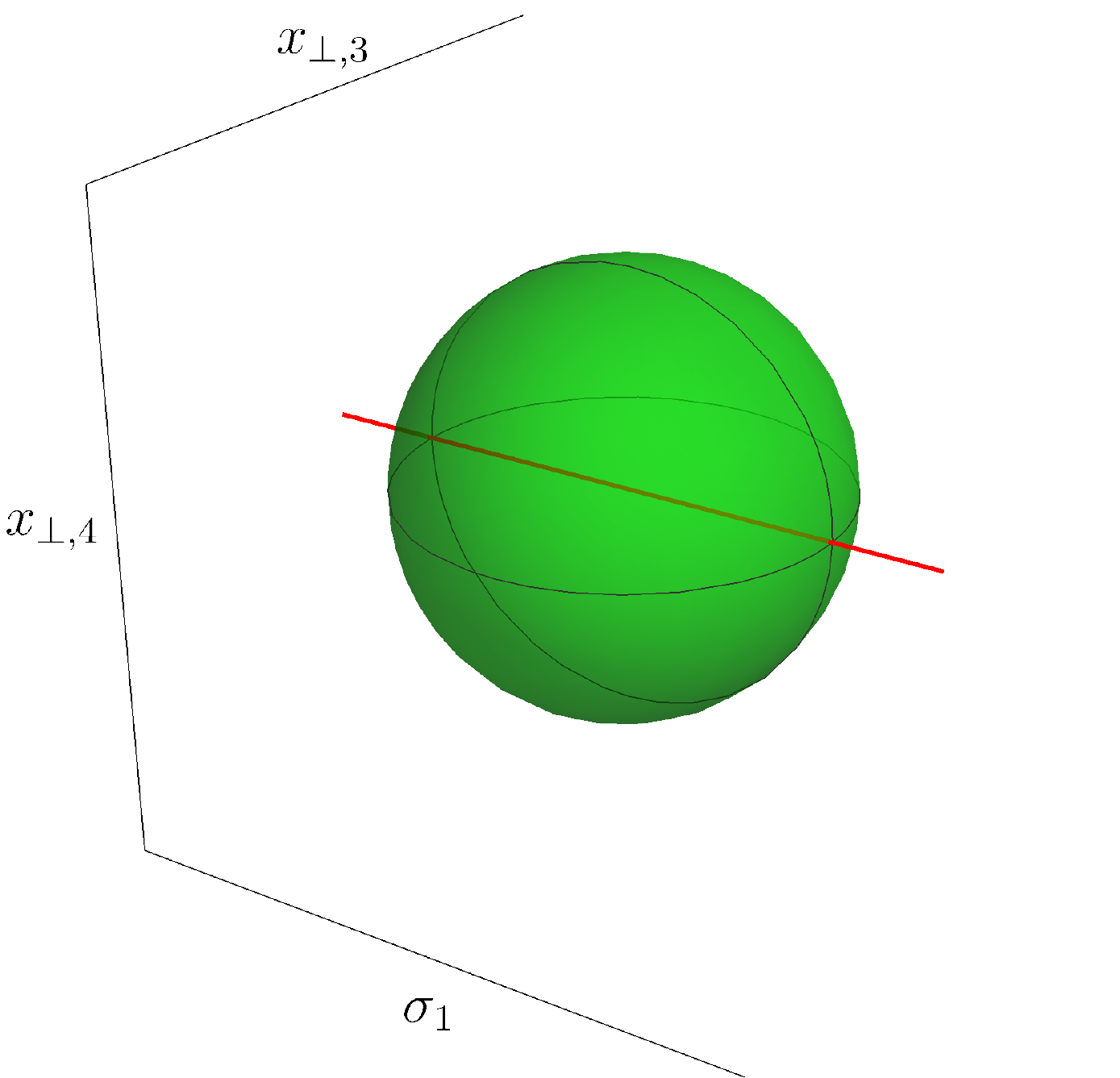}
	\end{subfigure}
	\caption{\textit{Left:} Sketch of the configuration we employ to compute the defect contribution to EE in $d=4$. The figure is an instance in time at $\tau =0$. The defect extends along the $\sigma_1$ direction, while the entangling surface extends along the remaining orthogonal directions and intersects the defect at $\sigma_1=0$. \\
	\textit{Right:} A spherical entangling surface centred on the defect on the time-slice $\tau=0$. This configuration corresponds to the one studied in \cite{Jensen:2018rxu}. Its defect contribution is given by twice \eq{SEE_log}.\label{fig:EE_plane}}
\end{figure}

In this section, we will compute the entanglement entropy contribution of the monodromy defect in the free complex scalar theory.
In the following, we assume that the defect is flat, and the region $A$ is the half space orthogonal to the defect. The entangling surface $\partial A$ is defined by $\tau=\sigma^1=0$ as depicted on the left-hand side of figure \ref{fig:EE_plane}. This configuration is the most symmetric case, and it is conformally related to a spherical entangling surface at $\tau = 0$ centred on the flat defect, as depicted on the right-hand side of figure \ref{fig:EE_plane}. We are interested in the $d=4$ case where the entangling surface intersects the monodromy defect in a single point and the EE shows a logarithmic divergence whose coefficient is given by \cite{Jensen:2018rxu}
\begin{equation}
\label{eq:SEE_log}
s_\text{\tiny log} = \frac{1}{6} \left( b- \frac{d_2}{3}\right).
\end{equation}

Here we employ the heat kernel method following ref.~\cite{Berthiere:2016ott} to show that the result is in perfect agreement with eq.~\eqref{eq:SEE_log}. 
As emphasised in \cite{Lewkowycz:2013laa, Herzog:2014fra, Lee:2014zaa, Herzog:2016bhv,Fursaev:2016inw}, in the case of the conformal free scalar, the entanglement entropy consists of two different contributions, one that can be computed from the heat kernel and a second one coming from the coupling to the scalar curvature~$\mathcal{R}$. To show this, we can consider the coupling to $\mathcal{R}$ as a deformation of the theory. The conical singularity leads to a scalar curvature of the form $\mathcal{R}=2 (n-1) \delta(\rho)/(n\rho)$ \cite{Solodukhin:1994yz} which implies the following contribution to the action
\begin{equation}
\label{eq:I_R_scalar}
I_{\text{\tiny scalar}}^{\mathcal{R}} [n]=  \pi\frac{d-2}{d-1} \frac{n-1}{n}\int_{\partial A} d^{d-2}\sigma \, \left| \varphi\right|^2\,.
\end{equation} 
Note that the delta function coming from $\mathcal{R}$ localises the integral over the entangling surface $\partial A$.
Thus the full partition function may be written as 
\begin{equation}
Z[n] = Z^{\mathcal{R}=0}[n] \left< e^{-I^{\mathcal{R}}_{\text{\tiny scalar}}} \right>,
\end{equation}
where $Z^{\mathcal{R}=0}[n]$ denotes the partition function of the theory without the coupling to $\mathcal{R}$. Applying the relation \eqref{eq:eelim}, we find
\begin{equation}
\label{eq:S_EE_total}
S_{A} = S_{A}^{\mathcal{R}=0} + S_{A}^{\mathcal{R}} \,,
\end{equation}
where $S_{A}^{\mathcal{R}=0}$ is the EE for the theory in the absence of the coupling to $\mathcal{R}$, and $S_{A}^{\mathcal{R}}$ is the contribution of eq.~\eqref{eq:I_R_scalar} which reads \cite{Lewkowycz:2013laa, Herzog:2014fra, Lee:2014zaa, Herzog:2016bhv}
\begin{equation}
S_{A}^{\mathcal R} = -\frac{\pi (d-2)}{ (d-1)} \int_{\partial A} d^{d-2} \sigma \left<|\varphi(x)|^2 \right>.
\end{equation}
This last contribution is straightforward to compute. Setting $d=4$ and using \eq{scalar-one-point-function-4d}, we find
\begin{equation}
\begin{split} \label{REEcont}
S_{A}^{\mathcal R} & =  \tfrac{(1-\alpha ) \alpha -2 \xi \alpha-2\tilde \xi (1-\a) }{12\pi}\int_{\partial A} d^{2} x \, \frac{1}{r^2} =  \tfrac{ (1-\alpha ) \alpha -2 \xi-2\tilde \xi (1-\a) \alpha }{12\pi} \int_0^{2\pi} d\phi\int_\varepsilon^L \frac{dr}{r}\\
& = \frac{(1-\alpha ) \alpha -2 \xi \alpha-2\tilde \xi (1-\a) }{6} \log \left( \tfrac{L}{\varepsilon}\right).
\end{split}
\end{equation}

The first contribution to \eqref{eq:S_EE_total} can be computed with the heat kernel method as done in \cite{Berthiere:2016ott} for the case of a real scalar in a bCFT.
Below we just give the definition of the heat kernel and we refer the reader to the detailed review \cite{Vassilevich:2003xt}.

Given the differential operator $D$, the corresponding heat kernel $K(s; x,y; D)$ is defined as
\begin{subequations}
\begin{eqnarray}
&& \left( \partial_s + D_x \right) K\left(s; x, y; D  \right) =0, \\
&& K(0; x, y; D)= \delta (x,y)\,.
\end{eqnarray}
\end{subequations}
The propagator then is
\begin{equation}
\label{eq:propagator_conical}
G(x,y) = \int_0^{+ \infty} ds \, K \left( s; x ,y; D \right),
\end{equation}
while the effective action may be written as
\begin{equation}
W = - \frac{1}{2} \int_0^{\infty} \frac{ds}{s} \text{Tr} K (s;  D)\,,
\end{equation}
where
\begin{equation}
\text{Tr} K (s; D) = \int d^d x \sqrt{g} \,K (s; x,x; D)\, .
\end{equation}
The strategy is then to compute the heat kernel in the presence of a conical singularity. To exploit the symmetry of the problem we consider polar coordinates around the entangling surface, writing $\tau= r \sin \phi$ and $\sigma^1 = r \cos \phi$ with $0 \le \phi \le 2\pi$. In this coordinate system the conical singularity is introduced by  making $\phi$ periodic with period $2\pi \, n$ (here $n$ is assumed to be real and $n > 1$). The heat kernel in the presence of such a conical singularity was found in \cite{Berthiere:2016ott,FURSAEV199453} as a deformation of $K(s; x,y; D)$:
\begin{equation}
\label{eq:heat_kernel_conical}
K_n (s; \phi, \phi'; D) = K (s; \phi - \phi'; D) + \frac{i}{4\pi n} \int_\Gamma  d \omega\, \cot \left(\frac{\omega}{2 n } \right)  K (s; \phi - \phi'+ \omega; D) \,,
\end{equation}
where the contour $\Gamma$ is given by two vertical lines going from $(-\pi + i \infty)$ to $(-\pi -i \infty)$ and from $(\pi - i\infty)$ to $(\pi + i \infty)$. It intersects the real axis twice between the poles of $\cot \omega/(2 \, n)$, once between $-2 \pi n $ and $0$, and once between $0$ and $2 \pi n$.

For simplicity we will illustrate the computation only for the regular part of the propagator, namely for $\xi=\tilde \xi=0$, being the generalisation to the singular part straightforward. The heat kernel of our case can be easily extracted from eq.~\eqref{eq:heat_kernel_prop_no_A}, and we have
\begin{equation}
K (s; x, x',\alpha) = \frac{1}{(4 \pi s)^{\frac{d}{2}}}\sum_m  e^{i m\theta}e^{-(\rho^2+\rho'^2+(\sigma-\sigma')^2)/(4 s)}   I_{|m-\alpha|} \left(\frac{\rho\,\rho'}{2s}\right) + c.c.\,,
\end{equation}
where $c.c.$ stands for complex conjugate. We now compute $\text{Tr}K_n$. By setting $\rho=\rho'$ and  $\sigma^a=\sigma'^a$ the integrals become
\begin{equation}
\begin{split}
&\text{Tr}K_n  = \frac{4\pi n}{(4 \pi s)^{\frac{d}{2}}}\int_0^\infty d \rho \int_0^{2\pi} d\theta \int_0^{\infty} dr \,   \int d^{d-4} \tilde \sigma \,  \rho \,  e^{-{\rho^2}/(2 s)}  \mathcal{I}^{(1)}_\alpha \left(\frac{\rho^2}{2s}\right)   \\
& + \frac{8\pi^2 n}{(4 \pi s)^{\frac{d}{2}}} \frac{i}{4\pi n} \int_\Gamma d\omega\cot\frac{\omega}{2n} \left[ \int_0^\infty d\rho \,\int_0^{\infty} dr \int d^{d-4}\tilde \sigma \,  \rho \, r  e^{-{\rho^2}/(2 s)- r^2 \sin^2 (\omega/2)/ s} \,  \mathcal{I}^{(1)}_\alpha \left(\frac{\rho^2}{2s}\right)  \right],
\end{split}
\end{equation}
where $\tilde \sigma$ parametrises the coordinates along the defect in common with the entangling surface, and the function $\mathcal{I}^{(1)}_\alpha$ was defined in eq.~\eqref{eq:I1_sum_m}.
After a change of variables we have
\begin{equation}
\label{eq:TrKn}
\begin{split}
\text{Tr}K_n  =\;&   \frac{4\pi n \, L^{d-2}}{(4 \pi s)^{\frac{d}{2}}} s  \int_0^{\frac{L^2}{2 s}} d \zeta  \,    e^{-\zeta}   \mathcal{I}^{(1)}_\alpha (\zeta)   \\
& +  \frac{8\pi^2 n \, L^{d-4}}{(4 \pi s)^{\frac{d}{2}}} s^2 \left[\frac{i}{8\pi n} \int_\Gamma d\omega\cot\left(\frac{\omega}{2n} \right) \frac{1}{\sin^2 \frac{\omega}{2}}\right] \left[  \int_0^{\frac{L^2}{2 s}} d \zeta     e^{-\zeta } \,  \mathcal{I}^{(1)}_\alpha \left(\zeta\right)  \right],
\end{split}
\end{equation}
where we introduced the cut-off $L$ to regulate the infrared divergence. 
The contour integral over $\omega$ can be evaluated exactly, and it gives
\begin{equation}
\frac{i}{8\pi n} \int_\Gamma d\omega\cot \left(\frac{\omega}{2n} \right)\frac{1}{\sin^2 \frac{\omega}{2}} = \frac{1}{6 \, n^2}(1-n^2)\,.
\end{equation}
The heat-kernel contribution to the entanglement entropy $S_{A}^{\mathcal{R}=0}$ can be found from the effective action as follows
\begin{equation}
S_{A}^{\mathcal{R}=0} =\left. \left( n \partial_n - 1 \right) W^{\mathcal{R}=0}\left[ n \right] \right|_{n=1}.
\end{equation}
The term linear in $n$ in eq.~\eqref{eq:TrKn} does not contribute. Thus, the whole contribution to the EE comes from the second one, and we find
\begin{equation}\label{eq:scalar-EE-heat-kernel-1}
S_{A}^{\mathcal{R}=0} = \frac{1}{6} \int_{\varepsilon^2}^{+\infty} ds\,  \left[\frac{8\pi^2  \, L^{d-4}}{(4 \pi s)^{\frac{d}{2}}} s    \int_0^{\frac{L^2}{2 s}} d \zeta     e^{-\zeta } \,  \mathcal{I}^{(1)}_\alpha \left(\zeta\right) \right],
\end{equation}
where we introduced the UV cutoff $\varepsilon$ to regulate the small $s$ behaviour of the integral. The resulting integrals are quite cumbersome but their computation is straightforward, and for $d=4$ we find
\begin{equation}
S_{A}^{\mathcal{R}=0}  =  \frac{L^2}{24\varepsilon^2}-\frac{(1-\alpha)\alpha}{6} \log \left( \frac{L}{\varepsilon}\right)+ \mathcal{O}(1)\,.
\end{equation}
With the singular modes of $\varphi$ we find instead
\begin{equation}
S_{A}^{\mathcal{R}=0}  =  \frac{L^2}{24\varepsilon^2}-\frac{(1-\alpha ) \alpha -2 \xi \alpha-2\tilde \xi (1-\alpha)}{6} \log \left( \frac{L}{\varepsilon}\right)+ \mathcal{O}(1).
\end{equation}
which precisely cancels eq.~\eqref{REEcont}. Thus, in the theory of a free complex scalar the contribution of the monodromy defect to the universal part of the entanglement entropy vanishes in $d=4$.  This is in agreement with the relation \eqref{eq:SEE_log} proven in \cite{Jensen:2018rxu}, and it confirms our findings in eqs.~\eqref{eq:b-free-scalar} and \eqref{eq:d2-free-scalar}.

\section{Free Fermion} \label{sec:fermion}


In this section, we compute $h$, $C_J$, and $C_\CD$ for monodromy defects in a theory of free Dirac fermions in arbitrary dimension $d\geq 3$.  As in the case of the free scalars, we will eventually specialise to $d=4$ in order to connect the results to the defect Weyl anomaly coefficients $b$, $d_1$ and $d_2$, which we compute for the monodromy defect and a conical singularity.  Lastly, we will compute defect EE as a check on $b$ and $d_2$ in $d=4$. 

Our starting point is the background geometry described in \eq{background-metric-R-d} on which we introduce frame fields $e^M$ with components $e^M=e^M{}_\mu \, dx^\mu$ for $M=0, \ldots, d-1 $.  We denote the determinant of the components $e^M{}_\mu$ by $|e|$, and their matrix inverse by $e^\mu{}_M$. The frame fields obey $g_{\mu\nu}= e^{M}{}_\mu e^{N}{}_\nu \eta_{MN}$ and $\eta_{MN} = e^\mu{}_M e^\nu{}_N g_{\mu\nu}$, where $\eta_{MN}$ is the flat $d$-dimensional Minkowski metric. In the following we will choose the frame 
\begin{align}\label{eq:frame-1}
e^{0}=dt\,,\quad e^{1} = d\rho\,, \quad e^{2} =\rho d\theta\,,\quad e^{\beta} = d\sigma^\beta\,,
\end{align}
where $\beta=3, \ldots, d-3$ labels the spatial directions along the defect with coordinates $\{\sigma^\beta\}$. 

On this background, we place a single free Dirac fermion $\psi$, which in $d$ dimensions has $2^{\lfloor\frac{d}{2}\rfloor}$ components.  Turning on a background gauge field $A$ for the vector $U(1)_V$ symmetry under which a Dirac fermion of unit charge is rotated by $\psi \to e^{i \vartheta}\psi$ and $\psi^\dagger \to e^{-i \vartheta}\bar\psi$, the Dirac action can then be written as
\begin{equation}\label{eq:dirac-action}
I_\mathrm{fermion}=-\int d^dx \, |e|\,  \bar\psi \Dslash \psi\,,
\end{equation}
where $\bar \psi = i\psi^\dagger \gamma^0$, and we denote the Dirac operator in the presence of a background gauge field $A_\mu$ as $\Dslash= \gamma^\mu(\nabla_\mu - i A_\mu)$. Here, $\nabla_\mu = \partial_\mu +\Omega_\mu$ with $\Omega_\mu = \frac{1}{8} \omega_\mu{}^{MN} \left[\gamma_{M},\gamma_{N}\right] $ and $\omega_{\mu}{}^{MN}$ being the spin-connection. We denote the $\gamma$-matrices in curvilinear coordinates by $\gamma_\mu = e^{M}{}_\mu \gamma_{M}$, which obey the Clifford algebra $\gamma_\mu\gamma_\nu+\gamma_\nu\gamma_\mu=+2 g_{\mu\nu}$.

\subsection{Mode expansion and propagator}

The fermion propagator can be computed by employing the same methods as in \sn{scalar} for the free scalar. We refer to appendix~\ref{app:fermion} for a more detailed discussion. In a suitable Clifford algebra basis, the Dirac equation can be solved by a spinor with $2^{\lfloor\frac{d}{2}\rfloor-1}$ components proportional to $J_{\pm (n-\alpha)}$ and the other $2^{\lfloor\frac{d}{2}\rfloor-1}$ components proportional to $J_{\pm (n+1-\alpha)}$ with $n\in \mathbb{Z}$. Note that for $n \neq 0$ with $0<\alpha<1$, one choice of sign results in a spinor with all components either regular or all divergent as $\rho \to 0$. For the divergent solutions at least one of the components diverges as $\CO(\rho^{-1})$ or worse, which makes them physically inadmissible. 

For $n = 0$, both $(\pm)$ solutions are admissible provided that $\alpha \neq 0,1$. In this case, half of the components are regular whereas the other half have divergent behaviour between $\CO(\rho^{-1})$ and $\CO(1)$. This tamer behaviour makes both solutions acceptable. Generally, one can introduce a parameter $\xi\in [0,1]$ which interpolates between these two solutions, and study the theory's correlation functions as a function of $\xi$. 

In order to write down the fermion mode expansion, we adopt complex coordinates in the directions normal to the defect $z=\rho e^{i\theta}$ and $\bar z=\rho e^{-i\theta}$. Moreover, it is convenient to perform a gauge transformation $A \to A - \alpha d\theta$ in order to remove the background gauge field $A$ from correlation functions at the expense of introducing an extra factor of $e^{-i\alpha \theta}$ in the fermion mode expansion. The mode expansion in this gauge can be written as 
\begin{equation}\label{eq:mode_decomposition}
\psi = \psi_{-\alpha} \left(\frac{z}{\rho}\right)^{-\alpha+\frac{1}{2}}+ \psi_\alpha \left(\frac{\bar{z}}{\rho}\right)^{\alpha-\frac{1}{2}}+\sum_{n =1}^\infty \psi_{n-\alpha} \left(\frac{z}{\rho} \right)^{n-\alpha+\frac{1}{2}}+\sum_{n = 1}^{\infty} \psi_{n+\alpha} \left(\frac{\bar{ z}}{\rho}\right)^{n+\alpha-\frac{1}{2}},
\end{equation}
where 
\begin{equation}\label{eq:n!=0_modes}
\psi_{n\pm\alpha}=\sum_{s=1}^{2^{\lfloor\frac{d}{2}\rfloor-1}}\int_{-\infty}^\infty d^{d-3}\vec{k}\int_0^\infty d k_\rho \left( \tilde{f}(k) a_{\mp n}^s(k) u^s_{\mp n,k}   + \tilde{f}^*(k) b_{\mp n}^{s \dagger}(k) v^s_{\mp n,k}   \right),
\end{equation}
and 
\begin{equation}
\tilde{f}(k)=\frac{e^{-i\omega t+i\vec{k}\cdot\vec{\sigma}}}{(\sqrt{2\pi})^{d-2}}\sqrt{\frac{k_\rho}{2^{\lfloor\frac{d}{2}\rfloor}\omega}} \,.
\end{equation}
At $n=0$, there are two solutions $\psi_{-\alpha}$ and $\psi_\alpha$ with $\xi \in [0,1]$ interpolating between them:
\begin{subequations}\label{eq:n=0-modes}
\begin{align}\label{eq:-a_mode}
\psi_{-\alpha}&=\sqrt{1-\xi}\sum_{s=1}^{2^{\lfloor\frac{d}{2}\rfloor-1}}\int_{-\infty}^\infty d^{d-3}\vec{k}\int_0^\infty d k_\rho \left( \tilde{f}(k) a_{ 0}^{(+) s}(k) u^{(+)s}_{ 0,k}   + \tilde{f}^*(k) b_{0}^{(+)s \dagger}(k) v^{(+)s}_{0,k}   \right),\\
\label{eq:a_mode}
\psi_{\alpha}&=\sqrt{\xi}\sum_{s=1}^{2^{\lfloor\frac{d}{2}\rfloor-1}}\int_{-\infty}^\infty d^{d-3}\vec{k}\int_0^\infty d k_\rho \left( \tilde{f}(k) a_{ 0}^{(-) s}(k) u^{(-)s}_{ 0,k}   + \tilde{f}^*(k) b_{0}^{(-)s \dagger}(k) v^{(-)s}_{0,k}   \right).
\end{align}
\end{subequations}
In the above equations, $u^s_{\mp n,k}$ and $v^s_{\mp n,k}$ for $n \geq 1$ are spinors whose components involve Bessel functions $J_{n\pm \alpha} (k_\rho \rho)$ and $J_{n\mp(1-\alpha)} (k_\rho \rho)$ and, in the frame we are using, are purely functions of $\rho$. For $n=0$, the spinors $u^{(\pm)s}_{0,k}$ and $v^{(\pm)s}_{0,k}$ have components $J_{\mp \alpha} (k_\rho \rho)$ and $J_{\pm(1-\alpha)} (k_\rho \rho)$. See appendix~\ref{app:fermion} for more details. Explicit solutions for the spinors in $d=4$ are given in eqs.~\eqref{eq:u} and \eqref{eq:v}. 

In order to compute the propagator, one can canonically quantise the mode expansions. However, this is cumbersome in general dimensions, and we will not present this method here. Instead, we will make use of the scalar propagator \eq{heat_kern_scal_gen_d} to directly achieve the same result.  It is well-known that in the absence of a monodromy, one can write the fermion propagator $G_F(x,x') = -\langle \bar\psi(x) \psi(x')\rangle$ in terms of the scalar propagator $G_S(x,x') = \langle{\vphi}^\dagger(x)\vphi(x')\rangle$. Schematically, $G_F (x,x{}^\prime) = -\Delslash G_S(x,x{}^\prime)$, where $\Delslash\equiv \gamma^\mu\nabla_\mu$ is explicitly with respect to the unprimed coordinates.

In the presence of a monodromy, the relation between $G_F$ and $G_S$ is modified as follows. Let $P_\pm = \frac{1}{2}(\mathds{1} \pm i \gamma_{1}\gamma_{2})$, where $\mathds{1}$ is the $2^{\lfloor\frac{d}{2}\rfloor}\times 2^{\lfloor\frac{d}{2}\rfloor}$ dimensional identity matrix. Then the fermion propagator in the presence of a monodromy defect takes the form\footnote{This is similar to the approach taken in~\cite{Herzog:2019bom}. We thank Christopher Herzog for pointing that out to us, and suggesting to use such a relation.}
\begin{align}\label{eq:fermion-propagator-1}
G_{F,\alpha, \xi}(x,x{}^\prime) = -\Delslash\left(P_- e^{i(\theta-\theta{}^\prime)/2}  G_{S,\alpha,1-\xi,0}(x,x{}^\prime) +P_+ e^{-i(\theta-\theta{}^\prime)/2}  G_{S,\alpha,0,\xi}(x,x{}^\prime)\right)\,,
\end{align}
where $G_{S,\alpha,\xi,\tilde{\xi}}(x,x{}^\prime)$ is the scalar propagator with singular modes as defined in \eq{two-point function_gaug_transf}. The factors of $e^{\pm i(\theta-\theta')/2}$ are due to working in the rotating frame \eq{frame-1} and ensure that the modes in the scalar Green's functions combine correctly into modes of the spinor Green's function. Notice that for any $\xi\in[0,1]$, one always needs a singular mode in at least one of the two scalar Green's functions because both $n=0$ modes of the fermion are singular.

In computing $h$, $C_J$, and $C_\CD$ we will only need to compute particular derivatives of~$G_S$ evaluated in either the coincident or defect limits, and so we will not display $G_F$ in full detail. Notice that in writing \eq{fermion-propagator-1} in terms of \eq{two-point function_gaug_transf} we have implicitly performed a Wick rotation to Euclidean signature $t\to-i\tau$.

\subsection{Correlation functions and central charges}

Equipped with the general form of the propagator in \eq{fermion-propagator-1}, we now compute the one-point functions of the stress tensor $T_{\mu\nu}$ and the $U(1)_V$ current $J_\mu$, as well as the two-point function of the displacement operator $\mathcal{D}_i$.

\subsubsection{One-point function of $\boldsymbol{T_{\mu\nu}}$}

We commence with an analysis of the components of the stress tensor one-point function from which we will extract $h$ in general $d\geq 3$. Specialising to $d=4$, we will thus obtain the defect Weyl anomaly coefficient $d_2$.  

The classical stress tensor can be computed by varying \eq{dirac-action} with respect to the frame fields $e^M{}_\mu$. The result is
\begin{align}
T_{\mu\nu} = \frac{1}{2}\bar{\psi}\gamma_{(\mu}\overset{\leftrightarrow}{D}_{\nu)}\psi\,,
\end{align}
where the parenthesis denotes symmetrisation over the indices, and 
\begin{align}
\bar{\psi}\gamma_\mu\overset{\leftrightarrow}{D}_\nu\psi = \bar{\psi}\gamma_\mu(\partial_\nu +\Omega_\nu -iA_\nu)\psi - (\partial_\nu\bar\psi)\gamma_\mu \psi + \bar\psi (\Omega_\nu -i A_\nu)\gamma_\mu \psi\,.
\end{align}
The one-point function of the stress tensor can then be written in terms of the fermion propagator $G_{F,\alpha,\xi}$ as follows:
\begin{align}
\langle T_{\mu\nu}(x)\rangle = -\frac{1}{2}\lim_{x^\prime\to x} \Tr\left[\left(\gamma_{(\mu}\partial_{\nu)}-\gamma_{(\mu}\partial^\prime_{\nu)}+\gamma_{(\mu}\Omega_{\nu)}+\Omega_{(\mu} \gamma_{\nu)}\right)G_{F,\alpha,\xi}(x,x^\prime)\right],
\end{align}
where $\partial^\prime_{\nu }$ denotes the derivative with respect to primed coordinates, and $A$ has been gauged away.

In order to compute $h$, it will suffice to analyse a single non-vanishing component of $\langle T_{\mu\nu}\rangle$ with others following from the tensor structures dictated by the remaining conformal symmetry.  To that end, let us consider 
\begin{equation}
\langle T_{\tau\tau}\rangle =-\frac{1}{2} \lim_{x' \to x} \Tr\left[ \gamma^0 \left(\partial_\tau -   \partial_{\tau^\prime} \right)G_{F,\alpha,\xi}(x,x')\right].
\end{equation}
To evaluate $\langle T_{\tau\tau}\rangle$, we need to substitute \eq{fermion-propagator-1} and compute the coincident limit of various combinations of derivatives acting on the scalar propagator. Each term contains at least one derivative of $G_S$ with respect to $\tau$ or $\tau'$. Using the integral representation given by \eq{heat_kern_scal_gen_d}, it is easy to see that all the terms with a single derivative with respect to $\tau$ or $\tau'$ vanish in the coincident limit $x \to x^\prime$ due to the appearance of $\eta$ in the integrand. The only terms that survive are
\begin{align}
\langle T_{\tau\tau} \rangle = \frac{1}{2}\sum_{\varsigma=\pm}\Tr(\gamma^{0}P_{\varsigma}\gamma^{0})(\partial_\tau^2 - \partial_{\tau^\prime}\partial_\tau)G_{\varsigma}\, ,
\end{align}
where we defined
\begin{align}
G_\varsigma \equiv \begin{cases} 
\lim_{x^\prime\to x} G_{S,\alpha,0,\xi}(x,x{}^\prime)\,, &\qquad \varsigma = +\\
\lim_{x^\prime\to x} G_{S,\alpha,1-\xi,0}(x,x{}^\prime)\,, &\qquad \varsigma = -\,.
\end{cases}
\end{align}

Evaluating the traces using standard $\gamma$-matrix identities and utilising the fact that $\partial_{\tau^\prime}\partial_\tau G_{\pm} = - \partial^2_\tau G_\pm$, one finds that
\begin{align}
\langle T_{\tau\tau} \rangle = 2^{\lfloor\frac{d}{2}\rfloor-1} \partial_\tau^2(G_+ + G_-).
\end{align}
All that remains is to compute the $\partial_\tau^2 G_\vsig$, which can be done straightforwardly by using the integral representation of the scalar propagator \eq{heat_kern_scal_gen_d}. Using the same scheme as in \sn{scalar}, one finds
\begin{equation}\label{eq:Dt_G+}
\begin{split}
\partial_\tau^2 G_+ &= -\frac{1}{2 (2 \pi )^{d/2} \rho ^d}\int_0^\infty d\zeta\,e^{-\zeta} \zeta ^{\frac{d}{2}-1}  (\mathcal{I}^{(1)}_\alpha (\zeta)+\xi (I_{-1+\alpha}(\zeta)-I_{1-\alpha}(\zeta)) \\
&=\frac{(2 d \xi-2 \alpha -d) \Gamma \left(\frac{d}{2}-\alpha \right) \Gamma \left(\frac{d}{2}+\alpha \right)\sin (\pi  \alpha ) }{2^{d+1} \pi ^{\frac{1}{2} (d+1)} d\, \Gamma \left(\frac{d+1}{2}\right)} \frac{1}{\rho ^{d}}\,.
\end{split}
\end{equation}
Computing $\partial_\tau^2 G_-$ follows similarly,
\begin{align}\label{eq:Dt_G-}
\partial_\tau^2 G_- &=\frac{  (2 \alpha -2 d \xi +d-2) \Gamma \left(\frac{d}{2}-\alpha +1\right) \Gamma \left(\frac{d}{2}+\alpha -1\right)\sin (\pi  \alpha )}{2^{d+1} \pi ^{\frac{1}{2} (d+1)} d\, \Gamma \left(\frac{d+1}{2}\right)}\frac{1}{\rho ^{d}}\,.
\end{align}
Combining $\partial_\tau^2 G_+ +\partial_\tau^2 G_-$, we thus arrive at
\begin{align}\label{eq:fermion-CT-d}
\langle T_{\tau\tau} \rangle = -\frac{((1-2 \alpha ) d \xi +\alpha  (2 \alpha +d-2)) \Gamma \left(\frac{d}{2}-\alpha \right) \Gamma \left(\frac{d}{2}+\alpha -1\right)\sin (\pi  \alpha )}{2^{d+1-\lfloor\frac{d}{2}\rfloor} \pi ^{\frac{1}{2} (d+1)} d\, \Gamma \left(\frac{d+1}{2}\right)} \frac{1}{\rho^d}\,.
\end{align}
The stress tensor one-point function for $\xi=0$ was previously computed in~\cite{Dowker:1987pk}. Using \eq{defect-T-one-point-function}, we find
\begin{align}\label{eq:fermion-h}
h = \frac{  ((1-2 \alpha ) d \xi +\alpha  (2 \alpha +d-2)) \Gamma \left(\frac{d}{2}-\alpha \right) \Gamma \left(\frac{d}{2}+\alpha -1\right)\sin (\pi  \alpha ) }{2^{d-\lfloor\frac{d}{2}\rfloor}\pi ^{\frac{1}{2} (d-1)} d\, \Gamma \left(\frac{d+1}{2}\right)}\,.
\end{align}
It is clear from the functional form that $h\geq 0$ for any value of $\alpha\in [0,1]$ and $\xi \in [0,1]$ in $d\geq 3$, and $h=0$ for all $\xi\in [0,1]$ at $\a =0$ and $\a=1$. To extract $d_2$, we set $d=4$ and use \eq{d2-h-4d}, which gives
\begin{align}\label{eq:fermion-d2}
d_2 = 3\alpha(1-\alpha )   \left(\alpha (1+\alpha)+2\xi(1-2 \alpha) \right).
\end{align}
Note that $d_2$ is positive, which is expected if the ANEC holds in the presence of a 2-dimensional defect \cite{Jensen:2018rxu}. 

\subsubsection{One-point function of $\boldsymbol{J_\theta}$}

In this subsection, we will analyse the current one-point function in order to extract $C_J$ in general $d\geq 3$ and then compute $b$ in $d=4$.  That is we will employ the same method to compute $b$ as in \sn{scalar} by integrating $C_J$.  From the form of the $U(1)_V$ current 
\begin{align}
J^\mu(x) = i\bar\psi \gamma^\mu \psi(x)\,,
\end{align}
and the fermion propagator in \eq{fermion-propagator-1}, we can express $\langle J_\theta\rangle$ as
\begin{align}\label{eq:fermion-J}
\langle J_\theta \rangle = -i\lim_{\theta\to\theta'}\sum_{\vsig=\pm} \left( \Tr[\gamma^\mu P_\vsig\gamma_\theta]\partial_\mu (e^{-i \vsig (\theta-\theta^\prime)/2}G_{\vsig})+\Tr[\gamma^\theta \Omega_\theta P_\vsig\gamma_\theta] G_{\vsig} \right),
\end{align}
where $\partial_\mu$ is again with respect to the unprimed coordinate. All other components of $\langle J_\mu \rangle$ will turn out to vanish, so we do not display them here.

To evaluate \eq{fermion-J}, we need to compute the coincident limit ($\eta \to 1$) of the scalar propagator $G_S$ and its first derivatives. This limit has already been computed for $G_+$ in \sn{scalar} and is given by the sum of eqs.~\eqref{eq:phisq_d_dim} and~\eqref{eq:phisq_d_dim-sing}. For convenience we reproduce the result here:
\begin{equation} \label{eq:G+_coinc}
\begin{split}
G_+&=\frac{1 }{2 (2 \pi )^{d/2} \rho ^{d-2}}\int_0^\infty d\zeta\, e^{-\zeta} \zeta^{\frac{d}{2}-2} (\mathcal{I}^{(1)}_\alpha (\zeta)+\xi (I_{-1+\alpha}(\zeta)-I_{1-\alpha}(\zeta))) \\
&=\frac{(2 \alpha +(d-2) (2 \xi -1)) \Gamma \left(\frac{d}{2}-\alpha -1\right) \Gamma \left(\frac{d}{2}+\alpha -1\right)\sin (\pi  \alpha )  }{2^{d} \pi ^{\frac{1}{2} (d+1)}  (d-2) \Gamma \left(\frac{d-1}{2}\right)} \frac{1}{\rho ^{d-2}}\,.
\end{split}
\end{equation}
The limit $G_-$ is again computed in the same manner as $G_+$ with $\alpha \to 1-\alpha$ and $\xi\to 1-\xi$ in the singular mode, which gives
\begin{align}
\begin{split}\label{eq:G-_coinc}
G_-&=\frac{ (2 \alpha -2 (d-2) \xi +d-2) \Gamma \left(\frac{d}{2}-\alpha -1\right) \Gamma \left(\frac{d}{2}+\alpha -1\right)\sin (\pi  \alpha ) }{2^{d} \pi ^{\frac{1}{2} (d+1)} (d-2) \Gamma \left(\frac{d-1}{2}\right) }\frac{1}{\rho ^{d-2}}\,.
\end{split}
\end{align}

We also need the coincident limit of first derivatives of the scalar propagator. Due to the appearance of $\eta$ in the integral representation of the propagator in \eq{heat_kern_scal_gen_d}, first derivatives along the defect vanish in the coincident limit. Derivatives along the transverse directions, however, are non-trivial. The $\rho$-derivative produces the same integral as $G_\pm$ up to a factor of $-(d-2)/(2\rho)$.

The $\theta$-derivative of $G_+$ produces a new integral, which has also been computed in \sn{scalar}. It is proportional to \eq{scalar-J-theta-one-point-function}. With the singular mode we find
\begin{equation}\label{eq:Dtheta_G+_coinc}
\begin{split}
\partial_\theta G_+&=\frac{i }{2 (2 \pi )^{d/2} \rho ^{d-2}} \int_0^\infty d\zeta\, e^{-\zeta } \zeta ^{\frac{d}{2}-2}  \left[\mathcal{I}^{(2)}_\alpha (\zeta )  -\xi (1-\alpha) (I_{-1+\alpha}(\zeta)-I_{1-\alpha} (\zeta))\right]\\
&=\frac{i ((2 \alpha -1) (2 \alpha +d-4)-4 (\alpha -1) (d-1) \xi ) \Gamma \left(\frac{d}{2}-\alpha \right) \Gamma \left(\frac{d}{2}+\alpha -2\right)\sin (\pi  \alpha ) }{2^{d+2} \pi ^{\frac{1}{2} (d+1)} \Gamma \left(\frac{d+1}{2}\right)}\frac{1}{\rho ^{d-2} }\,.
\end{split}
\end{equation}
Finally, we also have
\begin{equation}\label{eq:Dtheta_G-_coinc}
\partial_\theta G_-=\frac{i  (4 \alpha  (d-1) \xi -(2 \alpha +1) (2 \alpha +d-2)) \Gamma \left(\frac{d}{2}-\alpha -1\right) \Gamma \left(\frac{d}{2}+\alpha -1\right)\sin (\pi  \alpha ) }{2^{d+2} \pi ^{\frac{1}{2} (d+1)} \Gamma \left(\frac{d+1}{2}\right)}\frac{1}{\rho ^{d-2}}\,.
\end{equation}

With these results we return to evaluating \eq{fermion-J}. The $\gamma$-matrix traces can be performed using standard Clifford algebra identities. Recalling that derivatives of $G_\pm$ along the defect vanish in the coincident limit, we find
\begin{align}
\langle J_\theta \rangle  = 2^{\lfloor\frac{d}{2}\rfloor-1} \Big(i \partial_\theta (G_{+}+G_{-}\big)- \rho \partial_\rho (G_{+}-G_{-}\big)\Big).
\end{align}
Now using eqs.~\eqref{eq:G+_coinc} to \eqref{eq:Dtheta_G-_coinc}, the one-point function of the $U(1)_V$ current evaluates to
\begin{align}\label{eq:fermion-CJ}
\langle J_\theta\rangle = \frac{  (2(1-\alpha) -d  +2\xi (d-1)) \Gamma \left(\frac{d}{2}-\alpha \right) \Gamma \left(\frac{d}{2}+\alpha -1\right)\sin (\pi  \alpha ) }{2^{ d+1-\lfloor\frac{d}{2}\rfloor}\pi ^{\frac{1}{2} (d+1)}\Gamma \left(\frac{d+1}{2}\right) } \frac{1}{\rho^{d-2} }\,.
\end{align}
From \eq{CJ-d}, we can read off $C_J$ for the fermion monodromy defect directly from \eq{fermion-CJ}. Note that $C_J$ does not have a definite sign. For all $\a\in [0,1]$, $C_J<0$ when $\xi=0$ whereas $C_J>0$ for $\xi=1$. Moreover, $C_J \to -C_J$ when $\alpha \to 1-\alpha$ and $\xi \to 1-\xi$, which implies that $C_J=0$ when $\alpha=\xi=\frac{1}{2}$.

 Since derivatives along the defect vanish in the coincident limit, $\langle J_\beta\rangle = \langle J_\tau \rangle =0$. The only non-trivial component to check is $\langle J_\rho\rangle$ which indeed evaluates to zero.

At this point, we specialise to $d=4$ in order to compute $b$. In $d=4$, $C_J$ in \eq{fermion-CJ} reduces to
\begin{align}
\label{eq:fermion-CJ_d4}
C_J= \frac{(\alpha -1) \alpha  (\alpha -3 \xi +1)}{3 \pi ^2 }\,.
\end{align}
Note that the integral relation between $b$ and $\langle J_\theta\rangle$ in \eq{b-CJ-relation} picks up a sign due to Wick rotation.  With this in mind, we can now compute $b$ for a monodromy defect in theory of free fermions in $d=4$.  If we assume that $\xi$ itself is $\alpha$-independent, then 
\begin{align}\label{eq:fermion-b-0}
b =-12\pi^2 \int d\alpha \,C_J=   \alpha^2(2-\alpha^2-2\xi (3-2\alpha))+c(\xi)\,,
\end{align}
where $c(\xi)$ is an arbitrary $\a$-independent function.  Assuming $c(\xi)$ is linear in $\xi$, we can fix its form by requiring that $b=0$ when $\alpha=\xi=0$ and when $\alpha=\xi=1$. This gives
\begin{align}\label{eq:fermion-b-xi}
b =  \alpha^2(2-\alpha^2-2\xi (3-2\alpha))+\xi \,.
\end{align}
 In computing the defect EE in Section \ref{sec:EEfermion}, we will independently check $c(\xi) = \xi$ from the result for $d_2$ above and \eq{defect-EE-general}. Note that non-vanishing of the central charge at $\alpha \to 0,1$ suggests the existence of a decoupled sector of defect fermionic modes similar to the edge fermions in the (integer) quantum Hall effect. In fact we don't expect these fermions to be chiral for a monodromy generated by a vector $U(1)$ and thus the situation is more similar to the quantum spin Hall effect \cite{PhysRevLett.96.106802}, where a pair of fermions with opposite chirality emerges at the edge. 

\subsubsection{Two-point function of $\boldsymbol{\mathcal{D}_i}$}

In this subsection, we compute the displacement operator two-point function in general $d\geq 3$ and use its normalisation $C_\CD$ in $d=4$ to extract $d_1$. Working with the complex coordinates $z$, $\bar z$ adopted above, the frame fields become
\begin{align}
e^{M}{}_z = \frac{e^{-i\theta}}{2\rho}(\rho e^{M}{}_\rho-i e^{M}{}_\theta)\,,\qquad e^{M}{}_{\bar{z}} = \frac{e^{i\theta}}{2\rho}(\rho e^{M}{}_\rho +i e^{M}{}_\theta)\,,
\end{align}
and we can thus write
\begin{align}
\gamma_z = \frac{\bar z}{\rho}\gamma_{1}P_-\,,\qquad \gamma_{\bar{z}} = \frac{ z}{\rho}\gamma_{1} P_+\,.
\end{align}

The form of the displacement operator can be found by the same arguments as in \sn{scalar}. For $\xi=0$, $\mc{D}_i$ must be proportional to the $z,\bar z$-independent part of $J_\mu$,
\begin{equation}
\mathcal{D}_z \propto \left. \bar \psi \gamma_z \psi\right|_{z,\bar z=0} \,,\qquad  \mathcal{D}_{\bar z } \propto \left. \bar \psi \gamma_{\bar z} \psi\right|_{z,\bar z=0}\,.
\end{equation}
Including the additional divergent mode with $\xi \neq 0$ modifies the relationship between $\mc{D}_i$ and $J_\mu$. Using the mode expansion in eqs.~\eqref{eq:mode_decomposition}--\eqref{eq:n=0-modes}, we make the ansatz
\begin{subequations}
\begin{align}
\mathcal{D}_z &= c_1 \left[\bar\psi_{1+\alpha}\gamma_1 P_- \psi_{-\alpha} \right]+c_2 \left[\bar \psi_\alpha \gamma_1 P_- \psi_{1-\alpha}\right]\,,\label{eq:ansatz_Dz}\\
\mathcal{D}_{\bar{z} } &=  c_3 \left[\bar \psi_{-\alpha} \gamma_1 P_+ \psi_{1+\alpha} \right]+c_4 \left[\bar\psi_{1-\alpha}\gamma_1 P_+ \psi_{\alpha}\right] \label{eq:ansatz_Dzbar}\,,
\end{align}
\end{subequations}
where the coefficients $c_{1,\ldots, 4}$ are arbitrary complex numbers and the $[\cdot]$ denotes the defect operator defined by taking the defect limit $\rho\to0$. We can fix the coefficients by checking the Ward identity
\begin{equation}\label{eq:fermion_Ward}
\int_{-\infty}^\infty d^{d-2} \sigma\, \langle\bar\psi \gamma_{\bar z} \psi(z,\bar z,0) \mathcal{D}_z (\sigma)\rangle = \partial_z \langle\bar\psi\gamma_{\bar z} \psi(z,\bar z,0)\rangle=\partial_z\left( \frac{1}{2\bar z}\langle J_{\theta}\rangle \right).
\end{equation}
The correlator on the left-hand side can be written as
\begin{equation}
\begin{split}
\langle\bar\psi \gamma_{\bar z} \psi(z,\bar z,0) \mathcal{D}_z (\sigma)\rangle=\,&-c_1 \Tr\left[G_{F,\alpha}^{-\alpha} (x', x) \gamma_1 P_+ G_{F,\alpha}^{1+\alpha}(x,x')\gamma_1 P_- \right]\\
&-c_2\Tr\left[G_{F,\alpha}^{1-\alpha} (x', x) \gamma_1 P_+ G_{F,\alpha}^{\alpha}(x,x')\gamma_1 P_-\right],
\end{split}
\end{equation}
where $x = \{z,\bar{z},0\}$, $x' = \{0,0,\sigma\}$, and $G_{F,\alpha}^{\nu}$ denotes the two-point function of the defect operators labeled by $\nu$, i.e. the Wick contraction of $\hat{\bar\psi}_\nu$ and $\hat{\psi}_\nu$. The two-point function can be evaluated explicitly after performing the traces over $\gamma$-matrices. By then taking the result for $\langle \bar\psi \gamma_{\bar{z}}\psi(z,\bar z,0) \mc{D}_z (\sigma)\rangle$ and comparing to the right-hand side of \eq{fermion_Ward}, which can be easily computed from \eq{fermion-CJ}, $c_{1,2}$ are fixed uniquely. A similar analysis can be performed for $\mathcal{D}_{\bar{z}}$ such that eqs.~\eqref{eq:ansatz_Dz} and~\eqref{eq:ansatz_Dzbar} become
\begin{subequations}
\begin{align}
\mathcal{D}_z &= -2\pi  \alpha\left[ \bar\psi_{1+\alpha}\gamma_1 P_- \psi_{-\alpha} \right]+2\pi (1-\alpha)\left[\bar \psi_\alpha \gamma_1 P_- \psi_{1-\alpha} \right]\,, \\
  \mathcal{D}_{\bar{ z} } &= 2\pi   \alpha\left[\bar{\psi}_{-\alpha} \gamma_1 P_+ \psi_{1+\alpha} \right]- 2\pi(1-\alpha)\left[\bar\psi_{1-\alpha}\gamma_1 P_+ \psi_{\alpha}\right] \,.
\end{align}
\end{subequations}

Having found $\mathcal{D}_z$ and $\mathcal{D}_{\bar z}$ in the presence of singular modes with $\xi\neq 0$, we can now proceed with computing their two-point function, which takes the form
\begin{equation}
\begin{split}
\langle \mathcal{D}_{\bar z} (\sigma)\mathcal{D}_z(0)\rangle =\;  &4 \pi^2 \alpha^2(1-\xi)\,\Tr\left[G_{F,\alpha}^{-\alpha}(0,\sigma)\gamma_1 P_+ G_{F,\alpha}^{1+\alpha}(\sigma, 0)\gamma_1 P_-\right]  \\
&+ 4 \pi^2 (1-\alpha)^2 \xi\,\Tr\left[G_{F,\alpha}^{1-\alpha}( 0,\sigma)\gamma_1 P_+ G_{F,\alpha}^{\alpha}(\sigma, 0)\gamma_1 P_-\right].
\end{split}
\end{equation}
The $\gamma$-matrix traces can again be performed easily with Clifford algebra identities, and one is left with derivatives of the scalar propagator along the defect. Taking the defect limit, we find 
\begin{equation}\label{eq:fermion-CD-d}
\langle \mathcal{D}_{\bar z} (\sigma)\mathcal{D}_z( 0)\rangle =\frac{ ((1-2 \alpha  ) d \xi+\alpha  (2 \alpha +d-2)) \Gamma \left(\frac{d}{2}-\alpha \right) \Gamma \left(\frac{d}{2}+\alpha -1\right)\sin (\pi  \alpha )  }{2^{2-\lfloor\frac{d}{2}\rfloor}\pi ^{d-1} }\frac{1}{|\sigma^a|^{2d-2}}\,,
\end{equation}
where $C_\CD$ can be read off using \eq{defect-D-two-point-function}. Setting $p=d-2$ and $q=2$, on can see that $C_\CD$ and $h$ computed in eqs.~\eqref{eq:fermion-CD-d} and \eqref{eq:fermion-CT-d}, respectively, obey the conjectured relation in \eq{CD-h-general-p-q}.

Finally to compute $d_1$, we set $d=4$, which gives
\begin{align}
C_\CD = \frac{4 (1-\alpha) \alpha  \left(\alpha (1+\alpha)+2 \xi (1-2\alpha)\right)}{\pi ^2}\,.
\end{align}
Using the normalisation in \eq{d1-CD}, we arrive at
\begin{align}
d_1 = 3 (1-\alpha) \alpha  \left(\alpha (1+\alpha)+2 \xi (1-2\alpha)\right).
\end{align}
Comparing with \eqref{eq:fermion-d2}, we find that $d_1=d_2$ as was observed for the case of free scalars as well.

\subsection{Conical singularities}
In this subsection, we will follow the methods in \sn{scalar-renyi} and repeat the computation of central charges of free fermions in the presence of a conical singularity in $d=4$. We partition the background $\mathbb{R}^{1,d-1}$ into a region $A$ and its complement $\bar{A}$ on a surface of constant Euclidean time $\tau$ and place the theory of a single Dirac fermion on a branched $n$-sheeted cover over $A$. Since the theory on the $n$-sheeted cover is free, we can equivalently describe it by a system of $n$-free fermions $\Psi_i$ on a single sheeted background with a conical singularity at the boundary of $A$, see e.g. \cite{Casini:2005rm}.  We can diagonalise the boundary conditions at the branching surface $\vec{\Psi}(\tau = 0^+) = T\vec\Psi(\tau=0^-)$ where $T_{ij} = \delta_{i,j+1} +(-1)^{(n+1)}\delta_{i,1}\delta_{j,n}$ by defining
\begin{align} 
\psi_k = \sum_{j=1}^n e^{2\pi i \frac{k}{n}}\Psi_j\,,\qquad k =-\frac{n-1}{2}, \frac{n+1}{2},\,\ldots,\,\frac{n-1}{2}\,.
\end{align}
The fields $\psi_k$ pick up a phase $\alpha = k/n$ around the origin.   

We can now use the results of the preceding subsection for a single Dirac fermion with monodromy $\alpha$ to compute the defect central charges of a Dirac fermion in the presence of a conical singularity. The only subtlety that arises is that in our computation of $b$, $d_1$ and $d_2$ we assumed that $0 < \a <1$. The range of $\a$ can be shifted to include $-\frac{1}{2}<\a <\frac{1}{2}$ for all modes except for \eq{a_mode}, which diverges too strongly at the origin for $\a \leq 0$. To exclude it, we simply set $\xi=0$. The expressions for $b$, $d_1$ and $d_2$ are analytic in $\a$, so we can safely substitute $\alpha = k/n$, set $\xi=0$ and sum over $k$ to obtain the central charges for a Dirac fermion in the presence of a conical singularity.  

For $d_1$ and $d_2$, we can straightforwardly compute the sums 
\begin{equation}\label{eq:d1-d2-fermion-conical}
d_1=d_2 = \sum_{k=-\frac{n-1}{2}}^{\frac{n-1}{2}} 3\frac{k(n-k)}{n^4}\big(k^2+kn\big)=-\frac{(1-n^2)(17n^2 +7)}{80n^3}\,.
\end{equation}
We can perform the same sum in $b$ in a similar way
\begin{align}\label{eq:b-fermion-conical}
b=\sum_{k=-\frac{n-1}{2}}^{\frac{n-1}{2}} \frac{k^2(2n^2-k^2)}{n^4} = -\frac{(1-n^2)(37n^2+7)}{240n^3}\,.
\end{align}
Near $n= 1$, we find the behaviour
\begin{subequations}
\begin{align}
b &= \frac{11(n-1)}{30}-\frac{3}{10}(n-1)^2 +\CO(n-1)^3,\\
 d_1&=d_2 = \frac{3(n-1)}{5} -\frac{13}{20}(n-1)^2 + \CO(n-1)^3,
\end{align}
\end{subequations}
which is consistent with \eq{conical-trace-anomaly-n} using $a_{4d} = \frac{11}{360}$ and $c_{4d}= \frac{1}{20}$ for $d=4$ free Dirac fermions. We also see again that from eqs.~\eqref{eq:d1-d2-fermion-conical} and~\eqref{eq:b-fermion-conical} the relation $d_1=d_2= n(12 a_{4d}- \partial_n b)$ holds as expected \cite{Lewkowycz:2014jia}.

\subsection{Entanglement entropy}\label{sec:EEfermion}
 
In this subsection, we will compute the entanglement entropy for the free fermion in $d=4$ in the presence of a monodromy defect following the same methods used in \sn{scalar-EE}.  In fact, most of the results for the scalar EE can be directly brought to bear.  That is, since $\Dslash^2 = D^2$ on $\mathbb{R}^4$, the complex scalar and fermion heat kernels are directly related by
\begin{align}
K_F(s; x,\,x{}^\prime;D) =\frac{1	}{2} K_S(s;x,x{}^\prime;D) \,\mathds{1}_4\,,
\end{align}
where the factor of $1/2$ arises because we are considering a complex scalar. Thus, we will proceed with the computation by noting the relevant places where the approach for the free fermion differs slightly from the free scalar.

We start by partitioning the background into a region $A$ and its complement $\bar A$. We take $A$ to be the half-space orthogonal to the defect with the entangling surface $\partial A$ at $\tau = 0$ and $\sigma^1=0$. We will adopt polar coordinates in the $(\tau,\sigma^1)$-plane as $\tau = r\cos\phi$ and $\sigma^1 = r\sin\phi$. Rotations around the entangling surface are generated by $U(\phi) = \text{Exp}[\frac{1}{2}\gamma_{0}\gamma_{3}\phi]$.  In these new polar coordinates the fermionic heat kernel takes the form $K_F = \frac{1}{2} U(\phi) K_S$.  Note that now the heat kernel is anti-periodic around the entangling surface i.e. $K_F(s;x(\phi+2\pi),x{}^\prime;D) = - K_F(s;x(\phi),x{}^\prime;D)$.  Placing the system on the cone, the asymptotic expansion of the fermion heat kernel is modified as compared to the asymptotic expansion of the scalar heat kernel due to this anti-periodicity. The heat kernel for the Dirac fermion in the presence of the conical singularity has been found in \cite{Bergamin:2015vxa} (see also \cite{Fursaev:2011zz}) and it reads
\begin{align}
K_{F,n}(s;x,x{}^\prime;D)  &= K_F(s;x,x{}^\prime;D) +i \int_\Gamma \frac{d\omega}{8\pi n}\,\csc\frac{\omega-\Delta\phi}{2n} U(\omega) K_S (s;x(\omega),x{}^\prime;D) \,,
\end{align}
where $\Delta\phi \equiv \phi^\prime-\phi$.  The contour $\Gamma$ is the same as in the scalar heat kernel as discussed below \eq{heat_kernel_conical}.

In order to proceed, we will need to compute the trace
\begin{align}\begin{split}\label{eq:trace-fermion-heat-kernel-1}
\Tr K_{F,n}(s;x,x{}^\prime;D) =\;&\frac{L^2}{16\pi^2 s} \int_0^{\frac{L^2}{2s}} d\zeta \sum_m e^{-\zeta} I_{|m-\alpha|}(\zeta)\int_0^{2\pi n} d\phi\,\Tr U(\phi) \\
&+ \frac{i}{32\pi }\int_\Gamma d\omega\frac{\Tr U(\omega)}{\sin\frac{\omega}{2n} \sin^2\frac{\omega}{2}}\int_0^{\frac{L^2}{2s}}  d\zeta \sum_m e^{-\zeta} I_{|m-\alpha|}(\zeta)\,.
\end{split}\end{align}
From the form of $U(\omega)$ above, $\Tr\, U(\omega) =4 \cos \left(\frac{\omega}{2}\right)$.  The contour integral then evaluates to 
\begin{align}
\frac{i}{8\pi} \int_\Gamma d\omega \cos \left(\frac{\omega}{2}\right)\csc\frac{\omega}{2n} \csc^2\frac{\omega}{2} = \frac{n^2-1}{12 n}\,.
\end{align}
Plugging this into \eq{trace-fermion-heat-kernel-1}, we thus need to compute
\begin{align}
\Tr K_{F,n}(s;x,x{}^\prime;D) &= \left(\frac{L^2 \sin(n\pi)}{2\pi^2 s}+\frac{n^2-1}{12 n}\right)\int_0^{\frac{L^2}{2s}}  d\zeta \sum_m e^{-\zeta} I_{|m-\alpha|}(\zeta).
\end{align}
All of these sums and integrals were already encountered in \eq{scalar-EE-heat-kernel-1}, and so the rest of the computation of the fermion EE mirrors exactly the scalar computation. From \eq{fermion-propagator-1}, we see that the contribution of a single Dirac fermion with monodromy $\alpha$ to the defect EE contains terms coming from the regular modes and the two divergent $n=0$ modes coupled with $(1-\xi)$ and $\xi$. The result is
\begin{align}\label{eq:fermion-ee-alpha}
S_{A}  = \ldots +\frac{\alpha^2+(1-2\alpha)\xi}{6} \log \frac{L}{\e} +\CO(1),
\end{align}
where $\ldots$ contain the non-universal terms. From eqs.~\eqref{eq:fermion-d2} and \eqref{eq:fermion-b-xi}, we find $(b - d_2/3) = \alpha^2+(1-2\alpha)\xi$. This is precisely $6$ times the coefficient of the log-term in \eq{fermion-ee-alpha}, again in agreement with \eq{SEE_log}.  Note that the defect EE did not rely on any integration in parameter space and so confirms $c(\xi) = \xi$ in \eq{fermion-b-0}.

\section{Gukov-Witten Defects in $d=4$ Maxwell}\label{sec:maxwell}

In this section we consider a similar type of defect in Maxwell theory in $d=4$: the Gukov-Witten defect~\cite{Gukov:2006jk}. In the non-SUSY case, these defects are specified by a singularity in the gauge field along a 2-dimensional submanifold. The singularity is controlled by two parameters related by electric-magnetic duality. 

The electric version of the defect in Maxwell theory can be engineered by turning on a singular background configuration for the gauge field $A_\mu$ defined by eq.~\eqref{eq:A-monodromy}. More precisely we define the dynamical gauge field $\tilde a$ as
\begin{equation} \label{eq:Abackground}
\tilde{a}= a+ A \; ,
\end{equation}
where $a$ denotes non-singular dynamical fluctuations around the background $A$. Substituting \eq{Abackground} into the Maxwell action
\begin{equation}
\frac{1}{4} \int_{\CM_4} d\tilde{a} \wedge \star d\tilde{a}\,,
\end{equation}
one finds a term
\begin{equation} \label{eq:Eldefect}
- \frac{1}{2} 2 \pi \alpha\int_{\Sigma} \star da \; 
\end{equation}
in addition to the ordinary Maxwell term for $a$. Note that we have discarded the term quadratic in delta functions (i.e. we have assumed no self-intersection of the submanifold~$\Sigma$). With a suitable choice of coordinates, eq.~\eqref{eq:Eldefect} corresponds to the electric defect of Gukov and Witten \cite{Gukov:2006jk}.\footnote{Here we work in the Euclidean coordinate system defined below eq.~\eqref{eq:background-metric-R-d}, with the time running parallel to the defect. To obtain a physical electric defect we would have to make time perpendicular to the defect, which would not alter any of the discussion presented in this section. } Similarly we could obtain the magnetic defect 
\begin{equation}\label{eq:Magdefect}
\int_{\Sigma} da
\end{equation}
by introducing a $\int d\tilde{a} \wedge d\tilde{a}$ term in the Maxwell action. 

The defects specified by insertions of eqs.~\eqref{eq:Eldefect} and~\eqref{eq:Magdefect} are closely related to the monodromy defects studied in sections~\ref{sec:scalar} and~\ref{sec:fermion}. In particular, the defect eq.~\eqref{eq:Eldefect} can be obtained by gauging the $U(1)$ symmetry of those models on the background eq.~\eqref{eq:Abackground}. By giving the matter fields mass (and assuming it preserves the gauge symmetry) one obtains eq.~\eqref{eq:Eldefect} in addition to the Maxwell term for $a$ as an IR-effective action. The magnetic flux operator eq.~\eqref{eq:Magdefect} corresponds to a non-zero $\theta$-angle of a 2-dimensional theory living at the defect. As such it is a non-perturbative quantity.  

We now proceed to discuss the relevant central charges of the electric defect eq.~\eqref{eq:Eldefect}. In the absence of charged particles, the electric flux operator is the integral of a 2-form which is conserved by virtue of the equation of motion. Thus it generates a $U(1)$ one-form symmetry. The charged objects are the Wilson lines, and the charge operator counts the number of Wilson lines that link it~\cite{Gaiotto:2014kfa}. Thus, one would not expect this defect to affect any quantities sensitive to the metric.\footnote{For the magnetic Gukov-Witten defect the corresponding statement follows directly from its topological character.} We will now proceed to verify this statement by a direct computation. 

Since the defect couples to a term linear in the dynamical gauge field we can treat it as a delta function source and complete the square. This means that the propagator of $a$ remains unchanged after the introduction of the defect (more precisely it gets shifted by a constant proportional to $\alpha$). This is in contrast to the scalar and fermion in sections~\ref{sec:scalar} and~\ref{sec:fermion}, respectively. Thus, $\langle T_{\mu \nu} \rangle=0$, and so $d_2=0$.

The displacement operator can be deduced in the same way as before by using eq.~\eqref{eq:stress-conv-flux} with the gauge current for the dynamical field obtained by varying with respect to $A_\mu$. This yields
\begin{equation} \label{eq:MaxDisplacement}
\nabla^\mu T_{\mu \nu} = \nabla_\mu f^{\mu \alpha} F_{\alpha  \nu} \sim \delta(\Sigma)\nabla_\mu {f^{\mu}}_\nu \; ,
\end{equation}
where we have introduced field strength tensors $f$ and $F$ for $a$ and $A$, respectively. This equation can also be derived directly from the divergence of the stress tensor for $\tilde{a}$ and using integration by parts on one of the resulting terms involving a derivative of the delta function.\footnote{We again discard a term quadratic in delta functions as it doesn't affect correlation functions of $\CD_i$ at non-coincident points.} From eq.~\eqref{eq:MaxDisplacement} it follows that the displacement operator is proportional to the equation of motion for the field $a$. Physically this follows from the fact that the displacement of the vortex is equivalent to an infinitesimal shift in the gauge field. This also implies that the two-point function of the displacement operators has no contributions at non-coincident points, and therefore $d_1$ vanishes. 

At last we consider the contribution to the A-type anomaly, $b$. Unlike $d_1$ and $d_2$, one might expect that $b$ need not vanish as it is the coefficient of the defect's Euler density. However, this is not the case. We will show by explicit evaluation of the partition function that it, too, is zero. To this end, consider a spherical defect embedded in flat Euclidean space. Let $X^\mu$ with $\mu=1,\ldots,4$ be the Cartesian coordinates for the ambient space, and let the sphere be located in the $X^2=0$ plane at  
\begin{equation}\label{eq:CartSphere}
(X^1)^2+(X^3)^2+(X^4)^2 = R^2 \; .
\end{equation}
These are the coordinates defined in \App{spherical}. Eq.~\eqref{eq:Eldefect} can then be written as
\begin{equation} \label{eq:PolarCoupling}
2 \pi \alpha \int_{\mathbb{S}^2} d^2  X \; (\partial_2 a_{\mathscr{R}}- \partial_{\mathscr{R}} a_2) \;,
\end{equation}
where $\partial_2 \equiv \frac{\partial}{\partial X^2}$ and $\mathscr{R}$ is the radial coordinate for the sphere \eqref{eq:CartSphere} (more specifically it corresponds to the radial coordinate $\text{r}$ defined in \eq{Xr} restricted to the $\phi_1=\frac{\pi}{2}$ hyperplane). By rewriting eq.~\eqref{eq:PolarCoupling} as a coupling to a delta function source, completing the square and integrating out $a$ we arrive at the following contribution to the effective action
\begin{equation}\label{eq:Int1}
(2 \pi \alpha)^2 
\int_{\mathbb{S}^2}\int_{\mathbb{S}^2} d^2 X\, d^2 X^{\prime} \left[\left( \partial_2 \partial_2' + \partial_{\mathscr{R}} \partial_{\mathscr{R}}' \right) \Delta (X-X')\right]|_{X^2=X^2{}'=0, \mathscr{R}=\mathscr{R}'=R} \; ,
\end{equation}
where 
\begin{equation}
\Delta (s)=\frac{1}{(2 \pi)^4}\frac{1}{s^2}
\end{equation}
is the $d=4$ propagator restricted to the $2$-sphere.  After some algebra we can rewrite the integrand of \eqref{eq:Int1} as
\begin{equation}
\frac{3}{\vec{s}{}^{\,4}}+ \frac{R^2}{\vec{s}{}^{\,6}} \; ,
\end{equation} 
where $\vec{s}\equiv \vec{X}-\vec{X}{}^{\,\prime}=(X^1-X^1{}',X^3-X^3{}', X^4-X^4{}')$, and both points satisfy \eqref{eq:CartSphere}. 
To evaluate the integral in \eq{Int1}, we have to introduce a regulator to deal with the problematic $|\vec{s}|\to 0$ region. We choose dimensional regularisation so that the integral becomes
\begin{equation}\label{eq:IntDimReg}
\int_{\mathbb{S}^d}\int_{\mathbb{S}^d} d^d X \, d^d X^{\prime} \left(\frac{3}{\vec{s}{}^{\,4}}+ \frac{R^2}{\vec{s}{}^{\,6}} \right) \; ,
\end{equation}
where $d=2 - \epsilon$. This integral can be evaluated by expanding the integrand in spherical harmonics, applying the formula (c.f. \cite{Drummond:1975yc})
\begin{equation} \label{eq:SphereHarmExp}
\frac{1}{(\vec{X}-\vec{X}{}^{\,\prime})^{2\kappa}}= \sum_{m,n} \frac{(2R)^{-2 \kappa +d} \pi^{d/2} \Gamma(-\kappa + \frac{d}{2}) \Gamma(n+ \kappa)}{\Gamma(n+d-\kappa) \Gamma(\kappa)} Y_m^n(X)Y_m^n(X') \; .
\end{equation}
and using the property
\begin{equation} \label{eq:YmnInt}
\int_{\mathbb{S}^2} Y_m^n = R^{d/2}\delta_{0 m} \delta_{0n} \; ,
\end{equation}
where the normalisation with respect to the sphere radius $R$ was chosen so that the harmonics are orthonormal in $d$ dimensions. 

Rather curiously the divergent Gamma functions cancel between the numerators and denominators of \eqref{eq:SphereHarmExp} and we are left with a finite result:
\begin{equation}
\alpha^2 \frac{13}{8 \pi} \; .
\end{equation}
Finiteness implies that there is no conformal anomaly as this term can be removed by addition of a local counterterm on the defect proportional to 
\begin{equation}
\int_{\mathbb{S}^2} \mathcal{R}_{\mathbb{S}^2}\;,
\end{equation} 
and so $b=0$, too.
The corresponding result for magnetic defect vanishes identically as it involves integration over total derivatives on the sphere.

\section{Defect RG Flows}\label{sec:rg-flows}

In this section we revisit the monodromy defect in the theory of a free complex scalar and of a free Dirac fermion discussed in sections~\ref{sec:scalar} and~\ref{sec:fermion}, respectively. Relevant operator insertions on the defect trigger a defect RG flow which we study in detail.

\subsection{Scalar monodromy flows}

Let us begin with the monodromy defect in the free scalar theory introduced in section~\ref{sec:scalar}. As discussed there, only two singular modes are allowed to appear in the $\varphi$ defect OPE: $\hat O^-_{-\a}$ and $\hat O^-_{1-\a}$. Here, we consider the case where only the operator $\hat O^-_{-\a}$ is present (i.e. $\tilde \xi=0$ and $\xi\neq 0$), and we show that the IR fixed point corresponds to the dCFT with $\tilde\xi=\xi=0$. The other case will be completely analogous, and the IR fixed point is again the dCFT with $\tilde\xi=\xi=0$. Switching on both $\xi$ and $\tilde \xi$ would not make any conceptual difference in this derivation. 

Using the singular mode $\hat O^-_{-\alpha}(\sigma)$ we can construct the relevant quadratic deformation\footnote{Note that there exist other more general relevant deformations which are not quadratic. We will comment more on these operators in the discussion \sn{discussion}.}
\begin{equation}\label{relevantdef}
\lambda \int d^{d-2}\sigma\, \hat{O}^-_{-\alpha}(\sigma) \hat O^{\dagger -}_{-\alpha}(\sigma) \,.
\end{equation}
 Here, $\lambda$ is a relevant parameter with mass dimension $2\alpha$. Notice that this deformation is present only for $\xi\neq 0$. We would like to analyse the IR fixed point of the defect RG flow triggered by this deformation. Thus, we compute the correlator
\begin{equation}
  \braketbis{\varphi(x)\varphi^{\dagger}(x')}_\lambda\equiv\frac{\braketbis{\varphi(x)\varphi^{\dagger}(x') e^{-\lambda \int d^{d-2}\sigma\, \hat{O}^-_{-\alpha}(\sigma) \hat{O}^{\dagger -}_{-\alpha}(\sigma)}}}{\braketbis{ e^{-\lambda \int d^{d-2}\sigma\, \hat{O}^-_{-\alpha}(\sigma) \hat{O}^{\dagger -}_{-\alpha}(\sigma)}}}\,,
\end{equation}
where without loss of generality we set $x=\{\rho, \theta, \sigma\}$ and by normal rotational and defect translational symmetries $x' = \{\rho',0,0\}$. Expanding the exponential, performing the Wick contractions and the combinatorics we get
\begin{equation}
\begin{split}\label{afterWick}
\braketbis{\varphi(x)\varphi^{\dagger}(x')}_\lambda=\;&\braketbis{\varphi(x)\varphi^{\dagger}(x')}\\
 &+ \sum_{n=1}^\infty (-\lambda)^n  \int \prod_{i=1}^n d^{d-2} \sigma_{i} \frac{\braketbis{\varphi(x)\hat{O}^{\dagger -}_{-\alpha}(\sigma_{1} )}  \braketbis{\varphi^{\dagger}(x') \hat{O}^-_{-\alpha}(\sigma_{n} ) }}{\prod_{j=1}^{n-1}|\sigma_{j} -\sigma_{j+1} |^{d-2-2\alpha}}\,,
\end{split}
\end{equation}
where the denominator comes from the defect propagator $\braketbis{\hat{O}^-_{-\alpha}(\sigma_{k} )\hat{O}^{\dagger -}_{-\alpha}(\sigma_{k+1} )}$ in eq.~\eqref{defdefprop},  the propagator $\braketbis{\varphi(x)\varphi^{\dagger}(x')}$ is \eq{two-point function_gaug_transf} evaluated at $\tilde \xi=0$, and the bulk to defect propagator is 
\begin{align}
\braketbis{ \varphi(x) \hat{O}^{\dagger -}_{-\alpha}({\sigma}_1)}&=\frac{(c_{-\a}^-)^{1/2} e^{i \alpha \theta}}{\rho^{\a} (\rho^2+(\sigma-{\sigma}_1)^2)^{\frac{d}{2}-1-\a}}\,,
\end{align}
with $c_{-\a}^-$ given in eq.~\eqref{bulkdefcoeff1}.

In order to resum eq.~\eqref{afterWick}, it is useful to Fourier transform to momentum space in the directions along the defect. Consider the following Fourier representation 
\begin{align}
 \frac{1}{(\rho^2+\sigma^2)^{\frac{d}{2}-1-\alpha}}=\int \frac{d^{d-2} k}{(2\pi)^{d-2}} f(k \rho) k^{-2\a} e^{i k\cdot \sigma}\,,
\end{align}
where we slightly abuse notation by writing $k\equiv |k|=\sqrt{\delta^{ab}k_ak_b}$, and the function $f(k \rho)$ is given by
\begin{equation}
f(k \r)=k^{2\a} \int d^{d-2} \sigma \frac{e^{-i k\cdot \sigma}}{(\r^2+\sigma^2)^{\frac{d}{2}-1-\a}} =\frac{2\pi^{\frac{d}{2}-1}}{\Gamma(\tfrac{d}{2}-1-\a)} (2 k \r)^{\a}K_{\a}(k \r)\,,
\end{equation}
with $K_\nu(\zeta)$ denoting the modified Bessel function of the second kind. Inserting this into eq.~\eqref{afterWick} we get
\begin{align}
 &\hspace{-0.25cm}\braketbis{\varphi(x)\varphi^{\dagger}(x')}_\lambda=\braketbis{\varphi(x)\varphi^{\dagger}(x')}+\\
  &\hspace{0.75cm}\tfrac{c_{-\a}^- e^{i \a \theta}}{\rho^{\a}{\rho'}^{\a}}\sum_{n=1}^\infty (-\lambda)^n \int \prod_{i=1}^n d^{d-2} {\sigma}_i \prod_{j=1}^{n+1} \tfrac{d^{d-2} k_j k_j^{-2\a}}{(2\pi)^{d-2}} f(k_1 \rho) f(0)^{n-1} f(k_{n+1} {\rho'}) e^{i \sum_l k_l\cdot({\sigma}_{l-1}-{\sigma}_l)}\,, \nonumber
\end{align}
where it is understood that ${\sigma}_0\equiv \sigma$ and ${\sigma}_{n+1}=0$. The integration over the $n$ positions ${\sigma}_i$ gives $n$ delta functions which we use to perform $n$ momentum integrations. Finally, we perform the sum over $n$ leading to
\begin{equation}
 \braketbis{\varphi\varphi^{\dagger}}_\lambda=\braketbis{\varphi\varphi^{\dagger}}\\
 - \int \frac{d^{d-2} k}{(2\pi)^{d-2}} e^{i k\cdot \sigma} \frac{\lambda k^{-2\alpha}}{1+\lambda f(0) k^{-2\alpha}} c_{-\a}^-  \frac{f(k \rho)}{(k\rho)^{\a}} \frac{f(k {\rho'})}{(k\rho')^{\a}}e^{i \a \theta}\label{propresummom},
\end{equation}
where for brevity we have suppressed the coordinate dependence in the correlation functions and 
\begin{equation}\label{eq:f0def}
 f(0)= \frac{4^\a \pi^{\frac{d}{2}-1} \Gamma(\a)}{\Gamma(\frac{d}{2}-1-\a)}\,.
\end{equation}
 The position independent factor involving the coupling that appears in eq.~\eqref{propresummom} can be interpreted as a dimensionless effective coupling. We will study the beta function associated to this coupling in \sn{betafunction}. For the moment, we just point out that the IR fixed point can be formally reached by taking $\lambda\to \infty$ in eq.~\eqref{propresummom}, and the resulting propagator reads
\begin{equation}
 \braketbis{\varphi\varphi^{\dagger}}_{\rm{IR}}=\braketbis{\varphi\varphi^{\dagger} }
 - \xi\frac{ e^{i \a \theta} \sin \pi \a}{\pi^2}\int \frac{d^{d-2} k}{(2\pi)^{d-2}} e^{i k\cdot \sigma}  K_{\a}(k \rho)K_{\a}(k {\rho'})\,.
\end{equation}
The integral can be performed and the final result is
\begin{equation}\label{resulIR}
 \braketbis{\varphi\varphi^{\dagger}}_{\rm{IR}}=\braketbis{\varphi\varphi^{\dagger}} -\xi\frac{\Gamma\left( \frac{d}{2}-1-\a  \right)}{4\pi^{d/2}\Gamma\left(1- \a \right)}  F_{\hat{\Delta}^-, -\a}(\eta,\theta) +\xi\frac{\Gamma\left( \frac{d}{2}-1+\a  \right)}{4\pi^{d/2}\Gamma\left(1+ \a \right)}  F_{\hat{\Delta}^+, -\a}(\eta,\theta)\,
\end{equation}
with the defect blocks defined in eq.~\eqref{defectblock}. Comparing this result with the propagator \eq{two-point function_gaug_transf} and the coefficients eq.~\eqref{bulkdefcoeff1}, one easily sees that the second term on the right-hand side of eq.~\eqref{resulIR} precisely cancels the implicit $\xi$ dependence of the first term. This shows that starting with any value of $\xi$ in th UV, the RG flow triggered by the relevant deformation in eq.~\eqref{relevantdef} leads to the $\xi=0$ defect in the IR. 
\begin{figure}
	\begin{center}
		\includegraphics[scale=0.9]{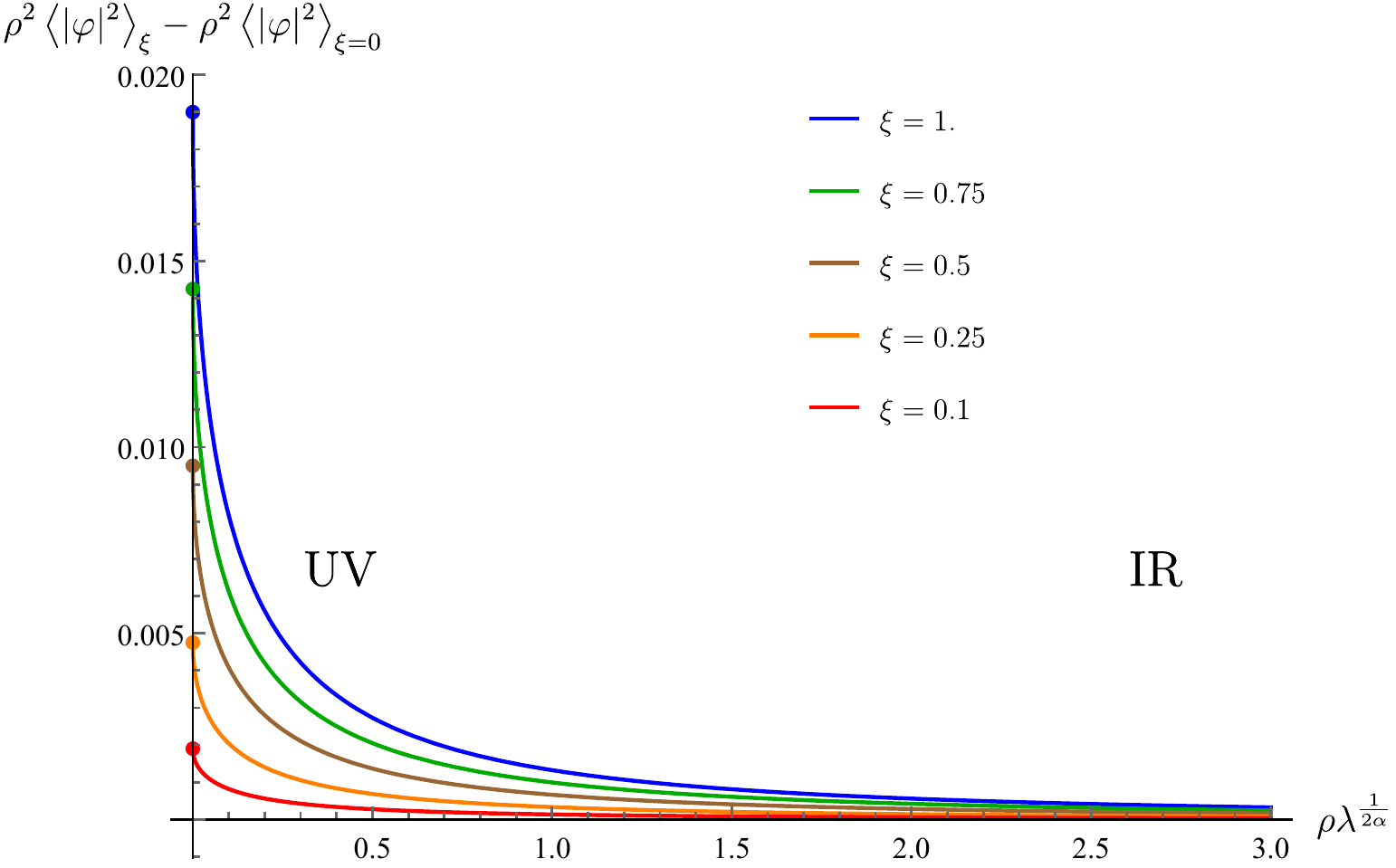}
		\caption{The quantity $\rho^2 \left< |\varphi|^2 \right>_{\xi}-\rho^2 \left< |\varphi|^2 \right>_{\xi=0}$ as a function of the dimensionless quantity $\rho \lambda^\frac{1}{2\alpha}$ for $d=4$, $\alpha = 0.75$, and different values of $\xi$. While in the UV the quantity depends on $\xi$, in the IR limit all the curves go to zero. \label{fig:phi2-plot}}
	\end{center}
\end{figure} 

We briefly consider the implications of the defect RG flow in $d=4$. Starting in the UV with arbitrary $\xi\in[0,1]$ and comparing to the IR with $\xi=0$, one can see that $b$, as given in \eq{b-free-scalar}, satisfies $b_{\rm{UV}}\geq b_{\rm{IR}}$ for any starting value of $\xi$. The inequality is saturated only for the trivial case of $\xi=0$ in the UV.  Thus, this flow obeys the $b$-theorem \cite{Jensen:2015swa}.  In arbitrary $d$, the RG flows for monodromy defects in free scalar CFTs can be analogously studied on the boundary $\partial \mathbb{H}^{d-1}$ of $\mathbb{H}^{d-1}\times S^1$ as in \cite{Giombi:2021uae} where a general defect $c$-theorem \cite{Nishioka:2021uef} was verified.

We conclude this subsection by computing the variation of the one-point function $\Delta \left< |\varphi|^2 \right> \equiv \left< |\varphi|^2 \right>_{\lambda}-\left< |\varphi|^2 \right>_{\lambda=0}$ induced by the relevant perturbation in eq.~\eqref{relevantdef}. To obtain this result we just need the coincident limit of the propagator eq.~\eqref{propresummom}. For simplicity, let us restrict our attention to $d=4$ where the result is
\begin{equation}
\Delta \braketbis{ |\varphi|^2 }= -\frac{2^{-1+2\alpha}}{\pi \Gamma(1-\alpha)^2} \frac{\xi }{\rho^2} \int_0^{+\infty} d\zeta \, \frac{\zeta \,\rho^{2\alpha}\lambda}{\zeta^{2\alpha}+\rho^{2\alpha}\lambda f(0)} K_{\alpha}(\zeta) K_{\alpha}(\zeta)\,.
\end{equation}
We notice that the deformation correctly goes to zero in the UV limit $\rho \lambda^\frac{1}{2\alpha} \rightarrow 0$, and it saturates to a constant value in the IR limit $  \rho \lambda^\frac{1}{2\alpha} \rightarrow +\infty$. In the IR limit, the integral can be solved analytically to give 
\begin{equation}
\label{eq:delta_phi_lim}
\Delta \left< |\varphi|^2 \right> \xrightarrow[\text{IR}] \,-\frac{\alpha  \xi }{4 \pi ^2 \rho ^2}\,,
\end{equation}
which is exactly the difference between the one-point functions in \eq{scalar-one-point-function-4d} with the values at $\xi =0$ and $\xi$. 

For generic values of $\rho \lambda^\frac{1}{2\alpha}$ the integral must be solved numerically. In figure \ref{fig:phi2-plot} we show the behaviour of
\begin{equation}
\rho^2 \left< |\varphi|^2 \right>_{\xi}-\rho^2 \left< |\varphi|^2 \right>_{\xi=0} = \Delta \left< |\varphi|^2\right> + \frac{\alpha  \xi }{4 \pi ^2 \rho ^2}
\end{equation}
for different values of $\xi$. In the UV regime  $\rho \lambda^{1/(2\alpha)} =0$ the curves depend on $\xi$ but they all reduce to zero in the IR limit $\rho \lambda^{1/(2\alpha)} \rightarrow +\infty$. This is again consistent with having an IR fixed point corresponding to $\xi=0$.

\subsubsection{Defect beta function}\label{sec:betafunction}

As an additional consistency check, we would like to find the beta function for the coupling $\lambda$ and verify the existence of an IR fixed point with the desired properties. Let us first analyse the resummed correlator eq.~\eqref{propresummom} in momentum space. By dimensional analysis, the momentum space two-point function of bulk scalars is scale-free, and, therefore, can only depend on dimensionless quantities. This also means that the explicit $k$-dependence of the expression
\begin{equation} \label{eq:coupdef1}
\frac{\lambda k^{-2\alpha}}{1+\lambda f(0) k^{-2\alpha}}  
\end{equation}
has to correspond to the resummed RG effects within the effective coupling at scale $k$, which is to be understood as an IR scale. 
On the other hand the coupling $\lambda$ appearing in \eq{coupdef1} is defined at an arbitrary UV cut-off scale $\Lambda$, which we can interpret as the defect thickness. The IR physics should be independent of this scale and eventually we will send $\Lambda\to\infty$. With this in mind we will modify \eq{coupdef1} as follows\footnote{At the computational level this modification amounts to reintroducing a cut-off into UV-finite integrals and thus can be seen as a choice of scheme. }
\begin{equation} \label{eq:coupdef2}
\frac{\lambda k^{-2\alpha}}{1+\lambda f(0)( k^{-2\alpha}- \Lambda^{-2\alpha})}  \; ,
\end{equation}
which leads to an expression that is equivalent for $\Lambda \to \infty$. 

Defining a dimensionless coupling 
\begin{equation}\label{eq:dim0lam}
\bar{\lambda}= \Lambda^{-2 \alpha} \lambda \; ,
\end{equation} 
we arrive at an expression for the effective coupling 
\begin{equation} \label{eq:barlam}
\bar{\lambda}(k)= \frac{\bar{\lambda}(\Lambda) \left(\frac{k}{\Lambda}\right)^{-2\alpha}}{1+\bar{\lambda}(\Lambda) f(0)\left( \left(\frac{k}{\Lambda}\right)^{-2\alpha}-1 \right)}\;.
\end{equation}
This expression reduces to $\bar{\lambda}(\Lambda)$ at $k=\Lambda$ as required by RG consistency.
By taking a logarithmic derivative of \eq{barlam} with respect to $k$ we get
\begin{equation} \label{eq:betaLam}
\beta_{\bar{\lambda}}= - 2\alpha \bar{\lambda}(k) + 2\alpha f(0) \bar{\lambda}(k)^2 \; .
\end{equation} 
 Rather curiously, eq.~\eqref{eq:betaLam} has the form of a one-loop exact beta function similar to the ones appearing in interacting SUSY gauge theories.  With the beta function \eq{betaLam} we can readily verify that \eq{barlam} is independent of the cutoff $\Lambda$ so we can safely take it to $\infty$. Furthermore \eq{betaLam} has an IR fixed point at
\begin{equation}\label{eq:FPallOrder}
\bar{\lambda}^*= \frac{1}{f(0)}= \frac{\Gamma(\frac{d}{2}-1-\a)}{4^\a \pi^{\frac{d}{2}-1} \Gamma(\a)} \; ,
\end{equation}
where we stress that the above expression holds to all orders in $\alpha$. We can compute the anomalous dimension of the composite operator $O_\lambda = \hat{O}^{\dagger -}_{-\alpha} \hat{O}^-_{-\alpha}$ at the IR fixed point\footnote{By anomalous dimension we mean its deviation from the `defect-free' ($\alpha=0$)  dimension. Namely, for the case at hand we have $\Delta_{O_{\lambda}}= d-2 + \gamma_{\lambda} $.}
\begin{equation}
\gamma_\lambda^*= \left.\frac{\partial \beta_{\bar{\lambda}}}{\partial\bar{\lambda}}\right|_{\bar{\lambda}= \bar{\lambda}^*}= +2\alpha \; ,
\end{equation}
which holds to all orders in $\alpha$. By invoking the free field character of the theory (in that the anomalous dimensions sum up) we conclude that the dimension of $\hat{O}^{-}_{-\alpha}$ has changed from $\hat{\Delta}^-_{-\alpha}$ to $\hat{\Delta}^+_{-\alpha}$ under the flow, where the dimensions were defined above \eq{diff_op}. This is effectively equivalent to the operator of dimension $\hat{\Delta}^+_{-\alpha}$ not existing at the IR fixed point as discussed below \eqref{resulIR}. \\
We can check this result by using conformal perturbation theory in the small $\alpha$ regime. The parameter $\alpha$ controls the UV scaling, so by keeping it small it will play the role of $\varepsilon$ in the $\varepsilon$-expansion. The leading order beta function in conformal perturbation theory can be obtained by the standard methods used in bulk CFTs (see e.g. \cite{Komargodski:2015grt})\footnote{The $d-$dependent factors in the following beta function that usually don't appear in the literature come from integrating over the loop integral along the defect $\int d^{d-2}x \,{(x^2)^{-\frac{d-2}{2}+\alpha}} \to \frac{\pi^{(\frac{d}{2}-1)}}{\Gamma(\frac{d}{2}-1)} \log{\Lambda}+ \mathcal{O}(\alpha)$.}
\begin{equation} \label{eq:beg0}
\beta_{\bar{\lambda}} = \frac{d\bar{\lambda}}{d \Lambda}= - 2\alpha \bar{\lambda} +  \frac{\pi^{\frac{d}{2}-1}}{\Gamma \left(\frac{d}{2}-1 \right)}  C_{O_{\lambda} O_{\lambda}}^{ O_{\lambda}} \bar{\lambda}^2 + \mathcal{O}(\lambda^3, \alpha^2) \; ,
\end{equation}
where $C_{O_{\lambda} O_{\lambda}}^{ O_{\lambda}}$ is the OPE coefficient of the composite operator $O_\lambda$ with itself. We can find this OPE coefficient by using the Wick contractions of $\hat{O}^-_{-\alpha}$ inside $O_\lambda$, which gives $C_{O_{\lambda} O_{\lambda}}^{ O_{\lambda}}=2$. Inserting this value for the OPE coefficient gives the perturbative beta function
\begin{equation} \label{eq:beg}
\beta_{\bar{\lambda}} = - 2\alpha \bar{\lambda} +  2\frac{\pi^{\frac{d}{2}-1}}{\Gamma \left(\frac{d}{2}-1 \right)}  \bar{\lambda}^2   + \mathcal{O}(\lambda^3, \alpha^2) \;,
\end{equation}
which agrees with \eq{betaLam} for small $\alpha$.
This perturbative beta function admits a fixed point 
\begin{equation} \label{eq:FPpert}
\bar \lambda^*= \alpha \frac{\Gamma(\frac{d}{2}-1)}{\pi^{\frac{d}{2}-1}} + \mathcal{O}(\alpha^2)\; .
\end{equation}
and an IR anomalous dimension for the composite operator
\begin{equation}
\gamma_\lambda^*=  2 \gamma_{\hat{O}^{-}_{-\alpha}}^* =\left.\frac{\partial \beta_{\bar{\lambda}}}{\partial \bar{\lambda}}\right|_{\bar{\lambda}=\bar{\lambda}^*} = +2\alpha \; ,
\end{equation}
which curiously holds to all orders in $\alpha$.

Before moving on, let us discuss the extremal cases of the RG flow defined by \eqref{eq:betaLam}. By taking the $\alpha \to 0$ limit, the second term in \eqref{eq:betaLam} dominates and we recover the beta function of the $\varphi^2$ defect deformation presented in \cite{Metlitski:2009iyg}, with the defect decoupling in the IR. Instead in the $\alpha \to 1$ limit, the second term vanishes, and the fixed point \eqref{eq:FPallOrder} moves to $\infty$. We are left with a flow analogous to a boundary mass deformation in a bCFT flowing from Neumann to Dirichlet boundary conditions in the IR. In this case the `Dirichlet' condition corresponds to the existence of a non-decoupled operator of dimension $d/2$, which is analogous to $\partial_\perp \phi$ in a bCFT.

\subsection{Fermion monodromy flows}
\label{sec:rg-flows_fermions}
In this subsection, we consider the defect RG flow on a monodromy defect in a theory of free Dirac fermions. In this case, we can construct two separate defect operators: ${\mathcal{O}_{-\alpha} \equiv \int d^{d-2} \sigma\, \hat{\bar{\psi}}_{-\alpha} \hat{\psi}_{-\alpha}}$ with dimension $\Delta = 1-2\alpha$ and $\mathcal{O}_{ \alpha}\equiv \int d^{d-2} \sigma\, \hat{\bar{\psi}}_{\alpha} \hat{\psi}_{\alpha}$ of dimension $\Delta = 2\alpha-1$. Here, $\hat \psi_{-\alpha} \equiv \lim_{\rho \rightarrow 0} \psi_{\alpha} \, \rho^{\alpha}/\sqrt{\xi}$ and $\hat \psi_{\alpha} \equiv \lim_{\rho \rightarrow 0} \psi_{\alpha}\, \rho^{1-\alpha}/\sqrt{1-\xi} $ are the defect operators associated to the divergent components of the modes $\psi_{-\alpha}$ and $\psi_{\alpha}$ defined in \eq{n=0-modes}.  The operator $\mathcal{O}_{-\a}$ is relevant when $\alpha > 1/2$, while $\mathcal{O}_{\a}$ is relevant for $\alpha < 1/2$, and both are marginal at $\alpha = 1/2$. 

The goal is then to show that, for $\a>1/2$, $\mathcal{O}_{-\alpha}$ triggers an RG flow toward a defect theory with $\xi = 1$, while, for $\a<1/2$, $\mathcal{O}_{\a}$ triggers a flow to $\xi =0$.
In the following we will focus on $1/2<\alpha<1$ and $0<\xi < 1$, and consider a flow triggered by $\mathcal{O}_{-\alpha}$. We repeat the same analysis performed above for the RG flow in the scalar theory to obtain the corresponding deformed propagator. Finally, we will show that sending the coupling constant to infinity one recovers the fermion propagator at $\xi = 1$. The other case follows completely analogously.  

To study the desired defect RG flow, we need to compute 
\begin{equation}
G^{(\lambda)}_{F,\alpha,\xi}(x,x') \equiv  \left< \psi (x) \psi^\dagger (x') \right>_{\lambda} i\gamma^0 =  \frac{\left< \psi (x) \psi^\dagger (x')  e^{\lambda \int d^{d-2}\sigma \,\hat{\bar{\psi}}_{-\alpha} \hat{\psi}_{-\alpha} } \right> i\gamma^0}{\left<  e^{\lambda \int d^{d-2}\sigma \,\hat{\bar{\psi}}_{-\alpha} \hat{\psi}_{-\alpha} } \right>}\,,
\end{equation}
where again without loss of generality we take $x = \{\rho,\,\theta,\,\sigma\}$ and by symmetry transformations set $x' =\{\rho',0,0\}$.  Ultimately, we want to obtain the variation of the propagator $\Delta G^{(\lambda)}_{F}(x,x') \equiv G^{(\lambda)}_{F,\alpha,\xi}(x,x')- G^{(0)}_{F,\alpha,\xi}(x,x')$.  By following the same steps as in the scalar case, we find
\begin{equation}
\label{eq:fermion:RG_series}
\begin{split}
G^{(\lambda)}_{F,\alpha,\xi}(x,x') & = \left< \psi (x) \psi^\dagger (x') \right> i\gamma^0  +\sum_{n = 1}^{\infty} \lambda^n \int \prod_{i=1}^{n} \, d^{d-2}\sigma_{i} i\left< \psi(x) \hat{\psi}_{-\alpha}^\dagger (\sigma_1)\right>i\gamma^0 \times \\
&\hspace{1cm}\times\prod_{k=1}^{n-1} i\left< \hat{\psi}_{-\alpha}(\sigma_{k})\hat{\psi}^\dagger_{-\alpha}(\sigma_{k+1}) \right>\gamma^0 \left< \hat{\psi}_{-\alpha}(\sigma_n)\psi^\dagger(x') \right>i\gamma^0 .
\end{split}
\end{equation}
As in the previous section, we need both the defect-defect propagator and the bulk-defect propagator.
These can be found by taking the defect limit of the fermion propagator \eq{fermion-propagator-1}. First of all, we observe that a mode of scalar propagator labelled by $\nu$ may be written in the following form
\begin{equation}
\label{eq:scalar_prop_IK}
G^{(\nu)}_S(x,x')= \int \frac{d^{d-2}  k}{(2\pi)^{d-1}}  \, e^{ik \cdot \sigma} K_{\nu} \left(k\,\rho\right) I_{\nu} \left(k\,\rho'\right), \qquad \rho> \rho'>0\,,
\end{equation}
and the normalisation for the ${m=0}$ and ${m=1}$ modes adopted in \eq{zero_mode_scalar} is understood. The above integral follows straightforwardly from performing the $k_\rho$-integral in equation \eq{prop_mode_int}.

The defect-defect propagator can be found by first taking the coincident limit in the orthogonal directions and then extracting the singular term proportional to $\rho^{-2\alpha}$ in the defect limit.  Taking this ordered pair of limits gives
\begin{equation}
\label{eq:fermion_defect_defect}
\left< \hat{\psi}_{-\alpha}(\sigma) \hat{\psi}^\dagger_{-\alpha}(0) \right> i\gamma^0 = -\frac{2^{2\alpha-1}\Gamma(\alpha)}{2 \pi\Gamma(1-\alpha)}  \int \frac{d^{d-2}k}{(2\pi)^{d-2}} \, e^{i k \cdot \sigma}k^{-2\alpha} i k_a\gamma^a P_{-}\,.
\end{equation}
The bulk-defect propagator can be similarly obtained by extracting the term in the propagator \eq{fermion-propagator-1} which diverges as $\rho'^{-\alpha}$ in the $\rho' \rightarrow 0$ limit. The result is
\begin{equation}
\label{eq:fermion_bulk_defect}
\begin{split}
\left< \psi(x) \hat{\psi}^\dagger_{-\alpha}(0) \right> i\gamma^0 = & \frac{2^{\alpha} \sqrt{1-\xi}}{2\pi\Gamma(1-\alpha)}  \int \frac{d^{d-2}k}{(2\pi)^{d-2}} \, e^{i k\cdot \sigma} e^{i \left(\frac{1}{2}-\alpha\right)\theta_1} k^{-\alpha} \times \\
& \quad\times \bigg{\{}\left[  i k_a \gamma^a K_{-\alpha}\left(k\rho \right)  P_{-} \right] + k K_{1-\alpha}\left(k\rho\right) \gamma^1 \left(P_{+}-P_{-}\right)  \bigg{\}}.
\end{split}
\end{equation}
By plugging eqs.~\eqref{eq:fermion_defect_defect} and \eqref{eq:fermion_bulk_defect} into \eq{fermion:RG_series}, and following the same steps as in the scalar case, we find
\begin{equation}
\label{eq:delta_prop_ferm}
\begin{split}
\Delta G^{(\lambda)}_{F}(x,x') &=  - \frac{2^{2\alpha}(1-\xi)}{4\pi^2\Gamma(1-\alpha)^2} \lambda \int \frac{d^{d-2}k}{(2\pi)^{d-2}} \, e^{i k\cdot \sigma} e^{i \left(\frac{1}{2}-\alpha\right)\theta_1}\frac{ k^{-2\alpha}}{1+C^2 k^{2-4\alpha} \lambda^2} \times \\
& \quad\times \left(\left[  i k_a \gamma^a K_{-\alpha}\left(k\rho \right)  P_{-} \right] - k K_{1-\alpha}\left(k\rho\right) \gamma^1 P_{-}\right)\left( 1-C k^{-2\alpha}\lambda i k_a\gamma^a\right) \times \\
&\quad \times \left(\left[ - i k_a \gamma^a K_{-\alpha}\left(k\rho' \right)  P_{-} \right] - k K_{1-\alpha}\left(k\rho'\right) \gamma^1 P_{+} \right),
\end{split}
\end{equation}
where $C \equiv 2^{2\alpha-1}\Gamma(\alpha)/(2 \pi \Gamma(1-\alpha))$. The contribution of \eq{delta_prop_ferm} to the fermion propagator corresponds to a non-conformal defect where the scale invariance of the defect is broken by the dimensionful coupling $\lambda$. It is easy to see that the IR fixed point is reached in the limit $\lambda \rightarrow +\infty$ where the propagator again describes a theory with a conformal defect. 

Now we show that the IR fixed point obtained by taking $\lambda \to \infty$ corresponds to $\xi = 1$. Namely, we need to prove that 
\begin{equation}
\label{eq:Ginf=G1-Gchi}
\Delta G^{\rm{IR}}_{F}(x,x') = G_{F,\alpha,1}(x,x')- G^{(0)}_{F,\alpha,\xi}(x,x')\,.
\end{equation}
Taking $\lambda\to\infty$ in \eq{delta_prop_ferm}, we obtain
\begin{equation}
\label{eq:delta_inf_prop_ferm}
\begin{split}
\Delta G^{\rm{IR}}_{F}(x,x') =\;&  \frac{(1-\xi)\sin \pi \alpha}{\pi^2}  \int \frac{d^{d-2}k}{(2\pi)^{d-2}} \, e^{i k\cdot \sigma} e^{i \left(\frac{1}{2}-\alpha\right)\theta_1}  \times \\
& \times \bigg\{\left[  i k_a \gamma^a K_{-\alpha}\left(k\rho\right) K_{-\alpha}\left(k\rho'\right) -k K_{1-\alpha}(k\rho)K_{-\alpha}(k\rho') \gamma^1 \right]P_{-} \\
&\;\;\;\, - \left[   i k_a \gamma^a K_{1-\alpha}\left(k\rho\right) K_{1-\alpha}\left(k\rho'\right) -k K_{-\alpha}(k\rho)K_{1-\alpha}(k\rho') \gamma^1  \right]P_+ \bigg\} .
\end{split}
\end{equation}
At this point, it is straightforward to check the relation \eq{Ginf=G1-Gchi} holds by computing the difference $G_{F,\alpha,1}-G_{F,\alpha,\xi}$ directly from \eq{fermion-propagator-1}. In the difference, only the modes $n=0,1$ contribute and we are left with
\begin{equation}
\begin{split}
G_{F,\alpha,1}(x,x')-G_{F,\alpha,\xi}(x,x') =& \frac{ \sin \pi \alpha (1-\xi)}{\pi^2} \Delslash \bigg\{  \int \frac{d^{d-2}k}{(2\pi)^{d-2}} \, e^{i k\cdot \sigma} e^{i\left(\frac{1}{2}-\alpha \right)(\theta_1-\theta_2)}  \times\\ 
&\,\,\times \left[  K_{1-\alpha}(k\rho) K_{1-\alpha}(k\rho') P_+- K_{\alpha}(k\rho) K_{\alpha}(k\rho') P_- \right] \bigg\} ,
\end{split}
\end{equation}
which gives exactly \eq{delta_inf_prop_ferm} after affecting the derivative $\Delslash$, thus proving \eq{Ginf=G1-Gchi}.

As a consistency check, we compute the change in the expectation value of the current $\braketbis{J_\theta}$ from the propagator \eq{delta_prop_ferm} at a generic value of~$\lambda$
\begin{equation}
\Delta \left< J^\theta (\rho) \right> = -\frac{i}{\rho} \lim_{\epsilon \rightarrow 0} \Tr \left[\Delta G^{(\lambda)}_{F,\alpha,\xi} (\rho+\eps,\rho) \gamma^2 \right],
\end{equation}
where all other coordinate dependence in $\Delta G^{(\lambda)}_{F,\alpha,\xi}(x,x')$ is suppressed as we set all but $\rho$ to $0$. As in the case of computing $\Delta\braketbis{|\vphi|^2}$ for the scalar RG flow above, we restrict our attention to $d=4$ where most of the contributions vanish, and we are left to compute the following integral
\begin{equation}
\label{eq:delta_current_ferm}
\Delta \left< J^\theta (\rho) \right> = \frac{2^{2\alpha+1}(1-\xi)}{4 \pi^3 \Gamma^2(1-\alpha)}\frac{C }{\rho^4} \int_0^{+\infty} d\zeta \, \frac{\zeta^{4-4\alpha}}{\rho^{2-4\alpha} \lambda^{-2}+\zeta^{2-4\alpha}C^2 } K_{-\alpha}(\zeta)K_{1-\alpha}(\zeta)\,,
\end{equation}
whose IR limit $\rho \lambda^{1/(2\alpha-1)} \rightarrow +\infty$ gives 
\begin{equation}
\label{eq:delta_diff_ferm_d4}
\Delta \left< J^\theta (\rho) \right> = \frac{(1-\xi )(1-\alpha)\alpha}{\pi^2} \frac{1}{\rho^4}\,.
\end{equation}
As expected, this is exactly the difference $\Delta C_J \equiv C_J(\xi = 1)-C_J(\xi)$ where $C_J$ is the coefficient of the one-point function of the current found in \eq{fermion-CJ_d4}.

\begin{figure}
	\begin{center}
		\includegraphics[scale=0.9]{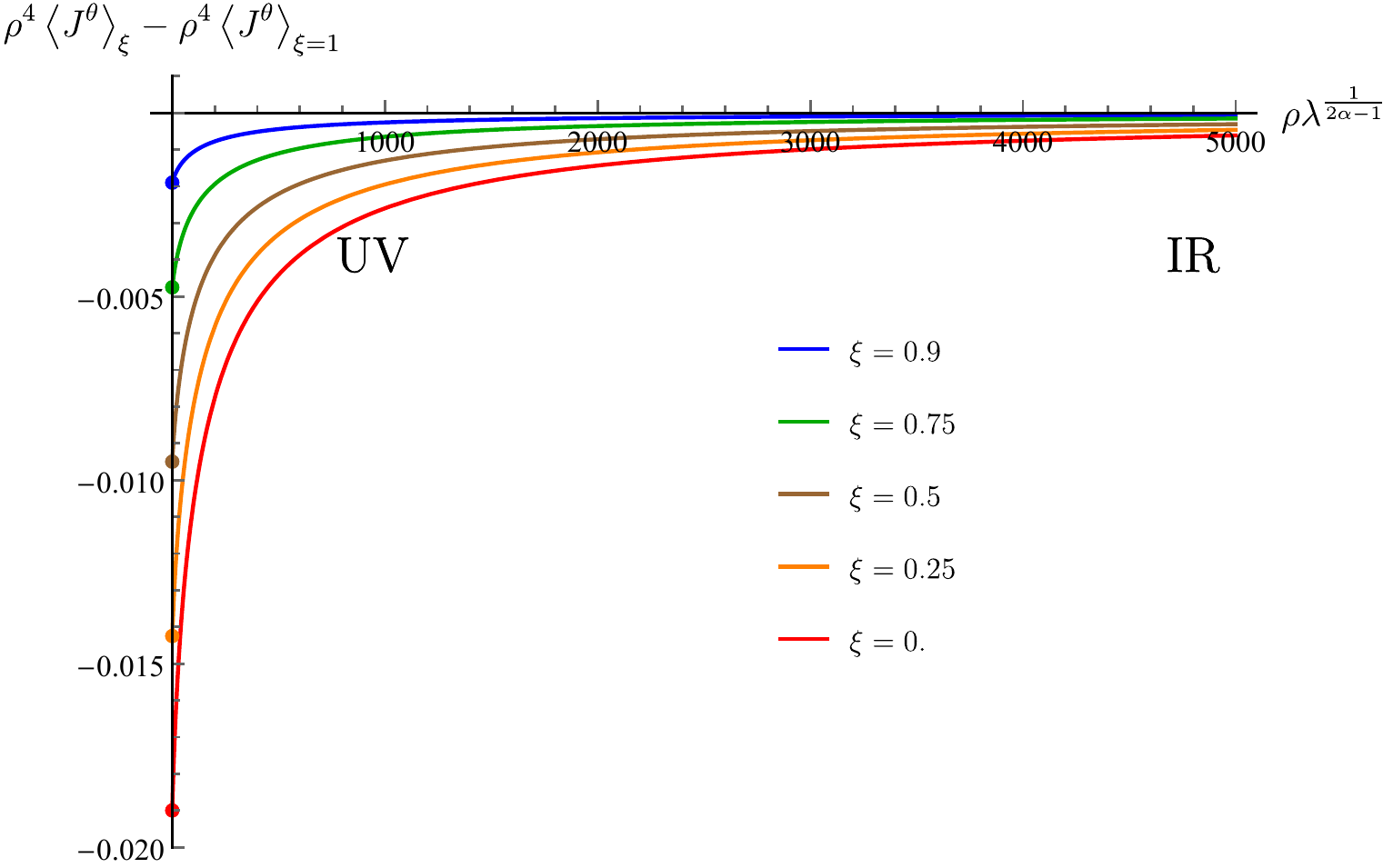}
		\caption{The quantity $\rho^4 \left< J^\theta  \right>_{\xi} - \rho^4 \left< J^\theta  \right>_{\xi =1}$ as a function of $\rho \lambda^{1/(2\alpha-1)}$ for $d=4$, $\alpha = 0.75$, and different values of $\xi$. While in the UV the quantity depends on $\xi$, in the IR limit all the curves go to zero. \label{fig:delta_J}}
	\end{center}
\end{figure}

For generic values of $\rho \lambda^{1/(2\alpha-1)}$ we need to solve the integral numerically. In \fig{delta_J} we show the quantity
\begin{equation}
\rho^4 \left< J^\theta  \right>_{\xi} - \rho^4 \left< J^\theta  \right>_{\xi =1} = \rho^4 \Delta \left< J^\theta  \right> - \frac{(1-\xi )(1-\alpha)\alpha}{\pi^2}
\end{equation}
as a function of $\rho \lambda^{1/(2\alpha-1)}$ for different values of $\xi$. In the UV regime  $\rho \lambda^{1/(2\alpha-1)} =0$ the curves depend on $\xi$ while they all go to zero in the IR limit $\rho \lambda^{1/(2\alpha-1)} \rightarrow +\infty$, confirming that the IR for all those cases corresponds to $\xi=1$.

Let us briefly take note of how $b$ behaves under defect RG flows.  Integrating $\Delta C_J$ from \eq{delta_diff_ferm_d4}, we find that $b$ is strictly positive, and at the IR fixed point where $\xi=1$ it vanishes identically.  Hence for any UV value $\xi\neq 1$, we have $b_{\rm{UV}}> b_{\rm{IR}}$, and so again the $b$-theorem is obeyed \cite{Jensen:2015swa}.  Further, when $0<\a<1/2$, the relevant operator $\CO_{\a}$ drives the flow. Following the same computations as above, one finds a conformal defect with $\xi=0$ at the IR fixed point and $\Delta C_J$ takes a form similar to that found in \eq{delta_diff_ferm_d4} but with $\a\to 1-\a$ and $\xi\to 1-\xi$.  Thus, since $\xi\in[0,1]$, any UV value of $\xi\neq 0$ flowing to $\xi=0$ in the IR has $b_{\rm{UV}}>b_{\rm{IR}}$, again in accord with the $b$-theorem.

We conclude this section with a comment on the limit $\alpha \rightarrow 1/2$, where both relevant deformations $\CO_{-\alpha}$ and $\CO_{\a}$ become marginal. The marginal case can be studied directly starting from \eq{delta_prop_ferm} for the deformation by $\CO_{-1/2}$. The result is a well-defined propagator without any scale. Also in this case, the limit $\lambda \rightarrow + \infty$ gives the propagator \eq{fermion-propagator-1} with $\xi = 1$. The difference between the marginal and the relevant case with $\alpha> 1/2$ is that now the defect theory is scale invariant for any value of $\lambda$. By a direct computation, it is not difficult to check that for $\alpha =1/2$, the perturbation does not affect the one-point function of the stress-tensor leading to the same value of $h$ for any value of $\lambda$, while the central charge $b$ is obviously invariant since the defect deformation is marginal. In addition, we observe that the coefficients $h$ in \eq{fermion-h}, $b$ in \eq{fermion-b-xi} and $C_\CD$ in \eq{fermion-CD-d} are in fact independent of $\xi$ precisely at $\alpha=1/2$. This is consistent with the fact that this marginal deformation reproduces the propagator \eq{fermion-propagator-1} with $\xi = 1$ at $\lambda \rightarrow + \infty$ even though it leaves invariant those charges. A similar discussion applies to the deformation by $\CO_{1/2}$.

\section{Discussion}\label{sec:discussion}

In this work we have presented an extensive study of the behaviour of monodromy defects in $d$-dimensional free CFTs.  By utilising the analytic methods available in free field theories, we have computed various important correlation functions of the stress tensor, conserved $U(1)$ currents, and the displacement operator, which in $d=4$ are related to the Weyl anomaly coefficients of a $p=2$ dCFT.  Further, we leveraged heat kernel methods to compute the universal part of the defect EE in $d=4$, and in doing so, we provided an explicit check on the values for $b$ and $d_2$.  In addition, we have used the analytic results obtained for monodromy defects in $d=4$ to find the defect central charges of free CFTs in the presence of conical singularities. We also studied an analogous system of defects in pure $d=4$ Maxwell theory, i.e. Gukov-Witten defects. By computing their defect central charges, we demonstrated that they are topological, as expected.

Beyond characterising defects through their central charges, we have also shown their behaviour under defect RG flows.  In particular, we have seen that monodromy defects in theories of free scalars and free Dirac fermions allow for peculiar mildly singular modes labelled by $\xi,\,\tilde{\xi} \in [0,1]$ that signal the presence of certain defect operators, $\hat{O}^-_{-\a}$ and $\hat{O}^-_{1-\a}$, appearing in the defect OPE.  We have used these novel operators to build relevant deformations that trigger defect RG flows.  In the case of scalar monodromy defects in particular, we have shown in detail using the defect OPE and the beta function that no matter what the values of $\xi,\,\tilde{\xi}\in[0,1]$ are in the UV, the IR fixed point of the defect flow is always a dCFT with $\xi=\tilde{\xi}=0$. Thus, the IR theory only retains the regular modes.

For the case of monodromy defects in free fermion CFTs, the story is a bit more subtle as the IR values of $\xi$ depend on the monodromy parameter, $\a$.  In the fermionic theory, there are always two paired singular modes $\psi_{-\a}$ and $\psi_{\a}$ characterised by a $\CO(\rho^{-\a})$ and $\CO(\rho^{\a -1})$  behaviour as $\rho\to0$ respectively.  When $0<\a<\frac{1}{2}$, the flow to the IR fixed point takes $\xi\to0$, i.e. the mode $\psi_{\a}$ disappears from the spectrum, whereas for $\frac{1}{2}<\a<1$, the theory flows to a dCFT with $\xi =1$, i.e. the one without $\psi_{-\a}$. An interesting feature, that is similar to the scalar case, is that the IR theory always retains the least singular mode. If the flux is set to $\a=\frac{1}{2}$ there are two exactly marginal deformations.  This is a rather interesting feature as non-trivial conformal manifolds without SUSY are not common. We leave a more extensive study of this manifold for future work.

Although in this work we only considered quadratic relevant deformations, this is not the only possibility. For instance, in the scalar theory it is possible to construct more general operators of the type $\left(\hat O^{-}_{-\alpha} \hat O^{\dagger -}_{-\alpha}\right)^n$ where $n\ge 1$ is an integer. A straightforward dimensional counting shows that these operators are relevant provided $\alpha> (n-1)(d-2)/(2n)$, where we remind the reader that $\alpha \in (0,1)$. Thus, we find non-quadratic relevant deformations if $2 <d< 6$. Interestingly, when $d=3$ or $d=4$ one can always restrict the range of $\alpha$ such that the operator is relevant for \textit{any} $n>1$.
In the fermionic case instead we find non-quadratic relevant deformations only if $2 < d <4$. This kind of operators may provide a dynamical mechanism for having non-trivial interacting fixed points with $\xi \ne 0,1$. We leave to future work a systematic study of such flows.

While the analysis in the main sections of this paper has focussed exclusively on non-SUSY monodromy defects, our free field computations have a direct connection to monodromy defects in a theory of $d=4$ free $\CN=2$ hypermultiplets.  In particular, we have shown that one can add $d_2$ (or $d_1$) for the monodromy defects of one free complex scalar with $\xi=\tilde\xi=0$, one free complex scalar with $\xi=1$, $\tilde{\xi}=0$, and a Dirac fermion with $\xi=0$ to obtain $6\a^2$. This exactly matches the value of $d_2=d_1$ for a free $d=4$ $\CN=2$ hypermultiplet at the defect SUSY preserving value $\xi=0$, which can be computed through spectral flow by $\a$ in the chiral algebra description \cite{Bianchi:2019sxz}.  Using the same strategy we can compute the value $b = 3\a^2$ for the A-type anomaly coefficient. 

Since $d_2=d_1$ can be found directly as the dimension of the defect identity module in the chiral algebra, one might wonder if there is a quantity in the chiral algebra that captures~$b$.  Given the special relationship that $b$ and $C_J(\a)$ have for monodromy defects in non-SUSY theories, we believe that an integrated defect three-point function $\braketbis{\varsigma_\a(\infty) J(\sigma) \varsigma_\a(0)}$ can be used to compute $b$. Here, $\varsigma_\a$ is the spectral flowed defect identity and $J$ is the affine Kac-Moody current associated to a preserved abelian flavour symmetry, i.e. for the Cartan $\widehat{U(1)}_f$ of the $\widehat{SU(2)}_f$ flavour symmetry.\footnote{The correlator $\braketbis{\varsigma_\a(\infty) J(\sigma) \varsigma_\a(0)}$ would give the one-point function of the moment map operator, i.e. the superprimary of the flavour current multiplet, which could then be obtained by supersymmetric Ward identities.} However, the formula in \eq{b-CJ-relation} was specifically derived for spherical Lagrangian defects, and it is unknown at this time if such a relation could be straightforwardly applied to the construction of SUSY-preserving monodromy defects in the $\mathcal{N}=2$ free hyper theory.  This is a question that merits further investigation.

\section*{Acknowledgements}
The authors would like to thank Christoper Herzog and Andy O'Bannon for their insightful, collaborative discussions throughout the duration of this work. V.P. would like to thank Alexander S\"oderberg for introducing him to the physics of monodromy defects. J.S. would like to thank Cl\'ement Berthiere for useful discussions on the heat kernel method.  L.B. is funded through
the MIUR program for young researchers ``Rita Levi Montalcini''. A.C. is funded by the Royal Society award RGF/EA/180098.  V.P. is funded by  Vetenskapsr$\mathring{\text{a}}$det under grant 2018-05572.   B.R. is funded by in part by the KU Leuven C1 grant ZKD1118 C16/16/005.  J.S. is funded in part by the Royal Society award RGF/EA/181020.
\hspace{0.51cm}

\appendix 
\section{Spherical Defects}\label{app:spherical}

In this section we outline the derivation of \eq{b-CJ-relation} for a spherical defect in $\mathds{R}^d$. Our strategy will be to apply a conformal transformation that leaves invariant the flat space $\mathds{R}^d$ but maps the support of the defect from the sphere $\mathbb{S}^{d-2}$ of radius $R$ to the plane $\mathds{R}^{d-2}$. This relates the spherical defect to the flat defect for which $J \cdot A = \a J^\theta$.

More concretely, we take the two directions transverse to the defect to have coordinates $x^1$ and $x^2$, and use the conformal transformation
\begin{equation}
\label{eq:mapping}
\begin{split}
X^1 & 
=R \frac{4 x^\mu x^\mu - R^2}{R^2 + 4 x^\mu x^\mu + 4 R \, x^1}\,,\quad
X^k = 4 R^2 \frac{x^k}{R^2 + 4 x^\mu x^\mu + 4 R \, x^1} \,,
\end{split}
\end{equation}
with inverse
\begin{equation}
\label{eq:mapping_inv}
\begin{split}
x^1 & 
=\frac{R}{2} \frac{R^2-X^\mu X^\mu}{\left( R-X^1 \right)^2+ X^k X^k}\,,\quad
x^k = R^2 \frac{X^k}{\left( R-X^1 \right)^2+ X^k X^k} \,,
\end{split}
\end{equation}
where $\mu=1,\ldots, d$ and $k=2,\dots,d$.
Under this transformation, the background flat metric of $\mathds{R}^d$ transforms to
\begin{equation}
\label{eq:transf_metric}
ds^2 = \Omega^2(X)\delta_{\mu\nu}dX^\mu dX^\nu\,, \qquad \Omega(X) = \frac{R ^2}{\left[(R -X^1)^2+ X^k X^k\right]}\,,
\end{equation}
where the support of the defect, i.e. the hyperplane $x^1=x^2=0$, is mapped to the sphere $\mathbb{S}^{d-2}$ defined by $X^\mu X^\nu \delta_{\mu\nu} = R^2$ and $X^2=0$.

In order to show \eq{b-CJ-relation}, we need to compute the integral on the right-hand side of \eq{log-generic-defect} when the defect is spherical. Namely, we need to evaluate the following integral
\begin{equation}
\label{eq:int_I_J}
I_J \equiv \int d^d X  \,  \delta^{\mu\nu} \left< J_\mu \left( X \right)\right>_{\text{\tiny spherical}} f_\nu \left( X \right) ,
\end{equation}
where the shape function $f_\mu(X)$, whose precise form we do not need, is the one of a spherical defect.

All we need is to find the integrand in \eq{int_I_J}. To do this, we first apply the transformation \eq{mapping}, and then we perform a Weyl transformation to remove the conformal factor in \eq{transf_metric}. Under such a transformation, a vector primary operator of dimension $\Delta_V$ transform as \cite{Ho_2011}
\begin{equation}
V_\mu'(x') = \Omega^{1-\Delta_V}(x) V_\nu(x) \frac{\partial x^\nu}{\partial x'^\mu}\,.
\end{equation}
Since the integrand is invariant under a change of coordinates, the only modification comes from the conformal factor. Thus, we simply obtain
\begin{equation}
 \delta^{\mu\nu} \left< J_\mu \left( X \right)\right>_{\text{\tiny spherical}} f_\nu \left( X \right) = \delta^{\mu\nu}  \left< J_\mu \left( x(X) \right)\right>_{\text{\tiny flat}} f_\nu \left( x(X) \right) \Omega^d (X) \,,
\end{equation}
where we used that  $\Delta_J=d-1$ for a conserved current, $g^{\mu\nu} = \Omega^{-2} \delta^{\mu\nu}$, and that $f$ does not contribute because $\Delta_A =1$. Finally, we are left with the following integral
\begin{equation}
I_J =  C_J \int d^d X \, \left[\frac{2 R }{\sqrt{(R^2-X^\mu X^\mu)^2 +4 R^2 (X^2)^2}} \right]^d .
\end{equation}
To solve this integral, it is convenient to employ spherical coordinates. We choose
\begin{equation}\label{eq:Xr}
X^2= \text{r} \cos \phi_1\,, \;\; X^1 = \text{r} \sin \phi_1 \cos \phi_2\,,\;\; \dots\,\; \;    X^d = \text{r} \sin\phi_1 \sin\phi_2 \dots \cos\phi_{d-1}\,,
\end{equation}
which gives
\begin{equation}
\label{eq:I_J_int}
\begin{split}
I_J & = C_J \, \text{Vol}\left( \mathbb{S}^{d-2} \right) \int_0^{+\infty} d\text{r} \int_0^\pi d\phi_1   \, \text{r}^{d-1} \sin^{d-2}\phi_1 \,\left[\frac{2 R }{\sqrt{(R^2-\text{r}^2)^2 +4 R^2 \text{r}^2 \cos^2 \phi_1}} \right]^d \\
& =  \frac{2^d \sqrt{\pi} \, \Gamma\left(\frac{d-1}{2}\right)}{\Gamma\left(\frac{d}{2}\right)} \text{Vol}\left( \mathbb{S}^{d-2} \right) C_J \int_0^{+\infty} d\text{r} \, R^d \frac{ \text{r}^{d-1}}{(R^2+\text{r}^2)\left| R^2-\text{r}^2 \right|^{d-1}}\,.
\end{split}
\end{equation}
The integral over $\text{r}$ converges in the range $0<d<2$ while is divergent for $d \ge 2$. We will adopt dimensional regularisation to obtain the value in our range of interest $d \ge 2$. Performing the integral is straightforward and we obtain
\begin{equation}
\label{eq:I_J_d_generic}
 I_J  = C_J\frac{2 \pi ^{\frac{d}{2}+1}}{\Gamma \left(\frac{d}{2}\right) \sin \left( \frac{\pi}{2} d \right)}\,.
\end{equation}
We note that the above result is well-defined for any value of $d  >0$ such that $ d \ne \text{even}$, while when $d$ is an even number the expression has a simple pole. This pole is expected since the free energy acquires an additional divergence due to the trace anomaly. In our case this corresponds to the derivative with respect to $\alpha$ of the A-type defect anomaly. We can extract the coefficient of the divergence by replacing $d \rightarrow d + \tilde \varepsilon$ where now $d$ is assumed to be a positive even integer and $0 <\tilde \varepsilon \ll 1$. Thus we find,
\begin{equation}
\label{eq:I_J_d_even}
 I_J  =(-1)^{d/2} C_J \frac{4  \pi^{d/2}}{\Gamma \left(\frac{d}{2}\right)} \, \frac{1}{\tilde \varepsilon} + \mathcal{O}(\tilde \varepsilon)\,, \qquad d = \, \text{even}\,.
\end{equation}
The pole in the dimensional regularisation maps to a logarithmic divergence when the integral \eq{I_J_int} is regularised by a UV cut-off $\varepsilon$, i.e. $1/\tilde \varepsilon \rightarrow \log(R/\varepsilon)$. More precisely, the integral diverges at the location of the defect $\text{r}=R$, and we must divide the integration region as follows $\text{r} \in (0,R-\varepsilon) \cup (R+\varepsilon, +\infty)$, where $\varepsilon$ is a UV cut-off. Performing the integral for fixed values of $d$, we find indeed that the universal part of $I_J$ matches exactly the equations \eqref{eq:I_J_d_generic} and \eqref{eq:I_J_d_even}, and so
\begin{equation}
\label{eq:der_logZ_d}
\frac{d}{d\alpha} \log Z[\alpha] = \begin{cases}
\displaystyle{ C_J\frac{2 \pi ^{\frac{d}{2}+1}}{\Gamma \left(\frac{d}{2}\right) \sin \left( \frac{\pi}{2} d \right)}} & d \ne \text{even}\,, \\
\phantom{a} \\
\displaystyle{ (-1)^{d/2} C_J \frac{4  \pi^{d/2}}{\Gamma \left(\frac{d}{2}\right)} \, \log \left( \frac{R}{\varepsilon} \right) } & d = \text{even} \,.
\end{cases}
\end{equation}
This straightforwardly gives the results in eqs. \eqref{eq:b-CJ-relation}, \eqref{eq:F_deriv}, and \eqref{eq:A-anom-der}.\footnote{In principle one might worry about other $\alpha$-dependent terms contributing to the partition function. Indeed in the presence of a bulk trace anomaly one gets a logarithmically divergent term proportional to $ \int F \wedge \star{F}$, where $F$ is the curvature of background gauge field \eqref{eq:bc-gauge-field}. This term, however, is quadratic in delta functions and hence doesn't contribute as explained in section \ref{sec:maxwell}.}

\section{The Scalar Propagator}\label{app:scalar}

Below we study the mode expansion and the propagator of a scalar field in the presence of a non-trivial monodromy in Lorentzian signature $(-,+,\dots,+)$. In the Minkowski space-time with metric \eq{background-metric-R-d}, the equation of motion for $\varphi$ reads
\begin{equation}
\label{eq:eq_mot_scalar}
\begin{split}
\frac{1}{\rho} \partial_\rho (\rho \partial_\rho \varphi) +\left[ -\partial_t^2 +\frac{1}{\rho^2}(\partial_\theta-i \alpha)^2 + \nabla_{||}^2 \right]\varphi =0\,,
\end{split}
\end{equation}
where $\nabla_{||}^2$ is the spatial part of the Laplacian operator in the direction parallel to the flux $\alpha$.
To find the general solution to the equation of motion, we employ the cylindrical symmetry of the problem and we write the ansatz
\begin{equation}
\label{eq:scalar_ansatz}
\varphi=e^{- i \omega t}e^{i \vec k \cdot \vec \sigma}e^{i n \theta}h(\rho)\,.
\end{equation}
Plugging \eq{scalar_ansatz} into \eq{eq_mot_scalar}, we obtain an equation for $h(\rho)$
\begin{equation}
\rho^2 h''+\rho \, h'+\left[\left(\omega^2-{\vec k^2}\right)\rho^2-\left(m-\alpha\right)^2\right]h=0\,,
\end{equation}
whose solutions are the Bessel functions 
\begin{equation}
h= J_{\pm (n -\alpha)} \left(k_\rho \, \rho \right), \qquad k_\rho = \sqrt{\omega^2 - \vec{k}^2}\,.
\end{equation}
While the large $\rho$ behaviour of the functions $J_\beta$ is physically reasonable for any $\beta$, their behaviour near $\rho=0$ needs to be discussed carefully. Expanding the Bessel function near zero, one finds
\begin{equation}
J_{\beta}(\zeta) = \zeta ^{\beta } \left(\frac{2^{-\beta }}{\Gamma (\beta +1)}-\frac{2^{-\beta -2} \zeta^2}{(\beta +1) \Gamma (\beta +1)}+\mathcal{O}\left(\zeta^3\right)\right).
\end{equation}
Depending on $\beta$, the function may be divergent as $\zeta \to 0$. 

If one asks for regularity of the scalar field in the limit $\rho \rightarrow 0$, one must require the order of the Bessel function to be non-negative, i.e. $\pm (n-\a) \geq 0$. This leads to the following solution
\begin{equation}
\begin{split}
\varphi= \sum_{n=-\infty}^{\infty} \int dk_\rho \int d^{d-3} {\vec k} \,  \Bigg[ & \tilde a_n ( k) e^{i n\theta}J_{|n-\alpha|} \left(k_\rho\,\rho\right)e^{-i \omega t+i {\vec k}\cdot {\vec \sigma}}  \\
&+\tilde b_n^*( k) e^{-in\theta}J_{|n+\alpha|} \left(k_\rho\,\rho\right)e^{i \omega t-i {\vec k}\cdot{\vec \sigma}}\Bigg],
\end{split}
\end{equation}
where $\tilde a_n ( k)$ and $\tilde b_n ( k)$ are undetermined functions, and we denoted $ k=(k_\rho, \vec k_{||})$. This choice of boundary conditions has been argued to a monodromy defect engineered by an infinitely thin and infinitely long solenoid carrying a magnetic flux $\alpha$ \cite{PhysRev.115.485,ALFORD1989140,deSousaGerbert:1988qzd}. 

To quantise the theory, we impose the canonical equal-time commutation relation $[\varphi(x,t),\dot \varphi^\dagger(x',t)]=i\delta^{(d-1)} (x-x')$. After the quantization, the coefficients $\tilde a_n ( k)$ and $\tilde b_n ( k)$ become operators, which are proportional to the canonical creation and annihilation operators $ a_n ( k)$ and $ b_n ( k)$ which satisfy\footnote{We found the following orthogonality property useful: $\int_{0}^\infty d\rho\, \rho J_\alpha(\rho v)J_\alpha(\rho u)=\frac{\delta(u-v)}{u}$.}
\begin{equation}
[a_n( k),a_{n'}^\dagger( k')]=[b_n( k),b_{n'}^\dagger( k')]=\delta(\vec k-\vec{k}')\delta(k_\rho- k'_\rho)\delta_{n,n'}\,.
\end{equation}
In terms of these, the correctly normalised mode expansion for $\varphi$ reads
\begin{equation}
\label{eq:scalar_field_modes}
\begin{split}
\varphi=   \sum_{n=-\infty}^{\infty} \int dk_\rho \int d^{d-3} {\vec k} \,  \frac{\sqrt{k_\rho}}{(\sqrt{2\pi})^{d-2}\sqrt{2\omega}}  \Bigg[&  a_n( k) e^{i n \theta}J_{|n-\alpha|} \left(k_\rho\,\rho\right)e^{-i \omega t+i {\vec k}\cdot{\vec{\sigma}}} \\
&+ b_n^\dagger ( k) e^{-i n\theta}J_{|n+\alpha|} \left(k_\rho\,\rho\right)e^{i \omega t-i {\vec k}\cdot{\vec{\sigma}}}\Bigg].
\end{split}
\end{equation}

More generally, we can relax the assumption that the defect corresponds to a solenoid. In this case we just require that the integral of the charge density $Q \sim \int \phi \dot \phi^\dagger$ is finite near $\rho \sim 0$, and one can allow for a mild singular behaviour $\varphi \sim \rho^{-1+\epsilon}$ with $\epsilon >0$. In particular, restricting the range of $\alpha$ to lie in the interval $\alpha \in (0,1)$, we can allow for the Bessel function $J_{-\alpha}$ (for the $n=0$ mode) in addition to $J_{+\alpha}$, and $J_{\alpha-1}$ (for $n=1$) in addition to $J_{1-\alpha}$. In the most general case there will be a specific ladder operator corresponding to each Bessel $J$, i.e. we will have $a_0^{(-)}, a_0^{(+)},a_1^{(-)}, a_1^{(+)}$ and analogously for the ladder operators $b$. The modes with the $+$ sign correspond to the regular modes while the $-$ sign to the mildly divergent ones.
The only non-trivial commutators are
\begin{equation}
\begin{split}
[a_0^{(\pm)}( k),a_0^{(\pm)\dagger}( k')]=[a_1^{(\pm)}( k),a_1^{(\pm)\dagger}( k')]=\delta(\vec k-\vec{k}')\delta(k_\rho- k'_\rho)\,,
\end{split}
\end{equation}
and similarly for the $b$ ladder operators. In order to respect the commutation relation $[\varphi(x,t), \dot \varphi(x',t)]=i \delta^{(d-1)}(x-x')$, we need to introduce a specific normalization for the modes $n=0,1$ described by two free parameters $\xi, \tilde \xi \in [0,1]$ as in equations \eq{zero_mode_scalar}. In particular, if we choose $\xi =0$, only the regular mode $J_\alpha$ will occur, while for $\xi=1$ we will have only the singular one. The same happens for the mode $n=1$ and $\tilde \xi$.

From the mode expansion \eq{scalar_field_modes} and the modification due to the singular mode, it is straightforward to write down \eq{mod_exp_salar_defect}.

\subsection*{The massless scalar propagator}
From the field solution \eq{scalar_field_modes} we can easily compute the propagator. The Euclidean two-point function can be written as
\begin{equation}\label{eq:scalar_prop_JJ-0}
\begin{split}
& G_{S,\alpha,\xi,\tilde \xi} (x,x')  =   \sum_{m =1}^{+\infty} e^{i (m-\alpha) (\theta-\theta')} G^{(m-\alpha)} (x,x')+ \sum_{m =0}^{+\infty} e^{-i (m+\alpha) (\theta-\theta')} G^{(m+\alpha)} (x,x') \\
 &+ \xi \, e^{-i \alpha (\theta-\theta')}\left[G^{(-\alpha)} (x,x') - G^{(\alpha)} (x,x')\right] + \tilde \xi \, e^{i (1-\alpha)(\theta-\theta')}\left[G^{(\alpha-1)} (x,x') - G^{(1-\alpha)} (x,x')\right] ,
\end{split}
\end{equation} 
where we defined
\begin{equation}\label{eq:prop_mode_int}
\begin{split}
G^{(\nu)}_S(x,x') \equiv   \int d^{d-3} {\vec k} \,dk_\rho\,d k_\tau \, \, \frac{k_\rho}{  (2\pi)^{d-1}} \frac{e^{-ik_\tau (\tau-\tau')+i {\vec k \cdot (\vec \sigma-\vec \sigma ')}}}{k_\rho^2+\vec k^2 + k_\tau^2} J_{\nu} \left(k_\rho\,\rho\right)J_{\nu} \left(k_\rho\,\rho'\right).
\end{split}
\end{equation}
By employing the identity
\begin{equation}
\frac{1}{\kappa^2}= \int_0^\infty ds\, e^{-\kappa^2 s}\,, \qquad \kappa^2>0\,,
\end{equation}
and performing the Gaussian integration over $\vec k$ and $k_\tau$, we obtain
\begin{equation}
\begin{split}
G^{(\nu)}_S (x,x') = \int_0^{+\infty} \,ds \int_0^{+\infty} d k_\rho \, \frac{k_\rho}{  2^{d-1} \pi^{d/2}} \frac{1}{s^{d/2-1}}e^{-\frac{(\vec \sigma-\vec \sigma')^2+(\tau-\tau')^2}{4 s}} e^{-k_\rho^2 s}&  \\
\times J_{\nu} \left(k_\rho\,\rho\right)J_{\nu} \left(k_\rho\,\rho'\right)&\,.
\end{split}
\end{equation}
Now, we integrate over $k_\rho$ by using the identity \eq{Bessel-integral-2}, which gives
\begin{equation}
\label{eq:heat_kernel_prop_no_A}
\begin{split}
G^{(\nu)}_S (x,x') 
&= \frac{1}{2^{d}  \pi^{d/2}}  \int_0^\infty ds \, s^{d/2-2} e^{-s(\rho^2+\rho'^2+(\sigma-\sigma')^2)/4 }     I_{\nu} \left(\frac{s\,\rho\,\rho'}{2}\right).
\end{split}
\end{equation}
This is the form of the propagator that we employed to obtain the desired correlation functions and entanglement entropy in sections \ref{sec:scalar_corr_func} and \ref{sec:scalar-EE}. Nonetheless, the integral over~$s$ can be performed analytically by noting the following relation
\begin{equation}
I_\alpha(z)=e^{\mp i\alpha \pi/2} J_\alpha\left(z e^{\pm i \pi/2}\right).
\end{equation} 
At this point the integral over $s$ in \eq{heat_kernel_prop_no_A} can be done by employing the identity \eq{Bessel-integral-1}, and the result is
\begin{equation}
\label{two-point function_gaug_transf}
\begin{split}
G^{(\nu)}_S (x,x')
&=   \frac{\Gamma\left( \frac{d}{2}-1+\nu  \right)}{4\pi^{d/2} \Gamma\left( 1 + \nu  \right)} \left(\frac{1}{\rho \rho'}\right)^{\frac{d}{2}-1} \left(\frac{\rho \rho'}{\rho^2+\rho'^2+(\sigma-\sigma')^2}\right)^{\frac{d}{2}-1+\nu} \\
&\times \phantom{}_2 F_1\left( \frac{d-2}{4}+\frac{\nu}{2},\frac{d-2}{4}+\frac{\nu}{2}+ \frac{1}{2}; \nu +1; \frac{4\rho^2 \rho'^2}{(\rho^2+\rho'^2+(\sigma-\sigma')^2)^2}  \right).
\end{split}
\end{equation}
It is straightforward to show that this reproduces exactly (up to the $\theta$ dependence) the defect blocks \eqref{defectblock} with the coefficients $c_s$ defined in eqs.~\eqref{bulkdefcoeff}, \eqref{bulkdefcoeff1} and \eqref{bulkdefcoeff2}.

\section{The Fermion Propagator}
\label{app:fermion}

In this appendix we provide concrete expressions for the fermion mode expansion in $d=4$, and explicitly compute the propagator \eq{fermion-propagator-1}. We will do so by imposing that the components of the Dirac fermion obey the canonical equal-time anti-commutation relations $\left\{\psi_A(x,t),\psi_B^\dagger(x',t)\right\}=\delta^{(d-1)}(x-x')\delta_{AB}$, where $A, B = 1, \ldots, 4$ are spinor indices. 

It will be convenient to use the following Clifford algebra representation 
\begin{align}\label{eq:Clifford_basis}
\gamma^0 &= \left( \begin{array}{cc} i\sigma^3 & 0\\ 0 & -i\sigma^3 \end{array} \right), &\gamma^1 &= \left( \begin{array}{cc} 0 & i\mathds{1}_2\\ - i\mathds{1}_2& 0 \end{array} \right),  & \gamma^2 &= \left( \begin{array}{cc} -\sigma^2 & 0\\ 0 & \sigma^2 \end{array} \right),& \gamma^3 &= \left( \begin{array}{cc} \sigma^1 & 0\\ 0 & -\sigma^1 \end{array} \right), 
\end{align}
where $\sigma^{1,2,3}$ are the Pauli matrices and $\mathds 1_2$ is the $2\times 2$ identity matrix. To solve the Dirac equation, we make the ansatz
\begin{align}\label{eq:psiAnsatz}
\psi = e^{-i\omega t}e^{im\theta}e^{ik_\parallel\sigma} \left( \begin{array}{c} \Psi(\rho) \\ \pm \Psi(\rho)  \end{array} \right)\,, 
\end{align} 
where $\Psi$ is a two-component spinor with $\rho$-dependence only, $m\in\mathds Z +\frac{1}{2}$, and $\omega,k_\parallel \in \mathds R$.\footnote{In this subsection only we denote the momentum in the $\sigma$-direction along the defect by $k_\parallel$.} The Dirac equation in the basis \eq{Clifford_basis} reduces to two coupled first-order ordinary differential equations for the components of $\Psi$
\begin{subequations}
\begin{align}
-i(\omega\pm k_\parallel)\Psi_2+\left(\frac{d}{d\rho}-\frac{\nu}{\rho}\right)\Psi_1&=0\,,\\
-i(\omega\mp k_\parallel)\Psi_1+\left(\frac{d}{d\rho}+\frac{\nu+1}{\rho}\right)\Psi_2&=0\,, 
\end{align}
\end{subequations} 
 where $\nu=m-\alpha-\frac{1}{2}$. These two equations can be combined into Bessel's equations for $\Psi_1$ and $\Psi_2$, and their solutions can be written in terms of Bessel functions of the first kind, $J$, as follows
\begin{equation}
\Psi = c^1 \left(
\begin{array}{c}
J_{\nu }(\rho\,k_\rho  )\\
iB_\pm  J_{\nu +1}(\rho\,k_\rho )\\
\end{array}
\right) + c^2  \left(
\begin{array}{c}
 J_{-\nu}(\rho \,k_\rho ) \\
-iB_\pm  J_{-(\nu +1)}(\rho\,k_\rho)\\
\end{array}
\right),
\end{equation}
where $k_\rho = \sqrt{\omega^2-k_\parallel^2}$, $B_\pm \equiv\frac{k_\rho}{\omega\pm k_\parallel}$, and $c^{1,2}$ are arbitrary integration constants. A field configuration is physically admissible if it is less divergent than $\rho^{-1}$ as we approach the defect at $\rho=0$. This requires setting either $c^1=0$ or $c^2=0$ for all $n\equiv m-\frac{1}{2} \in \mathds Z\setminus\{0\}$. In the case of $n=0$, both solutions are admissible. For the solution with coefficient $c^1$, the first component $\Psi_1\sim \rho^{-\alpha}$ as $\rho \to 0$, whereas $\Psi_2$ is regular. For the solution with coefficient $c^2$, the second component $\Psi_2\sim \rho^{-1+\alpha}$, whereas $\Psi_1$ is regular. Note that both solutions have one component that diverges at the defect. As shown in~\cite{ALFORD1989140}, it is the former that corresponds to an infinitely long and infinitely thin solenoid. The most general solution keeps both modes with a parameter $\xi\in[0,1]$ interpolating between them as in \eq{n=0-modes}.

Taking a linear combination, one obtains the general solution to the Dirac equation
\begin{equation}
\begin{split}
\psi =& \sum_{n=-\infty}^\infty\hspace{-0.15cm}
{\vphantom{\sum}}'\sum_{s=1}^2 \int_{-\infty}^\infty \frac{d k_\parallel}{2\pi} \int_0^\infty d k_\rho\, \sqrt{\frac{k_\rho}{4\omega}}\, e^{-i\omega t}e^{+ik_\parallel \sigma} e^{+in\theta}e^{i\theta/2} a^s_n(k) u^s_{n,k}  \\
&+\sum_{n=-\infty}^\infty\hspace{-0.15cm}
{\vphantom{\sum}}'\sum_{s=1}^2 \int_{-\infty}^\infty \frac{d k_\parallel}{2\pi} \int_0^\infty d k_\rho\,\sqrt{\frac{k_\rho}{4\omega}}\, e^{+i\omega t}e^{-ik_\parallel\sigma}e^{+in\theta}e^{i\theta/2} b^{s\,*}_n(k) v^s_{n,k} \,,\label{eq:solDirac2}
\end{split}
\end{equation}
where now $\omega$ is understood to be a function of $k=(k_\rho, k_\parallel)$. The spinors are
\begin{align}
u^1_{n,k}&=\left( \begin{array}{c} C_+ \, J_{\varsigma_n(n-\alpha) }(\rho\,k_\rho  ) \\ i\varsigma_n C_-\, J_{\varsigma_n(n+1-\alpha)}(\rho\,k_\rho ) \\ C_+ \, J_{\varsigma_n(n-\alpha) }(\rho\,k_\rho  ) \\ i\varsigma_nC_- \, J_{\varsigma_n(n+1-\alpha)}(\rho\,k_\rho ) \end{array} \right),& u^2_{n,k}&=\left( \begin{array}{c} C_- \, J_{\varsigma_n(n-\alpha) }(\rho\,k_\rho  ) \\ i\varsigma_n C_+ \, J_{\varsigma_n(n+1-\alpha)}(\rho\,k_\rho ) \\ -C_- \, J_{\varsigma_n(n-\alpha) }(\rho\,k_\rho  ) \\ -i\varsigma_n C_+ \, J_{\varsigma_n(n+1-\alpha)}(\rho\,k_\rho ) \end{array} \right),\label{eq:u}
\end{align}
and 
\begin{align}
v^1_{n,k}&=\left( \begin{array}{c} C_+ \, J_{\varsigma_n(n-\alpha) }(\rho\,k_\rho  ) \\ -i\varsigma_n C_- \, J_{\varsigma_n(n+1-\alpha)}(\rho\,k_\rho ) \\ C_+\, J_{\varsigma_n(n-\alpha) }(\rho\,k_\rho  ) \\ -i\varsigma_n C_-\, J_{\varsigma_n(n+1-\alpha)}(\rho\,k_\rho )  \end{array} \right),& v^2_{n,k}&=\left( \begin{array}{c} C_-\, J_{\varsigma_n(n-\alpha) }(\rho\,k_\rho  ) \\ -i\varsigma_n C_+ \, J_{\varsigma_n(n+1-\alpha)}(\rho\,k_\rho ) \\ -C_-\, J_{\varsigma_n(n-\alpha) }(\rho\,k_\rho  ) \\ i\varsigma_n C_+ \, J_{\varsigma_n(n+1-\alpha)}(\rho\,k_\rho ) \end{array} \right),\label{eq:v}
\end{align}
where $C_\pm\equiv \sqrt{\omega \pm k_\parallel}$. In the above, $\varsigma_n=+1$ for $n\geq 1$ and $\varsigma_n=-1$ for $n\leq -1$, and $\sum'$ is an instruction to sum over both $n=0$ modes. More concretely, denote the spinors with $\varsigma_0=\pm 1$ by $u^{(\pm)s}_{0,k}$ and $v^{(\pm)s}_{0,k}$, and the ladder operators by $a_0^{(\pm)s}(k)$ and $b_0^{(\pm)s*}(k)$. As in the scalar case, one can introduce a parameter $\xi\in[0,1]$ which interpolates between the two $n=0$ modes. Then $\sum'$ means: sum the $\varsigma_0=+1$ mode with an extra overall factor of $\sqrt{1-\xi}$, and the $\varsigma_0=-1$ mode with an extra factor of $\sqrt{\xi}$. Promoting $a^{1,2}_{n}(k)$ and $b^{1,2}_{n}(k)$ to operators whose non-zero anti-commutators are 
\begin{equation}
\left\{a^s_{n}(k),a^{s'\,\dagger}_{n'}(k')\right\}=\left\{b^s_{n}(k),b^{s'\,\dagger}_{n'}(k')\right\}=\delta(k_\parallel-k_\parallel')\delta(k_\rho - k_\rho') \delta_{nn'}\delta^{ss'}\,,
\end{equation}
the components of the Dirac spinor then obey the canonical equal time commutation relations.

Using the explicit mode expansions, the fermion propagator
\begin{equation}
G_{F,\alpha,\xi}(x,x')_{AB}=\begin{cases} \langle\psi_A (t,\sigma,\rho,\theta) \bar \psi_B (t',\sigma',\rho',\theta') \rangle & \text{if } t>t'\,,\\
-\langle\bar \psi_B (t',\sigma',\rho',\theta') \psi_A (t,\sigma,\rho,\theta)  \rangle & \text{if } t'>t\,,
\end{cases}
\end{equation}
can be straightforwardly computed. Assuming $t>t'$,
\begin{multline}
G_{F,\alpha,\xi}(x,x')_{AB} =i\sum_n{}^{'} \int \frac{d k_\parallel}{2\pi} \int \frac{d k_\rho}{2\pi} \,\frac{k_\rho}{4\omega}\,
e^{-i\omega(t-t')}e^{ik_\parallel(\sigma-\sigma')}e^{in(\theta-\theta')}e^{i(\theta-\theta')/2} \\
\left((u^1_{n,k})_A (u^{1\dagger}_{n,k}\gamma^0)_{B}+(u^2_{n,k})_A (u_{n,k}^{2\dagger}\gamma^0)_{B}\right),
\end{multline}
and similarly for $t<t'$. Using the scalar mode expansion. It is straightforward to verify that 
\begin{multline}
G_{F,\alpha,\xi}(x,x')_{AB} =  - \gamma^\mu (\partial_\mu +\Omega_\mu -i A_\mu ) \\
\left( P_-\sum_n{}^{''}\mathcal{I}_{\varsigma_n(n-\alpha)} e^{in(\theta-\theta')}+P_+ \sum_n{}^{''}\mathcal{I}_{\varsigma_n(n+1-\alpha)} e^{in(\theta-\theta')}\right),
\end{multline}
where
\begin{equation}
\mathcal I_\nu=\int \frac{d k_\parallel}{2\pi} \int \frac{d k_\rho}{2\pi}  \, \frac{k_\rho}{2\omega} J_\nu (\rho k_\rho) J_{\nu} (\rho' k_\rho)\,e^{-i\omega(t-t')}e^{ik_\parallel(\sigma-\sigma')}e^{i(\theta-\theta')/2}\,, \label{eq:calI}
\end{equation}
and $\sum''$ is an instruction to sum over both $n=0$ modes, one with $\varsigma_0=+1$ and an extra overall factor of $1-\xi$, and the other with $\varsigma_0=-1$ and an extra factor of $\xi$. Comparing with \eq{scalar_prop_JJ-0}, one identifies 
\begin{subequations}
\begin{align}
\sum_n{}^{''}\mathcal{I}_{\varsigma_n(n-\alpha)} e^{in(\theta-\theta')} &= e^{i(\theta-\theta')/2} G_{S,\alpha,1-\xi,0}(x,x')\,,\\
\sum_n{}^{''}\mathcal{I}_{\varsigma_n(n+1-\alpha)} e^{in(\theta-\theta')} &= e^{-i(\theta-\theta')/2} G_{S,\alpha,0,\xi}(x,x')\,,
\end{align}
\end{subequations}
which gives precisely \eq{fermion-propagator-1} after applying the gauge transformation \eq{gauge_transf}.

\section{Useful Formulae}\label{app:formulae}

In this appendix, we collect identities that were used in the main body of the text.  The following integral identity involving a single Bessel-$J$ function
\begin{align}\label{eq:Bessel-integral-1}
\int_0^\infty ds\, s^{\lambda-1}\,e^{-p s}J_\alpha(a s) &= \left(\frac{a}{2p}\right)^\alpha\frac{\Gamma(\lambda+\alpha)}{p^\lambda\Gamma(\alpha+1)}{}_2F_1\left(\frac{\lambda+\alpha}{2},\frac{\lambda + \alpha +1}{2};\alpha+1;-\frac{a^2}{p^2}\right),
\end{align}
which is valid for Re$(\alpha+\lambda)>0$ and Re$(p\pm ia)>0$ was used in  the evaluation of the scalar propagator. In the same computation, we also encountered integrals involving products of $J_\alpha(a s)$, which required the following integral identity
\begin{align}\label{eq:Bessel-integral-2}
\int_0^\infty ds\, s\,e^{-p^2s^2}J_\alpha(a s)J_\alpha(bs)&= \frac{1}{2p^2} e^{-\frac{a^2+b^2}{4p^2}}I_\alpha\left(\frac{ab}{2p}\right),
\end{align}
where $I_\alpha(s) = e^{i\pi\alpha/2}J_{\alpha}(is)$ is the modified Bessel function of the first kind.  

Finally, we used the following identities involving sums of Bessel functions:
\begin{align}\label{eq:Bessel-sums-1}
\sum_{k=0}I_{k+\nu}(z) = \frac{1}{2(1-\nu)}\Big(e^{z}\int_0^z dt \, e^{-t}I_{\nu-1}(t) -z (I_{\nu-1}(z) + I_{\nu}(z))\Big),
\end{align}
and
\begin{align}\label{eq:Bessel-sums-2}
\sum_{k=1}^\infty (k+\nu)I_{k+\nu}(z) = \frac{z}{2}(I_{\nu+1}(z) + I_{\nu}(z))\,.
\end{align}

\bibliographystyle{JHEP}
\bibliography{monodromy-defects}
\end{document}